\documentclass[a4paper,11pt]{article}
\pdfoutput=1
\usepackage{jheppub}
\usepackage{graphicx,color}
\usepackage{amsmath,autobreak}
\usepackage{array}
\usepackage{ulem}
\newcolumntype{P}[1]{>{\centering\arraybackslash}p{#1}}
\allowdisplaybreaks
\RequirePackage{ifpdf}
\usepackage{amsmath}
\usepackage{mathtools}

\usepackage{soul} % for strikethrough
\usepackage{multirow, caption, booktabs} % for multirow
\usepackage[final]{pdfpages}
\usepackage{ifpdf}
\usepackage{slashed}

\usepackage{hyperref}

\usepackage{color}
\usepackage{graphics}

\usepackage{etoolbox}
\usepackage{fixmath}

\usepackage{caption}
\usepackage{subcaption}
\usepackage{amsfonts}

\usepackage{multirow}
\usepackage{epstopdf}

\usepackage{relsize}
\usepackage{float}

\usepackage[table]{xcolor}
\usepackage{tabularx}

\allowdisplaybreaks
\newcommand{\ben}{\begin{enumerate}}
\newcommand{\een}{\end{enumerate}}
\newcommand{\beq}{\begin{equation}}
\newcommand{\eeq}{\end{equation}}
\newcommand{\bal}{\begin{align}}
\newcommand{\eal}{\end{align}}

\newcommand{\bea}{\begin{eqnarray}}
\newcommand{\eea}{\end{eqnarray}}
\newcommand{\nn}{\nonumber}
\newcommand*\oline[1]{%
   \vbox{%
     \hrule height 0.5pt%                  % Line above with certain width
     \kern0.25ex%                          % Distance between line and content
     \hbox{%
       \kern-0.2em%                        % Distance between content and left side of box, negative values for lines shorter than content
       \ifmmode#1\else\ensuremath{#1}\fi%  % The content, typeset in dependence of mode
       \kern-0.05em%                        % Distance between content and left side of box, negative values for lines shorter than content
     }% end of hbox
   }% end of vbox
}

\newcommand{\Nb}{\oline{N}}

\newcommand{\dFAoNA }{\frac{d_F^{abcd}d_A^{abcd}}{N_A}}
\newcommand{\dFFoNA }{\frac{d_F^{abcd}d_F^{abcd}}{N_A}}

\newcommand{\as}{a_s}

\def\z#1{\zeta_{#1}}

\newcommand{\Ca}{C_A}
\newcommand{\Cf}{C_F}
\newcommand{\nf}{n_F}
\newcommand{\CA}{C_A}
\newcommand{\CF}{C_F}
\newcommand{\NF}{n_f}

\def\Dm1{{{\delta(1-z)}}}

\def\nf{{n^{}_{\! f}}}

\def\g0#1DY{{g_{0#1}^{DY}}}

\def\lnmW1{{{\ln (1-\omega)}}}

\usepackage{tikz}
\usetikzlibrary{positioning,arrows}
\usetikzlibrary{decorations.pathmorphing}
\usetikzlibrary{decorations.markings}
\usetikzlibrary{shapes.geometric}
\usepackage{endnotes}
\tikzset{
    vector/.style={decorate, decoration={snake}, draw},
    provector/.style={decorate, decoration={snake,amplitude=2.5pt}, draw},
    antivector/.style={decorate, decoration={snake,amplitude=-2.5pt}, draw},
    fermion/.style={draw=black,
      postaction={decorate},decoration={markings,mark=at position .55
        with {\arrow[draw=black]{>}}}},
    fermionbar/.style={draw=black, postaction={decorate},
                       decoration={markings,mark=at position .55 with {\arrow[draw=black]{<}}}},
    fermionnoarrow/.style={draw=black},
    gluon/.style={decorate, draw=black,decoration={coil,amplitude=4pt, segment length=6pt}},
    scalar/.style={dashed,draw=black,
      postaction={decorate},decoration={markings,mark=at position .55
        with {\arrow[draw=black]{>}}}},
    scalarbar/.style={dashed,draw=black,
      postaction={decorate},decoration={markings,mark=at position .55
        with {\arrow[draw=black]{<}}}},
    scalarnoarrow/.style={dashed,draw=black},
    electron/.style={draw=black,
      postaction={decorate},decoration={markings,mark=at position .55
        with {\arrow[draw=black]{>}}}},
    bigvector/.style={decorate, decoration={snake,amplitude=4pt}, draw},
}

\usepackage{amsmath}
\usepackage{autobreak}
\allowdisplaybreaks
\usepackage{ulem}

\usepackage{etoolbox}
\usepackage{fixmath}

\allowdisplaybreaks
\usepackage[latin1]{inputenc}

\def\z#1{\zeta_{#1}}

\def\Dm1{{{\delta(1-z)}}}

\def\nf{{n^{}_{\! f}}}

\def\g0#1DY{{g_{0#1}^{DY}}}

\def\lnmW1{{{\ln (1-\omega)}}}

\usepackage{tikz}
\usetikzlibrary{positioning,arrows}
\usetikzlibrary{decorations.pathmorphing}
\usetikzlibrary{decorations.markings}
\usetikzlibrary{shapes.geometric}

\usepackage{endnotes}
\title{Next-to SV resummed Drell-Yan cross section beyond leading-logarithm}
\author{A. H. Ajjath,}
\author{Pooja Mukherjee,}
\author{V. Ravindran,}
\author{Aparna Sankar,}
\author{Surabhi Tiwari,}

\affiliation{The Institute of Mathematical Sciences, HBNI, IV Cross Road, Taramani, Chennai 600113, India}
\emailAdd{ajjathah@imsc.res.in}
\emailAdd{poojamukherjee@imsc.res.in}
\emailAdd{ravindra@imsc.res.in}
\emailAdd{aparnas@imsc.res.in}
\emailAdd{surabhit@imsc.res.in}
\abstract{We present the resummed predictions for inclusive cross section for  Drell-Yan (DY) 
production up to next-to-next-to leading logarithmic ($\rm \overline{ NNLL}$) accuracy taking into account both
soft virtual (SV) and next-to SV (NSV) threshold logarithms.   We restrict ourselves
to resummed contributions only from quark anti-quark ($q \bar q$) initiated channels.   The resummation is performed in Mellin-$N$ space.  We derive the $N$-dependent coefficients  and the  $N$-independent constants to desired accuracy
for our study.  The resummed results are matched through the minimal prescription procedure 
with the fixed order results.   We find that the resummation, taking into account the NSV terms, appreciably increases the cross section while decreasing the sensitivity to renormalisation scale. \textcolor{black}{We observe that, at 13 TeV LHC energies, the SV+NSV resummation at $\rm \overline{ NLL}  (\rm \overline{NNLL})$ gives about 8\% (2\%) corrections respectively to the NLO (NNLO) results for the considered $Q$ range: 150-3500 GeV}.  In addition, the absence of quark gluon initiated 
contributions to NSV part in the resummed terms leaves large
factorisation scale dependence indicating their importance
at NSV level.  We also study
the numerical impact of $N$-independent constants and explore the ambiguity involved in exponentiating them.  
Finally we present our predictions for the neutral Drell-Yan process at various center of mass of energies. 
}

\begin{document} 
%%%Preprint
\preprint{IMSc/2021/07/05}
\maketitle

\setstcolor{red}
%%%Introduction
\section{Introduction} \label{intro}

%%general intro
Standard Model (SM) has been extremely successful in describing the physics of 
elementary particles. Thanks to precise predictions of various observables from SM 
and their measurements at the collider experiments with unprecedented accuracy,
we could validate the SM  and at the same time set stringent constraints on the parameters
present in various beyond SM (BSM) scenarios.     
While there have been strenuous efforts in search of  new physics signatures at the large hadron collider (LHC), it is important to improve the level of precision in SM and BSM predictions to arrive at sensible conclusions.   Precise predictions of observables
require the use of complex mathematical techniques and a deeper understanding 
of the underlying theory. The spin-offs include new developments in
various branches of mathematics and other fields, and in addition, the perturbative predictions
dealing with Feynman loop and phase space integrals demonstrate rich
mathematical structure in gauge theories.  In particular, these results have shed light
on the underlying structure of the ultraviolet (UV) and infrared (IR) sectors of the SM.     

Among innumerous final states produced in hadron collisions, leptons are relatively easy to observe
due to the clean environment 
and the corresponding measurements are less plagued by experimental uncertainties.  
The production of a pair of leptons, called Drell-Yan (DY) production 
is customarily used  for luminosity monitoring at the hadron colliders. 
Theoretically, for very long, the observables in DY production belong to the category of 
``well studied" quantities in the SM and as well as in various BSMs.    
Note that the next-to-next-to leading order (NNLO) quantum chromodynamics (QCD) 
correction \cite{Hamberg:1990np,Matsuura:1990ba,Harlander:2002wh} to this
process was computed more than three decades ago, see also \cite{Altarelli:1978id,Altarelli:1979ub,Matsuura:1987wt,Matsuura:1988nd,Matsuura:1988sm,Matsuura:1990ba,Hamberg:1990np,vanNeerven:1991gh,Harlander:2002wh,Moch:2005ky,Ravindran:2006cg,deFlorian:2012za,Ahmed:2014cla,Ahmed:2015qda,Catani:2014uta,Li:2014afw,Duhr:2020seh}.  Similar results are also available in certain BSMs, see
\cite{Ahmed:2016qhu,Banerjee:2018vvb,Majhi:2010zg}.
More recently, a series of results on inclusive cross sections for the production of
a pair of leptons, single Z/$W^\pm$  at N$^3$LO in perturbative QCD has become available \cite{Duhr:2020seh}.  
These corrections \cite{Duhr:2020seh} are already found to be tiny, and at the invariant mass $Q=150$ GeV of a
pair of leptons, they reduce the cross section by   
$1\%$.  The renormalisation and factorisation scale uncertainties as well as the uncertainties from the choice of PDFs give about $2.5\%$. 
 
Dedicated efforts like in \cite{Duhr:2020seh} to obtain 
perturbative QCD results provide a theoretical laboratory  
to understand the structure of the perturbation series.  
Due to the complexity involved in performing many body phase space integrals in higher order computations,
one resorts to the method of threshold expansion.  For example, at every perturbative order in the strong coupling
constant, the Feynman diagrams are computed as a series expansion around the threshold region denoted by 
$Q^2\approx \hat s$, where 
 $Q$ is the invariant mass of the pair of leptons produced in the partonic reaction whose center of mass energy is
$\sqrt{\hat s}$.   Such an expansion not only provides reliable estimates of the higher order effects but also 
shed light on the logarithmic structure in higher order perturbative results. 
The leading terms in the threshold region contain
contributions from virtual subprocesses as well as from soft gluons from real emissions.  These are often called
soft plus virtual contributions (SV). 
The SV terms at third order were known for some time, see
\cite{Moch:2005ky,Ravindran:2005vv,Ravindran:2006cg,deFlorian:2012za,Ahmed:2014cha, Kumar:2014uwa,
Ahmed:2014cla,Catani:2014uta,Li:2014bfa}.  In addition, 
using the resummation framework developed in \cite{Sterman:1986aj,Catani:1989ne} for threshold logarithms in SV contributions,  
several numerical studies were carried out to NNLL accuracy to improve the predictions,  see \cite{Moch:2005ba,Bonvini:2010ny,Bonvini:2012sh,Catani:2014uta}.
In \cite{Ajjath:2020rci}, we reported the numerical impact of threshold corrections within the resummation framework.   
We found that the inclusion of large threshold logarithms to N$^3$LL accuracy further reduces theoretical uncertainties. 

The subleading terms in the threshold expansion contain logarithms of the form $\ln^j(1-z),j\ge0$
and numerically they are found to be as important as leading SV terms in the expansion, see \cite{Anastasiou:2014lda,Bonvini:2014joa,Das:2020adl}
in the context of Higgs production. These are called next-to-soft virtual (NSV) logarithms.
There have been several dedicated studies to understand the structure of these logarithms in inclusive reactions
at higher orders and efforts to resum them like one does for SV terms, see 
\cite{Laenen:2008ux,Grunberg:2009yi,Moch:2009hr,Laenen:2010kp,Laenen:2010uz,Bonocore:2014wua,Bonocore:2015esa,Beneke:2019oqx,Beneke:2019mua,Bonocore:2016awd,DelDuca:2017twk,deFlorian:2014vta,Das:2020adl}. 
Using the resummation framework of NSV terms at LL proposed in \cite{Laenen:2008ux},   their numerical impact was studied  in \cite{vanBeekveld:2021hhv} taking into account SV terms at N$^3$LL for DY and Higgs boson productions.
Similar studies were done for the scalar and pseudo scalar Higgs boson productions in \cite{Bonvini:2014joa,Ahmed:2016otz}.   All these studies were at LL level as far as NSV logarithms are concerned  and also
restricting to diagonal partonic channels, namely quark anti-quark for Drell-Yan, gluon fusion or bottom quark annihilation for Higgs boson productions.  

Recently, in \cite{Ajjath:2020ulr}, we set up a formalism for the first time to study all order structure
of these NSV logarithms in order to go beyond LL approximation.   While NSV logarithms show
up both in diagonal and off-diagonal partonic channels, we have restricted to only to the former.
We found that
unlike SV logarithms, the NSV ones were controlled in addition to the process independent anomalous dimensions, the functions that depend on the process under consideration through certain differential
equations.  The latter allowed  us to systematically resum NSV logarithms in Mellin $N$ space to all orders along with SV ones to obtain results at $\overline{\rm{N^nLL}}, n\ge 0$ accuracy.   In order to distinguish between SV and SV+NSV resummed results, we denote the NSV included results by $\overline{\rm N^nLL}$. In this article, we study the numerical impact of NSV logarithms in the invariant mass distribution of a pair of leptons in DY process at the LHC up to $\overline{\rm{NNLL}}$ accuracy. 
%In this article, we report their numerical impact on the invariant mass distribution of pair of leptons in DY process at the LHC.  

\textcolor{black} {
%The goal of the present article is to study the numerical impact of NSV logarithms on the invariant mass distribution of a pair of leptons in DY process at the LHC, at NNLL accuracy and match it to the known NNLO results.
The paper is structured as follows. In  Sec.\ \ref{sec:DY}, we briefly describe the theoretical framework for computing the invariant mass distribution of a pair of leptons in DY process,
%. We also discuss the threshold expansion of partonic coefficient functions (CFs) 
taking into account the NSV effects. Further in Sec.\ \ref{sec:resummation}, we review the formalism given in \cite{Ajjath:2020ulr} for computing the SV+NSV resummed cross section of di-lepton production in DY process. In addition, we also discuss different resummation prescriptions to explore the ambiguity involved in exponentiating the $N$-independent constants.  In  Sec.\ \ref{sec:pheno}, we study the phenomenological aspects of NSV logarithms in great detail and present our findings %along with the estimation of scale uncertainties
and finally we conclude in Sec.\ \ref{sec:concl}}.

\section{{\color{black}Theoretical Framework}}\label{sec:DY}
%\section{Drell-Yan process}\label{sec:DY}
In the QCD improved parton model, the invariant mass distribution of a pair of leptons produced in 
hadron colliders can be
expressed  as a convolution of perturbatively calculable coefficient
functions (CFs), $\Delta_{ab}$, and non-perturbative
flux $\tilde \Phi_{ab}$. That is,
\begin{eqnarray}
{d \sigma \over dQ}(q^2,\tau) = \sigma^{(0)}_{DY} \int_{\tau}^1 {d z \over z } \tilde \Phi_{ab}\left({\tau \over z},\mu_F^2 \right) \Delta_{ab}(q^2,\mu_F^2.z) \,.
\end{eqnarray}
Here $a,b=q,\overline q,g$ refer to incoming partonic states
and $\sigma^{(0)}_{DY}$ is the born cross section:
\begin{align}
\sigma_{DY}^{(0)} &= \frac{2\pi }{n_c } \bigg[ \frac{Q}{S} {\cal F}^{(0)} \bigg]\,, 
%\sigma_Z^{(0)} &=  \frac{2\pi }{n_c } \bigg[ \frac{\pi\alpha}{8s_w^2 c_w^2 S}  \bigg]\,, \nn\\
%\sigma_{W^\pm}^{(0)} &= \frac{2\pi }{n_c } \bigg[ \frac{\pi\alpha}{4 s_w^2 S}  \bigg]\,,
\end{align}
with $Q=\sqrt{q^2}$ being the invariant mass of the lepton pairs and $n_c=3$ in QCD. The factor ${\cal F}^{(0)}$ is found to be
\begin{align}
\begin{autobreak}
{\cal F}^{(0)} =
{4 \alpha^2 \over 3 q^2} \Bigg[Q_q^2
- {2 q^2 (q^2-M_Z^2) \over  \left((q^2-M_Z^2)^2
+ M_Z^2 \Gamma_Z^2\right) c_w^2 s_w^2} Q_q g_e^V g_q^V
+ {q^4 \over  \left((q^2-M_Z^2)^2+M_Z^2 \Gamma_Z^2\right) c_w^4 s_w^4}\Big((g_e^V)^2
+ (g_e^A)^2\Big)\Big((g_q^V)^2+(g_q^A)^2\Big) \Bigg]\,.
\end{autobreak}
\end{align}
with $\alpha$ being the fine structure constant and $c_w,s_w$ are respectively the sine and cosine of Weinberg angle. $M_Z$ and $\Gamma_Z$ are the mass and the decay width of the $Z$-boson. Also,
\begin{align}
 g_a^A = -\frac{1}{2} T_a^3 \,, \qquad g_a^V = \frac{1}{2} T_a^3  - s_w^2 Q_a \,,
\end{align}
where $Q_a$ being the electric charge and $T_a^3$ is the weak isospin of the electron or quarks.
The flux $\tilde \Phi_{ab}$  is defined in terms of parton distribution functions (PDF) $f_a,f_b$ of incoming partons $a$ and $b$ respectively at the
factorisation scale $\mu_F$:
\begin{eqnarray}
\tilde \Phi_{ab}\left({\tau \over z},\mu_F^2\right) = \int_{\tau \over z}^1 {d y \over y } f_a(y,\mu_F^2) f_b\left({\tau \over z y},
\mu_F^2\right) 
\end{eqnarray}
where $\tau=q^2/S$ is the hadronic scaling variable with $S$ being the square of hadronic center of mass energy.   The fits of non-perturbative PDFs are available to NNLO level while the 
CFs are perturbatively calculable in powers of renormalized strong coupling constant, $a_s = g_s^2/16 \pi^2$:
\begin{eqnarray}
\Delta_{ab}(q^2,\mu_F^2,z) = \sum_{i=0}^\infty a_s^i(\mu_R^2) \Delta_{ab}^{(i)}(q^2,\mu_R^2,\mu_F^2,z)
\end{eqnarray}
where $g_s$ is the QCD coupling constant and $\mu_R$ is the renormalisation scale. Higher order corrections from  QCD are inevitable at hadron colliders
as they are often large and they reduce uncertainties resulting from the scales $\mu_R, \mu_F$, the choice
of PDFs and $a_s$.
Note that DY predictions are  stable with respect to factorization ($\mu_F$) and renormalisation ($\mu_R$) scales 
already at NNLO in QCD. One finds $2\%$ uncertainty in the predictions at NNLO for a canonical 
variation of factorization and renormalisation scales compared to NLO where it is about $9.2\%$. \textcolor{black}{Similarly,
the K-factor improves marginally from $1.25$ at NLO to $1.28$ at NNLO. 
At third order, the scale uncertainties can be determined
from previous order DY results.  However,  
such estimates make sense only if  we include the scale independent parts which originate genuinely at third order. 
The recent third order results  \cite{Duhr:2020seh} predict decrease in the cross section by less than $1\%$ for $Q=150$ GeV and the
uncertainties from scales and from the choice of PDFs give about $2.5\%$}. While this is a great improvement in the predictions,
it is important to estimate other missing higher order effects.

\subsection{Threshold expansion}
Computing CFs beyond N$^3$LO are highly challenging, hence one can look for alternative approaches to identify the dominant contributions to the CFs.  
In the case of leptons with high invariant mass, the threshold expansion of
CFs around $z \approx 1$ was observed to be a good alternative to exact computation. 
Here, $z$ is the partonic scaling variable defined by 
$z=q^2/\hat s$, with $\hat s$ the {square of} partonic center of mass energy. At the threshold, we decompose the CFs as 
\begin{eqnarray}\label{deltadecompose}
\Delta_{ab}(q^2,\mu_F^2,z) =  \delta_{a b} \Delta_{a\overline a}^{SV}(q^2,\mu_F^2,z) + \Delta_{ab}^{reg}(q^2,\mu_F^2,z)\,.
\end{eqnarray}
Here $\Delta_{a\overline a}^{SV}$ denotes the soft-virtual (SV)            
corrections which comprises pure virtual contributions from $q +\overline q \rightarrow l^+ l^-$ and
leading threshold contributions from quark anti-quark initiated partonic channels with at least one emission of on-shell parton. 
The former depends on the scale $z$ through $\delta(1-z)$ while the latter through both $\delta(1-z)$
and plus distributions ${\cal D}_k(z)$ defined by 
\begin{eqnarray}
{\cal D}_k(z) = \left( {\ln^k(1-z) \over 1-z} \right)_+ \,,
\end{eqnarray}
and are integrable with any regular function $f(z)$:
\begin{eqnarray}
\int_0^1 dz ~f(z) \left( {\ln^k(1-z) \over 1-z} \right)_+  = \int_0^1 dz ~\big(f(z)-f(1)\big) \left( {\ln^k(1-z) \over 1-z} \right)\,.
\end{eqnarray}
The second term in \eqref{deltadecompose}, $\Delta_{ab}^{reg}$, refers to regular CFs, which contain 
%terms other than $\delta(1-z)$ and plus distributions, which include
terms of the form $(1-z)^m \ln^k(1-z) ,m,k=0,1,\cdots,\infty$ with rational and irrational constants. 
Expanding both SV and regular CFs perturbatively in powers of $a_s$,
\begin{eqnarray}
\Delta^J_{ab}(q^2,\mu_F^2,z) = \sum_{i=0}^\infty a_s^i(\mu_R^2) \Delta_{ab}^{J,(i)}(q^2,\mu_R^2,\mu_F^2,z)
,\, \quad \quad J=SV,{ reg}
\end{eqnarray}
we find
\begin{eqnarray}\label{SV}
\Delta^{SV,(i)}_{ab}(z) = \delta_{ab}\left( \Delta_{a \overline a,\delta}~ \delta(1-z)
+ \sum_{k=0}^{2i-1}  \Delta^{(i)}_{ a \overline a,{\cal D}_k} {\cal D}_k(z) \right)\, ,
\label{delsvexp}
\end{eqnarray}
and 
\begin{eqnarray}
\Delta^{reg,(i)}_{ab}(z) =  
%\sum_{k=0}^{2i-1}  \Delta^{reg,(i)}_{ a b,k} \ln^k(1-z) +
\sum_{k=0}^{2i-1} \sum_{l=0}^\infty  \Delta^{reg,(i)}_{ a b,l,k} (1-z)^l \ln^k(1-z) .
\label{delregexp}
\end{eqnarray}
The above expansion is called threshold expansion.
The systematic threshold expansion of CFs in partonic scaling variable $z$ will be useful provided
the partonic flux $\tilde \Phi_{ab}(\tau/z)$ that multiplies them to give hadronic cross sections also dominates in the 
same region for a given hadronic scaling variable $\tau$.

% on SV
In general, the CFs in inclusive cross sections such as DY and Higgs productions, 
the energy scales $q^2,\mu_R^2$ and $\mu_F^2$ appear 
as logarithms, in addition to the  partonic scaling variable $z$ appearing through
$\delta(1-z)$, plus distributions ${\cal D}_k(z)$ and regular functions of $z$.
The coefficients of these terms are perturbatively computable and are controlled by 
set of differential    
equations that depend on the UV and IR anomalous dimensions.  
The solutions to these equations
demonstrate rich universal structure which can be exploited to understand the structure of the coefficients to all orders
in perturbation theory.   The IR structure of multi-loop amplitudes 
beyond two loops \cite{Becher:2009cu,Becher:2009qa,
Gardi:2009qi,Catani:1998bh} (see \cite{Ajjath:2019vmf,H:2019nsw} for a QFT with mixed gauge groups),
of inclusive cross sections to third order
\cite{Hamberg:1990np,Harlander:2002wh,Duhr:2020seh,Anastasiou:2015vya, Mistlberger:2018etf,Duhr:2019kwi}
provide better understanding of CFs. 
For complete list of Higgs production in gluon fusion see \cite{Georgi:1977gs,Graudenz:1992pv,Djouadi:1991tka,Spira:1995rr,Catani:2001ic,Harlander:2001is,Anastasiou:2002yz,Harlander:2002wh,Catani:2003zt,Ravindran:2003um,Moch:2005ky,Ravindran:2006cg,deFlorian:2012za,Bonvini:2014jma,deFlorian:2014vta,Anastasiou:2014vaa,Idilbi:2005ni,Li:2014afw,Anastasiou:2015yha,Anastasiou:2015vya,Das:2020adl} 
and \cite{Altarelli:1978id,Altarelli:1979ub,Matsuura:1987wt,Matsuura:1988nd,Matsuura:1988sm,Matsuura:1990ba,Hamberg:1990np,vanNeerven:1991gh,Harlander:2002wh,Moch:2005ky,Ravindran:2006cg,deFlorian:2012za,Ahmed:2014cla,Ahmed:2015qda,Catani:2014uta,Li:2014afw,Duhr:2020seh} for Drell-Yan production.

Among the aforementioned terms that contribute to CFs, the SV terms are known for several observables. In particular for DY and Higgs productions beyond second order,
see \cite{Moch:2005ky,Ravindran:2005vv,Ravindran:2006cg,deFlorian:2012za,Ahmed:2014cha, Kumar:2014uwa,
Ahmed:2014cla,Catani:2014uta,Li:2014bfa}.
We obtain these results from process dependent pure virtual subprocesses  and soft gluons from real emissions in the threshold region. The latter is a universal quantity in such a sense that they do not depend on the hard process under study, but only on the nature of incoming states. The soft and collinear modes in a scattering process can be captured at the Lagrangian level using effective theory approach.  For example, soft-collinear effective theory (SCET)  \cite{Bauer:2000yr,Bauer:2001yt,Bauer:2002nz}  provides a convenient framework to compute the SV results order by order in perturbation theory.  In addition, the intrinsic scales in the theory can be used
to set up renormalisation group equations whose solutions sum up large logarithms from threshold regions to all orders.

{\color{black}When SV terms are convoluted with the appropriate PDFs to obtain hadronic cross sections, 
one finds that they 
%not only dominate over other contributions but also 
give large contributions at every order,  
questioning the reliability of the predictions from the truncated series}.
This was successfully resolved in the seminal works by
Sterman \cite{Sterman:1986aj} and Catani and Trentedue \cite{Catani:1989ne}
through reorganisation of the large logarithms to all order in the perturbative series, called
the threshold resummation.  There is a vast literature on this which is applied to variety of 
processes, see  
\cite{Catani:1996yz,Moch:2005ba,Bonvini:2012an,Bonvini:2014joa,Bonvini:2014tea,Bonvini:2016frm} for Higgs production in gluon fusion,
\cite{Bonvini:2016fgf,H:2019dcl} for bottom quark annihilation, for DY \cite{Moch:2005ba,Bonvini:2010ny,Bonvini:2012sh,
H.:2020ecd,Catani:2014uta} and for DIS and SIA of $e^+ e^-$ \cite{Cacciari:2001cw}.  
Threshold resummation is conveniently performed in Mellin space where the conjugate variable to $z$ is $N$.  In Mellin space, all the
$z$-space convolutions become normal products. 
The threshold limit in Mellin space is when $N$ goes large, which corresponds to $z \rightarrow 1$ in $z$-space.  Due to smallness of $a_s(\mu_R^2)$,  one finds that the exponent in the $N$ space at each order 
in $a_s(\mu_R^2)$ contains ${\mathcal O}(1)$ terms defined by   
$\omega = 2 a_s(\mu_R^2) \beta_0 \ln N$.  Such terms spoil the truncation of the perturbative series.  
Using renormalisation group improved solution to RG of $a_s$, one can reorganise the perturbative series
in the exponent wherein $\omega$ terms are summed up at every order in $a_s$.      
Following \cite{Sterman:1986aj,Catani:1989ne} one finds
that for $\Delta_{c \overline c,N} = \int_0^1 dz z^{N-1} \Delta_{c \overline c}(z)$,
\begin{eqnarray}
\label{resumgen}
\lim_{N\rightarrow \infty} \ln \Delta_{c\overline c,N}^{SV} =
 \ln \tilde g^c_0(a_s(\mu_R^2))+
\ln N g^c_1(\omega) + \sum_{i=0}^\infty a_s^i(\mu_R^2) g^c_{i+2}(\omega) \,,
\end{eqnarray}
where $\tilde g^{c}_0(a_s(\mu_R^2))$ is $N$ independent.
Inclusion of  successive terms in \eqref{resumgen} predicts the leading-logarithms (LL),
next-to-leading (NLL) etc. logarithms to all orders in $a_s$.
The exponents $g^{c}_i(\omega)$ depend on process independent/universal
IR anomalous dimensions while the constant  $\tilde g^c_0$ depends on the specific hard process.
Results for the resummation of threshold logarithms in $N$ space up to third order are available for variety of
inclusive processes such as DY and Higgs productions to 
perform threshold resummation to next-to-next-to-next-to leading logarithmic (N$^3$LL) accuracy \cite{Catani:2014uta,Moch:2005ba,H.:2020ecd,H:2019dcl}.
Threshold resummation is also found to play important role for differential observables like rapidity \cite{Catani:1989ne,Westmark:2017uig,Banerjee:2017cfc,Banerjee:2018vvb,Lustermans:2019cau}. Inclusion of these effects are  
shown to improve the fixed order results.
\subsection{Next-to SV}
Perturbative predictions of beyond SV terms are available for partonic sub processes up to third order for a  variety of 
hadronic cross sections, namely Drell-Yan production and bottom quark as well as gluon initiated Higgs boson 
productions at the hadron colliders.  Like SV terms, these results not only play an important role to precisely predict
the respective observables, but also shed light on the structure of beyond SV terms in the
threshold expansion at higher orders.  Among these,  let us consider a class of leading terms:
%{\color{red}
%\begin{eqnarray}
%\label{DeltaR}
%\Delta_{ab}^{NSV}(z) = \sum_{k=0}^\infty \Delta_{ab}^{reg,(k)} \ln^k(1-z) \,.
%\end{eqnarray}}
\begin{eqnarray}\label{NSV}
\Delta_{ab}^{NSV}(z) = \sum_{i=0}^\infty a_s^i(\mu_R^2) \Delta_{ab}^{NSV,(i)}(z)
,\, 
\end{eqnarray}
where $\Delta_{ab}^{NSV,(i)}(z)$ is defined by setting $l=0$ in \eqref{delregexp},i.e.,
\begin{eqnarray}
\label{DeltaR}
\Delta_{ab}^{NSV,(i)}(z) = \sum_{k=0}^{2 i-1} \Delta_{ab,0,k}^{reg,(i)} \ln^k(1-z) \,.
\end{eqnarray}
%\qquad See def given in 2.12 and 3.2\\}
These contributions are often called  next-to SV (NSV) or next to leading power (NLP) contributions.  
There have been several studies to understand the NSV terms 
in inclusive processes \cite{Laenen:2008ux,Laenen:2010kp,Laenen:2010uz,Bonocore:2014wua,Bonocore:2015esa,Beneke:2019oqx,Beneke:2019mua,Bonocore:2016awd,DelDuca:2017twk}.
The physical evolution equation  was exploited earlier in the work by \cite{Grunberg:2009yi} to understand the effect of these terms.
A remarkable development was made by Moch and Vogt
in \cite{Moch:2009hr} (and \cite{deFlorian:2014vta,Das:2020adl})
using  the  physical evolution kernels (PEK) and the fixed order results that are available for DIS, 
semi-inclusive $e^+ e^-$ annihilation and Drell-Yan production of
a pair of leptons in hadron collisions.
They found that in kernels that govern the physical equations, 
there is an enhancement of single-logarithms at large $z$ up to third order.
Conjecturing that it will hold true to all orders around $z=1$,
the logarithms were systematically resummed to all orders exactly like the way of SV resummation.
In addition, absence of certain powers of $\ln(1-z)$ terms in the kernel at a given order in $a_s$ 
can be used \cite{Moch:2009hr} to predict certain next-to SV logarithms at  higher orders.

Recently, in \cite{Ajjath:2020ulr}, we investigated  the structure of NSV terms present in the 
quark anti-quark initiated channels in the inclusive production of pair of leptons in Drell-Yan process and 
gluon/bottom anti-bottom initiated ones for Higgs boson production. \textcolor{black}{We also analyzed the all-order perturbative structure of the NSV logarithms in the
coefficient functions of deep inelastic scattering (DIS) and semi-inclusive e$^+$e$^-$ annihilation
(SIA) processes in \cite{Ajjath:2020sjk}. The formalism is even extended in the context of rapidity distributions to study the all-order behaviour of the NSV terms in addition to the SV distributions in the aforementioned threshold processes, namely Drell-Yan  and Higgs production through gluon fusion and bottom quark annihilation in \cite{Ajjath:2020lwb}.} 
We used the well known factorisation properties and renormalisation group invariance along with certain universal structure
of real and virtual contributions obtained through Sudakov K+G equation.
Like, SV terms, NSV terms do demonstrate rich perturbative structure with certain universal anomalous dimensions. 
We found that the NSV logarithms in Mellin space can also be resummed in a systematic fashion to all orders in perturbation theory.
Fixed order results known up to third order for DY productions can be used to determine the threshold exponents from the NSV logarithms
with third order logarithmic accuracy.  The present article explores the numerical impact of these resummed results taking into account
both SV and NSV logarithms in the quark anti-quark initiated channels for the DY process at the LHC.

%%%%%%%%%%%%%%%%%%%%%%%%%%%%%%%%%%%%%%%%%%%%%%%%%
\section{Resummation of SV+NSV}\label{sec:resummation}
%%%%%%%%%%%%%%%%%%%%%%%%%%%%%%%%%%%%%%%%%%%%%%%%
In \cite{Ajjath:2020ulr}, some of us have developed a theoretical formalism to systematically resum the NSV contributions 
in the diagonal channels of inclusive cross sections of Drell-Yan and Higgs boson productions at the hadron colliders.
For completeness, we briefly describe the formalism \cite{Ajjath:2020ulr}, which shows 
how the building blocks of perturbative results in the threshold region can be organised using their factorisation
properties and the logarithmic structure.  Thanks to a set of
differential equations that govern these building blocks, it is possible to sum up certain class of threshold logarithms 
to all orders in perturbation theory.  
In particular, the solutions to such equations lead to a compact integral representation in $z$ space that captures
SV and NSV terms of inclusive rates in these processes.
The integral representation can be conveniently used in Mellin $N$ space to resum ${\mathcal O}(1)$ terms that show up in 
large SV and NSV contributions at every order to obtain reliable theoretical predictions at colliders.

We begin by defining $\Delta_q$ as a sum of SV and NSV contributions to diagonal channels: 
\begin{eqnarray}\label{eq:Deltaqfull}
\Delta_q(z) = \Delta_{q \overline q}^{SV}(z) +
\Delta_{q \overline q}^{NSV}(z)
\end{eqnarray}
where the SV part is defined in \eqref{SV} and NSV in \eqref{NSV}.
%The NSV part of the $\Delta_{q\overline q}^{reg}$, denoted by $\Delta_{q \overline q}^{NSV}$,is obtained by setting $l=0$ in the  right side of \eqref{delregexp}:  
%{\color{red}
%\begin{eqnarray}
%\Delta_{q \overline q}^{NSV,(i)}(z) = \sum_{k=0}^{2i-1} \Delta^{reg,(i)}_{q \overline q,k} \ln^k(1-z) .
%\end{eqnarray}}
Using the mass factorisation that separates collinear singular part from the bare partonic cross sections 
and the UV and IR  renormalisation group equations that various building blocks satisfy  
we can cast the CFs of inclusive cross sections in dimensional regularisation ($n=4+\epsilon$) 
as
\begin{eqnarray}
\label{masterq}
\Delta_q(q^2,\mu_R^2,\mu_F^2,z) = 
\mathcal{C}\exp \bigg( \Psi^q\big(q^2,\mu_R^2,\mu_F^2,z,\epsilon\big)\bigg)\bigg |_{\epsilon=0} \,,
\end{eqnarray}
where $\Psi^q$ is finite in the limit $\epsilon \rightarrow 0$
and is  given by
\begin{eqnarray}\label{Psi}
    \Psi^q\big(q^2,\mu_R^2,\mu_F^2,z,\epsilon\big) = &\Bigg( \ln \bigg( Z_{UV,q}\big(\hat{a}_s,\mu^2,\mu_R^2,\epsilon\big)\bigg)^2 +   \ln \big| \hat{F}_{q}\big(\hat{a}_s,\mu^2,-q^2,\epsilon\big)\big|^2\Bigg) \delta\big(1-z\big)\nonumber \\
    &+2 \mathrm{\Phi}_q\big(\hat{a}_s,\mu^2,q^2,z,\epsilon\big) - 2\mathcal{C} \ln \Gamma_{qq}\big(\hat{a}_s,\mu^2,\mu_F^2,z,\epsilon\big) \,,
\end{eqnarray}
where $Z_{UV,q}$ is the overall renormalisation constant which is unity for vector/axial vector interactions in quark
anti-quark initiated channels.  The bare strong coupling constant, $\hat a_s=\hat g_s^2/16 \pi^2$, with $\hat g_s$ the bare QCD coupling constant and the scale $\mu$ results from dimensional regularisation.  The square of the form factor (FF), $\hat F_q$, encodes pure virtual contributions 
to $q + \overline q \rightarrow l^+ l^-$ while the soft-collinear function, $\mathrm{\Phi}_q$, contains contributions from remaining partonic subprocesses
normalised by square of the form factor.  Thanks to the fact that mass factorisation terms required for
the SV and NSV contributions to diagonal channels depend only on diagonal kernels $\Gamma_{q\overline q}$ where
we need to keep only diagonal splitting functions $P_{q q}$, the logarithm of these kernels completely decouples from the rest. 

The symbol ``$\mathcal{C}$" refers to convolution, which acting on any exponential of a function $f(z)$ takes the
following expansion::
\begin{eqnarray}
        \mathcal{C}e^{f(z)} = \delta(1-z) + \frac{1}{1!}f(z) + \frac{1}{2!}\big(f\otimes f\big)(z) + \cdots
\end{eqnarray}
Since we have restricted ourselves to SV+NSV contributions to $\Delta_{q}$,
we keep only those terms that are proportional to SV distributions namely $\delta(1-z)$,
${\cal D}_i(z)$ and NSV terms $\ln^i(1-z)$ with $i=0,1,\cdots$ and drop rest of the terms resulting from
the convolutions.

The form factor, soft-collinear function and Altarelli-Parisi (AP) kernels that contribute to $\Delta_q$ 
are computable order by order in $a_s$ in perturbation theory.
One finds that each of them demonstrates rich infrared structure through certain differential equations. 
For example, the form factor satisfies Sudakov's K+G differential equation while the mass factorisation kernels satisfy
AP evolution equations.  In addition, they are independently renormalisation group invariants. 
These differential equations are controlled by universal UV and IR anomalous dimensions that are perturbatively calculable.
Thanks to these differential equations and the fact that $\Delta_q$ is finite, one finds
that soft-collinear function ${\mathrm \Phi}_q$ also satisfies K+G like differential equation.
The solution to the form factor is expressed in terms
of cusp ($A^q$), soft ($f^q$), collinear ($B^q$) anomalous dimensions and process dependent constants, while for the mass factorisation
kernels, one finds the solution 
in terms of diagonal AP splitting functions which contain only $\delta(1-z),{\cal D}_0(z)$ and $\ln^j(1-z),j=0,1$ terms.  
Unlike the form factor and AP kernels, the solution to ${\mathrm {\Phi}}_q$ is hard to obtain without the knowledge of
their kernels $\overline {\rm K}$ and $\overline {\rm G}$ (see \cite{Ajjath:2020ulr}).  The singular kernel $\overline {\rm K}$ can be determined from singular terms of FF and AP kernels while
the finite part is obtained from the fixed order results of $\Delta_q$. 
We use the perturbative results known to third order
to parametrise the kernels in terms of $\delta(1-z)$, plus distributions and $\ln(1-z)$ in
dimensional regularisation.  
The resulting $z$ dependent solution of ${\mathrm \Phi_q}$ depends on process independent anomalous dimensions
$A^q,f^q, C^q$ and $D^q$ and certain process dependent quantities.
Combining the solutions from all the differential equations one obtains an all order exponentiation of 
SV and NSV contributions to $\Delta_q$ as given in \eqref{masterq}.  While each piece contains both UV and IR divergences as 
poles in $\epsilon$, the divergences cancel among themselves when  $\epsilon \rightarrow 0$, leaving finite $\Psi^q$.   In \cite{Ajjath:2020ulr}, 
an integral representation for the function $\Psi^q$ in terms of $z$ was obtained:
\begin{eqnarray}
\label{resumz}
\Delta_q(q^2,\mu_R^2,\mu_F^2,z)= C^q_0(q^2,\mu_R^2,\mu_F^2)
~~{\cal C} \exp \Bigg(2 \Psi^q_{\cal D} (q^2,\mu_F^2,z) \Bigg)\,,
\end{eqnarray}
where
\begin{eqnarray}
\label{phicint}
\Psi^q_{\cal D} (q^2,\mu_F^2,z) &=& {1 \over 2}
\int_{\mu_F^2}^{q^2 (1-z)^2} {d \lambda^2 \over \lambda^2}
        P^{\prime}_{qq} (a_s(\lambda^2),z)  + {\cal Q}^q(a_s(q^2 (1-z)^2),z)\,,
\end{eqnarray}
with
\begin{eqnarray}
\label{calQc}
{\cal Q}^q (a_s(q^2(1-z)^2),z) &=&  \left({1 \over 1-z} \overline G^q_{SV}(a_s(q^2 (1-z)^2))\right)_+ + \varphi_{f,q}(a_s(q^2(1-z)^2),z).
\end{eqnarray}
The coefficient $C_0^q$ is $z$ independent and is expanded in powers of $a_s(\mu_R^2)$ as
\begin{eqnarray}
\label{C0expand}
C_0^q(q^2,\mu_R^2,\mu_F^2) = \sum_{i=0}^\infty a_s^i(\mu_R^2) C_{0i}^q(q^2,\mu_R^2,\mu_F^2)\,,
\end{eqnarray}
The results for $C^q_{0}$ can be found in \cite{Catani:2014uta} and the coefficients $ C_{0i}^q$ are given in Appendix \ref{app:C0q}.
The splitting function $P^\prime_{qq}$ is related to the AP splitting functions
$ P_{q q}\big(z,\mu_F^2\big)$. 
Expanding the latter around $z=1$ and dropping those terms that do not contribute to SV+NSV, we find 
\begin{eqnarray}
        P_{qq}\big(z,a_s(\mu_F^2)\big) &=& 2 B^q(a_s(\mu_F^2)) \delta(1-z) + P^{\prime}_{qq}\big(z,a_s(\mu_F^2)\big)\,,
\end{eqnarray}
where,
\begin{eqnarray}
        P^{\prime}_{qq}\big(z,a_s(\mu_F^2)\big) &=& 2  \Bigg[ A^q(a_s(\mu_F^2)) {\cal D}_0(z)
%\nonumber\\&&
                      + C^q(a_s(\mu_F^2)) \ln(1-z) + D^q(a_s(\mu_F^2)) \Bigg]
%\nonumber
\end{eqnarray}
The constants $C^q$ and $D^q$ can be obtained from the
the splitting functions $P^{\prime}_{q q}$ which are known to three loops in QCD \cite{Moch:2004pa,Vogt:2004mw}
(see \cite{GonzalezArroyo:1979df,Curci:1980uw,Furmanski:1980cm,Hamberg:1991qt,Ellis:1996nn,Moch:2004pa,Vogt:2004mw,Soar:2009yh,Ablinger:2017tan,Moch:2017uml} for the lower order ones).
The cusp, soft and the collinear anomalous dimensions and  the constants $C^q$ and $D^q$ are expanded
in powers of $a_s(\mu_F^2)$:
\begin{eqnarray}
X^q(a_s(\mu_F^2)) = \sum_{i=1}^\infty a_s^i(\mu_F^2) X_i^q, \quad \quad X = A,f,B,C,D
\end{eqnarray}
where $X^q_i$ to third order are available in \cite{Moch:2004pa,Vogt:2004mw} and are listed in appendix \ref{app:anodim}.
The function 
$\overline{G}^q_{SV}\big(a_s(q^2(1-z)^2)\big)$ is related to the threshold exponent $\textbf{D}^q\big(a_s(q^2(1-z)^2)\big)$ via Eq.(46) of \cite{Ravindran:2006cg} (see appendix \ref{app:anodim}).
The function $\varphi_{f,q}$ in powers of $a_s$ is given by
\begin{eqnarray}
\label{varphiexp}
\varphi_{f,q}(a_s(q^2(1-z)^2),z) = \sum_{i=1}^\infty a_s^i(q^2(1-z)^2) \sum_{k=0}^i \varphi_{q,i}^{(k)} \ln^k(1-z) \,,
\end{eqnarray} 
The coefficients $\varphi_{q,i}^{(k)}$ are known to third order and are listed in  Appendix \ref{app:phic} (See also \cite{Ajjath:2020ulr}).
%The next to SV corrections to various inclusive processes were also studied 
%in a series of papers \cite{Laenen:2008ux,Laenen:2010kp,Laenen:2010uz,
%Bonocore:2014wua,Bonocore:2015esa,Beneke:2019oqx,Beneke:2019mua} and
%lot of progress have been made leading to a better understanding of the underlying physics.

The resummation of threshold logarithms can be conveniently done in Mellin space, where  
 $z \rightarrow 1$ translates to large $N$ limit. 
In the latter, ${\mathcal O}(1)$ terms from $\omega=2 \beta_0 a_s(\mu_R^2) \ln N$ show up at every order in $a_s$ spoiling the truncation of perturbative
series in the exponent. This can be resolved by reorganising the series using the integral representation \eqref{phicint} and 
the resummed strong coupling
constant.  
To include SV and  NSV terms in the resummation in Mellin space, we need to keep $\ln N$ as well as ${\cal O}(1/N)$
terms in the large $N$ limit.
We find that  
\eqref{phicint} can correctly predict
only SV and NSV terms while the predictions beyond the NSV terms
namely ${\cal O}((1-z)^n \ln^j(1-z));n,j\ge 0$ in $z$ space and terms of
${\cal O}(1/N^2)$ in $N$ space will not be correct!.
The Mellin moment of $\Delta_q$ was obtained in \cite{Ajjath:2020ulr}
and is given by 
\begin{eqnarray}
\label{DeltaN}
\Delta_{q,N}(q^2,\mu_R^2,\mu_F^2) = C_0^q(q^2,\mu_R^2,\mu_F^2) \exp\left(
\Psi_{N}^q (q^2,\mu_F^2)
\right)\,,
\end{eqnarray}
where
\begin{eqnarray}\label{eq:Mellin}
\Psi_{N}^q(q^2,\mu_F^2) = 2 \int_0^1 dz z^{N-1}\Psi_{\cal D}^q (q^2,\mu_F^2,z) \,.
\end{eqnarray}
{\color{black}
Also, $C_0^q$ are $N$-independent constants coming from FF and the $\delta(1-z)$ part of soft-collinear function and AP kernels. Note that the $\delta(1-z)$ pieces in $z$-space translates to $N$-independent pieces in Mellin space. And the Mellin transformation of the plus distributions, given in \eqref{eq:Mellin}, give rise to $\ln N$ and $N$-independent constants.
Hence expressing $\Psi_{N}^q$ as
\begin{eqnarray}
\label{eq:Psi}
\Psi_{N}^q = \Psi_{\rm{SV},N}^q + \Psi_{\rm{NSV},N}^q
\end{eqnarray}
where $\Psi_{\rm{SV},N}^q$ contains  $\ln^j N, j=0,1,\cdots$ terms and 
while $\Psi_{{\rm NSV},N}^q$ contains terms of the form $(1/N) \ln^j N, j=0,1,\cdots$.
We find that $\Psi_{\rm{SV},N}^q$ takes the form:
\begin{eqnarray}
\label{PsiSVN}
        \Psi_{\rm{SV},N}^q = \ln(g_0^q(a_s(\mu_R^2))) + g_1^q(\omega)\ln N + \sum_{i=0}^\infty a_s^i(\mu_R^2) g_{i+2}^q(\omega) \,,
\end{eqnarray}
Here the $g^q_i$ coefficients are universal and they depend only on the initial partons. The constants $\ln g_0^q$ are the $N$-independent pieces obtained after Mellin transformation of $\Psi_{\cal D}^q$ and they satisfy the condition $\Psi_{\rm{SV},N}^q - \ln(g_0^q) = 0$ when $N=1$. Expanding them in powers of $a_s$ we get,
\begin{align}
\label{lng0}
\ln g_0^q(a_s(\mu_R^2)) = \sum_{i=1}^\infty a_s^i(\mu_R^2)~ g_{0,i}^q \,.
\end{align}
These exponents agree with those given in \cite{Catani:1989ne,Moch:2005ba,H:2019dcl},
and they are listed in the Appendices \ref{app:g0q} and \ref{app:gN}. 
%can be found in the  Appendix of \cite{H.:2020ecd}. 
% While the Mellin transformation of the plus distributions as given in \eqref{} give rise $\ln N $ and $N$-independent constants. The latter $N$-independent pieces are denoted by $\ln (g_0^q)$ which is given in the exponent \eqref{}. And those comes from form factor and $\delta(1-z)$ part of soft distribution function and AP kernels are collectively represented by $C_0^q$, see \eqref{}. 
In standard $N$-approach we absorb the $N$-independent pieces $g_{0,i}^q$ into $C_0^q$ and collectively define it as
\begin{equation}
\label{g0t}
\tilde g_0^q(q^2, \mu_R^2, \mu_F^2) = C_0^q(q^2,\mu_R^2,\mu_F^2) \  g_0^q(a_s(\mu_R^2)) \,.
\end{equation}
Thus the resulting quantity comprises of $\delta(1-z)$ contributions from
the form factor, soft-collinear function, AP kernels and $N$-independent part
of the Mellin moment of the distributions in $\Psi^q_{\cal D}(q^2,\mu_F^2,z)$. 
%This implies an ambiguity in
%treating the $N$ independent terms in the resummed results.  
%We also provide $g_0^q(a_s(\mu_R^2))$ in the ancillary files of \cite{Ajjath:2020ulr}.
% {\color{black} The $N$-independent coefficients $C^q_0$ and $g_0^q$ are related to the coefficients $\tilde g_0^q$ given in the paper \cite{H:2019dcl,H.:2020ecd} using the relation,
% % \begin{equation}
% % \tilde g_0^q(q^2, \mu_R^2, \mu_F^2) = C_0^q(q^2,\mu_R^2,\mu_F^2) \  g_0^q(a_s(\mu_R^2)) \,.
% % \end{equation}
% Expanded them in $a_s(\mu_R^2)$ gives,
% \begin{equation}
% \label{eq:g0t}
%   \tilde g_0^q(a_s(\mu_R^2)) = \sum_{i=0}^\infty a_s^i(\mu_R^2) \tilde g^q_{0,i}\quad \,,
% \end{equation}
The coefficients $\tilde g^q_{0,i}$ are listed in the Appendix \ref{app:g0t}}.

The function $\Psi_{{\rm{NSV}}, N}^q$ in \eqref{eq:Psi} is given by
\begin{align}
\label{PsiNSVN}
 \Psi_{{\rm{NSV}},N}^q = {1 \over N} 
\sum_{i=0}^\infty a_s^i(\mu_R^2) \bigg ( \bar g_{i+1}^q(\omega)
%\nonumber\\
%&& 
%+ {1 \over N} \Bigg(h^q_{00}(\omega) + h^q_{01}(\omega) \ln N 
+ h^q_{i}(\omega,N) \bigg)\,,
\end{align}
with 
%  \Bigg)\,,
%\nonumber\\
\begin{align}\label{hNSV}
h^q_i(\omega,N) = \sum_{k=0}^{i} h^q_{ik}(\omega)~ \ln^k N.
\end{align}
where $\bar g^q_i(\omega)$ and $h^q_{ik}(\omega)$ are presented in %the appendices of \cite{Ajjath:2020ulr} and they are also listed in 
the Appendices \ref{app:gbN} and \ref{app:hN} respectively. 
{\color{black} In each exponents, 
$g_i^q(\omega)$, $\overline g_i^q(\omega)$ and $ h_{ik}^q(\omega)$, we resum ${\cal O}(1)$ term $\omega$ in Mellin space to all orders in perturbation theory.}  
This is possible because of the argument in the coupling constant $a_s (q^2(1-z)^2)$ 
resulting the integrals and from the function ${\cal Q}^q$. 

%As the resummed results differ depending on how we treat the $N$-independent constants,
%we have proposed various schemes in \cite{H:2019dcl}  to study their numerical impact.
%We briefly describe them below for completeness:
In the SV part of the resummed result, the intrinsic ambiguity that exists while dealing with
what needs to be exponentiated gives scope to explore their impact.
Among different prescriptions, the standard approach is to exponentiate only large-$N$ pieces coming
from the threshold region. Also, for large $N$, the expansion of Euler Gamma functions gives {\it Euler-Mascheroni} constant, $\gamma_E$ and considering these large effects, one can exponentiate $\overline{N}$, which is defined as $ \overline{N} = N \exp(\gamma_E)$, instead of $N$ without disturbing the fixed order predictions.  Numerically, however this can make a difference
at the leading logarithmic accuracy as was already seen in \cite{Das:2019btv} where the perturbative convergence was shown to improve
with $\overline{N}$ terms.   In a different scheme, called {\it Soft exponentiation}
one exponentiates the Mellin moment of soft-collinear function \cite{Ravindran:2005vv,Ravindran:2006cg} which contains all the plus distributions and 
$\delta(1-z)$ terms.
Alternatively, one can also exponentiate the complete form factor along with the soft-collinear function in the Mellin space. This
approach was explored in \cite{Bonvini:2014joa,Bonvini:2016frm} to study the inclusive Higgs boson production in gluon fusion.
It was found to predict results that are less sensitive to the unphysical scales compared to the standard threshold approach.
This approach is theoretically justified because the form factor satisfies the Sudakov K+G type equation \cite{Sudakov:1954sw, Mueller:1979ih, Collins:1980ih, Sen:1981sd, Ravindran:2005vv,Ravindran:2006cg} whose solution  is
an exponential of $N$ independent constant.
For the numerical study of DY production this approach was used in \cite{Eynck:2003fn}.
%In the present article, we study the numerical effects of threshold and NSV logarithms at NNLL accuracy matched to 
%the known NNLO results. We restrict ourselves to neutral DY production at the LHC energies. 
We give the expressions for different resummation schemes below.
\begin{itemize}
\color{black}
\item {\textbf{\textit{Standard} $N${\textit{ exponentiation}}}:}  In this scheme, we exponentiate only the large-$N$ pieces that contribute to the CFs and the exponent is devoid of $N$-independent pieces. Note that this will only change the $\Psi_{{\rm SV},N}^q$. The $\Psi_{{\rm NSV},N}^q$ will remain same as it contains only $N$-dependent terms. Hence  we write:
\begin{eqnarray}
\Delta_{q,N}(q^2,\omega) = \tilde g_0^q(q^2) \exp\left(
G_{{\rm SV}, N}^q (q^2,\omega) + \Psi_{{\rm NSV}, N}^q (q^2,\omega) 
\right)\,,
\end{eqnarray}
where NSV part is defined in \eqref{PsiNSVN}. And the SV part is given as,
\begin{equation}
    G_{{\rm SV}, N}^q (q^2,\omega) = \Psi_{{\rm SV}, N}^q (q^2,\omega) - \ln(g_0^q) 
\end{equation}
which can be obtained from the Mellin transformation of $\Psi_{\cal D}^q$ and keeping only those terms that vanish when $N=1$. This is the standard approach adopted all along this paper and we compare other prescriptions with the $N$-exponentiation scheme in later sections.
%The coefficient $g_0^q$ is obtained by setting $\Psi^q_{{\rm SV},N} - \ln g_0^q=0$ when $N=1$.

\item {\textbf{\textit {Standard} $\Nb$ {\textit {exponentiation}}}:} Here, the large logarithms are expressed in functions of $\Nb$ instead of $N$, where  $\Nb = N\exp(\gamma_{E}^{})$. Hence along with large $N$-pieces, the $\gamma_E$ terms are also taken into exponentiation as they also contribute to ${\cal O}(1)$ terms. The resulting exponent takes the form:
\begin{align}
\Delta_{q,\overline N}(q^2,\mu_R^2,\mu_F^2) =  \bar{ \tilde{g}}_0^q(q^2,\mu_R^2,\mu_F^2) \exp\left(
G_{{\rm SV},\overline N}^q (q^2,\omega) + \Psi_{{\rm NSV}, \overline N}^q (q^2,\omega) 
\right) \,.
\end{align}
Here the $\bar{ \tilde{g}}_0^q(q^2,\mu_R^2,\mu_F^2)$ are given by 
\begin{equation}
\label{g0nb}
\bar{ \tilde{g}}_0^q(q^2,\mu_R^2,\mu_F^2) = C_0^q(q^2,\mu_R^2,\mu_F^2) \  \bar{{ g}}_0^q(a_s(\mu_R^2)) \,,
\end{equation}
where $\overline g_0^q$ results from the $N$-independent part of the Mellin transformation of the distributions and  the  $\gamma_E$'s are all absorbed into the $\overline N$-dependent functions.  In doing so, the $\overline N$-dependent part of SV, i.e., $G_{{\rm SV},\overline N}^q$, is obtained by setting $\Psi^q_{{\rm SV},\overline N} - \ln \overline g_0^q=0$ 
when $\overline N=1$.
Hence, $G^q_{{\rm SV}, \overline{N}} $ and $\Psi_{{\rm NSV}, \overline N}^q$ are given by
\begin{align}\label{eq:gnb}
G^q_{{\rm SV}, \overline{N}} &= g^q_1(\overline \omega) \ln \overline{N} + \sum_{i=0}^\infty a_s^i(\mu_R^2)~ g^q_{i+2}(\overline{\omega})\,,
% \end{align}
% \begin{align}
\\
  \Psi_{\rm{NSV},\overline N}^q &= {1 \over N} 
\sum_{i=0}^\infty a_s^i(\mu_R^2) \bigg ( \bar g_{i+1}^q(\overline \omega)
+ h^q_{i}(\overline \omega,\overline N) \bigg)\,,
\end{align}
with 
\begin{align}\label{hnb}
h^q_i(\overline \omega,\overline N) = \sum_{k=0}^{i} h^q_{ik}(\overline \omega)~ \ln^k \overline N.
\end{align}
The expansion parameter $\overline{\omega} = 2 \beta_0 a_s(\mu_R^2) \ln \overline{N}$.  Numerically, this can make difference at every logarithmic accuracy. For the SV, the perturbative convergence was shown to improve with $\overline N$ terms in \cite{H.:2020ecd}. We will address the impact of the same in NSV case in next section.
\item {\textbf {\textit {Soft exponentiation}}:} Another scheme one can adopt is the \textit{Soft exponentiation}, where we exponentiate the complete finite part of soft-collinear function $\rm \Phi_q$. This includes the $N$-independent pieces arising after the Mellin transformation of $\Psi_{\cal D}^q$, along with the $N$-dependent ones which we used in Standard $N$-exponentiation. For this scheme the resummed result takes the form:
%is different from the standard $N$ exponentiation approach in the following way. In latter we only exponentiate the large $N$-pieces. But in principle, the whole 
%finite part of soft-collinear function $\rm \Phi_q$  can be taken in to exponent, which is done in Soft exponentiation by relaxing the condition $\Psi^q_{N}-\ln g^q_0=0$ when $N=1$. We obtain  
\begin{align}\label{eq:resN}
\Delta_{q,N}(q^2,\omega) = \tilde{g}^{q,\rm {Soft}}_0(q^2) \exp \Big( \Psi^{q,\rm {Soft}}_{{\rm SV}, N}(q^2,\omega) + \Psi^{q}_{{\rm NSV}, N}(q^2,\omega)\Big) \,.
\end{align}
with
\begin{align}\label{eq:gnbsoft}
\Psi_{{\rm SV},N}^{q,\rm {Soft}}(q^2,\omega) = \ln N ~g^{q,\rm Soft}_1(\omega) + \sum_{i=0}^{\infty}  a_s^i g^{q,\rm Soft}_{i+2}(q^2,\omega) \,.
\end{align}
Here the complete finite part coming from the soft-collinear function is absorbed into the resummed exponents, which thereby gives rise to the exponents, $g_i^{q,\rm Soft}$. 
The remaining $N$-independent terms coming from  finite part of form factor and AP kernels contribute to $\tilde{g}^{q,\rm{Soft}}_0$, whose expansion in powers of $a_s$ is given as:
\begin{align}\label{eq:g0bsoft}
\tilde{g}^{q,\rm{Soft}}_{0i}(q^2) = 1+\sum_{n=1}^{\infty} a_s^i ~ \tilde{g}^{q,\rm{Soft}}_{0i}(q^2) \,.
\end{align}
%$\Psi_{\Nb}^{q,\rm Soft}$ and $ \gb_0^{\rm Soft}$ have similar expansion as eq.\ (\ref{eq:resGNb}) 
%and eq.\ (\ref{eq:gnb}) respectively and the corresponding coefficients are calculated and 
%presented in Appendix  of \cite{H:2019dcl} . 
The $N$-independent constants $\tilde{g}^{q, \rm Soft}_{0i}(q^2)$ and the resummed exponents $g_i^{q, \rm Soft}(\omega)$ are 
listed in appendix \ref{app:softexp}.
\item {\textbf {\textit {All exponentiation}}:}  
In light of \eqref{masterq}, which came out as a consequence to the first-order differential equations satisfied by each of the building blocks results in an all-order exponential structure for  $\Delta_{q,N}$. Hence it is natural to study the numerical impact of entire contribution taken in to Mellin space, which is done in {\it All exponentiation} scheme. The resummed result then takes the form:
%In this scheme, as the name suggests, the entire contributions of $\Delta_{q,N}$ is taken into the Mellin space and the resummed exponent changes accordingly. 
%
% we compute the Mellin moment of $\Psi^q$ by exponentiating the entire
% contribution  as 
\begin{align}\label{eq:resall}
\Delta_{q,N}(q^2,\omega) = \exp \Big( \Psi_{{\rm SV},N}^{q,\rm All}(q^2,\omega) +  \Psi_{{\rm NSV},N}^{q}(q^2,\omega)  \Big) \,,
\end{align}
where \begin{align} \label{gnall}
\Psi_{{\rm SV},N}^{q,\rm All}(q^2,\omega) = \ln N ~g_1^{q,\rm All}(q^2) + \sum_{i=0}^{\infty} a_s^i g_{i+2}^{q,\rm All}(q^2,\omega)  \,.
\end{align}
The resummed exponents $g_i^{q, \rm All}(\omega)$ contain both $N$-dependent and independent terms and are listed in appendix \ref{app:Allexp}. 
This scheme was explored in \cite{Bonvini:2014joa,Bonvini:2016frm}
to study the inclusive cross section for the production of Higgs boson in gluon fusion at the LHC. 
For similar study for the DY in DIS and $\overline{\rm MS}$ schemes, see \cite{Eynck:2003fn}. 
%The relevant resummed exponent has been 
%provided in Appendix of \cite{H:2019dcl}.
\end{itemize}

%In \cite{Das:2019btv}, it was found that the $\overline{N}$-exponentiation shows a faster convergence compared to 
%the $N$-exponentiation for the charged and neutral DIS processes. 
%In the $\overline{N}$-exponentiation, the entire $\gamma_{E}^{}$ dependent terms 
%are exponentiated through the variable $\overline{N} = N \exp(\gamma_E^{})$; 
%while in the $N$-exponentiation these terms are distributed among the exponent 
%and the $N$ independent term $g_0$. As a result the deviation starts already at the LL accuracy. 
In \cite{Ajjath:2020rci}, we had studied how various schemes discussed so far can affect the predictions
of invariant mass distribution of lepton pairs, inclusive $Z$ and $W^\pm$ production rates. 
In the present paper we extend this analysis in the presence of resummed NSV exponent and study the
numerical impact on the production of a pair of leptons in DY process at the LHC.
At NNLO level, we have the contributions from all the channels. 
%\At N$^3$LO only SV+NSV contributions
%are publicly available,  hence, our numerical predictions will be based on fixed order N$^3$LO${}_{\rm {SV+NSV}}$
%results for the CFs and on parton distribution functions known to NNLO accuracy.
\textcolor{black}{Our numerical predictions are based on fixed order NNLO
results for the CFs and on parton distribution functions known to NNLO accuracy.}
The resummed results are matched to the fixed order result 
in order to avoid any double counting of threshold logarithms. 
The resummed result at a given accuracy, say $\overline {\rm N^nLL}$, is
computed by taking the difference between the resummed result and the same truncated 
up to order $a_s^n$. Hence, it contains contributions from the SV and NSV 
terms to all orders in perturbation theory starting from $a_s^{n+1}$:   
\begin{align}\label{eq:matched}
\sigma_N^{\rm {N^nLO+\overline {\rm N^nLL}}} &= 
\sigma_N^{\rm {N^nLO}} +
\sigma^{(0)} 
\sum_{ab\in\{q,\bar{q}\}}
  \int_{c-i\infty}^{c+i\infty} \frac{dN}{2\pi i} (\tau)^{-N} \delta_{a\overline b}f_{a,N}(\mu_F^2) f_{b,N}(\mu_F^2) \nn\\
&\times \bigg( \Delta_{q,N} \bigg|_{\overline {\rm {N^nLL}}} - {\Delta_{q,N}}\bigg|_{tr ~\rm {N^nLO}}     \bigg)  \,.
\end{align}
where $\sigma_N$ is the Mellin moment of $d\sigma/dQ$. 
{\color{black}
To distinguish between SV and SV+NSV resummation, all along the paper, we denote the former by N$^n$LL and the latter by $\overline{\rm  N^nLL}$ for the $n^{\rm th}$ level logarithmic accuracy.}
The Mellin space PDF ($f_{i,N}$) can be evolved using QCD-PEGASUS \cite{Vogt:2004ns}. 
Alternatively,  we use the technique described in \cite{Catani:1989ne,Catani:2003zt} to directly
deal with PDFs in the $z$ space. The contour $c$ in the Mellin inversion  
can be chosen according to {\it Minimal prescription} \cite{Catani:1996yz} procedure. 
In the above \eqref{eq:matched} the second term represents the resummed result truncated 
to N$^n$LO order. \textcolor{black}{  In Table \ref{tab:res}, we list the resummation coefficients which are required to obtain the resummed $\Delta_{q,N}$ at a given logarithmic accuracy.   
In the next section we present the numerical results along with the scale uncertainties for the neutral DY process 
at the LHC. }
% \begin{table}[h!]
% \centering
% \begin{center}
% \begin{small}
% {\renewcommand{\arraystretch}{1.6}
% \begin{tabular}{|P{2.9cm}|P{3.2cm}||P{3.2cm}|}
%  \hline
%  \multicolumn{2}{|c||}{GIVEN} & \multicolumn{1}{c|}{Resummed, $\Delta_{q,N}$}\\
%  \hline
%  \hline
%  \rowcolor{lightgray}
%  Logarithmic Accuracy & Resummed Exponents    &  $\Delta_{q,N}\big|_{\overline{\rm N^nLL}}$
%     \\
% %   \rowcolor{lightgray}
% %                   Accuracy &   Exponents  & \\
%  \hline
% $\overline{\rm LL}$&	$\tilde g^q_{0,0},g^q_1,\overline g^q_1,h^q_0$              & $\Delta_{q,N}\big|_{\overline{\rm {LL}}}$ \\
%  \hline
% $\overline{\rm NLL}$ &$\tilde g^q_{0,1},g^q_2,\overline g^q_2, h^q_1$   & $\Delta_{q,N}\big|_{\overline{\rm {NLL}}}$\\
%   \hline
% $\overline{\rm NNLL}$&$\tilde g^q_{0,2},g^q_3,\overline g^q_3, h^q_2$    & $\Delta_{q,N}\big|_{\overline{\rm {NNLL}}}$ \\
%   \hline
%  %\hline
% \end{tabular}}
% \end{small}
% \end{center}
% 	\caption{\label{tab:res} The all order predictions for  $\Delta_{q,N}$ for a given set of resummation coefficients \Big\{$\tilde g^q_{0,i},g^q_i(\omega),\overline{g}^q_i(\omega), h^q_{i}(\omega)$\Big\} at a given order.  }
% \end{table}

 \begin{table}[h!]
 \centering
 \begin{center}
 \begin{small}
 {\renewcommand{\arraystretch}{1.6}
 \begin{tabular}{|P{3.5cm}||P{3.5cm}|}
  \rowcolor{lightgray}
  Logarithmic Accuracy & Resummed Exponents    
    \\
 %   \rowcolor{lightgray}
 %                   Accuracy &   Exponents  & \\
  \hline
 $\overline{\rm LL}$&	$\tilde g^q_{0,0},g^q_1,\overline g^q_1,h^q_0$     \\
  \hline
 $\overline{\rm NLL}$ &$\tilde g^q_{0,1},g^q_2,\overline g^q_2, h^q_1$  \\
  \hline
 $\overline{\rm NNLL}$&$\tilde g^q_{0,2},g^q_3,\overline g^q_3, h^q_2$   \\
   \hline
  %\hline
 \end{tabular}}
 \end{small}
 \end{center}
 	\caption{\label{tab:res}  The set of resummation coefficients \Big\{$\tilde g^q_{0,i},g^q_i(\omega),\overline{g}^q_i(\omega), h^q_{i}(\omega)$\Big\} which is required to compute $\Delta_{q,N}$ at a given logarithmic accuracy.  }
 \end{table}

% I need the following expressions  (in terms of color factors for qqbar channels) in tex format:

% \begin{itemize}
% \item  $C_0^q$ in 2.6
% \item $\overline G^c_{SV}$ in eqn 2.8
% \item $A,B,C,D,f$ to third/fourth order for $q$
% \item $\varphi_{q,i}^k$ in 2.13
% \item
% all $g_0, g_i$ in 2.17
% \item
% all $\overline g_i$ , $h_{ik}^q$ in 2.20 and 2.21
% \item 
% expressions for various resummation schemes
% \end{itemize}

%% Pheno Results
\newpage
\section{Phenomenology} \label{sec:pheno}
\textcolor{black}{
In this section, we peform a detailed numerical study on  the impact of resummed soft virtual plus next-to-soft virtual (SV+NSV) results for the production of di-leptons in 
neutral DY process at the LHC. We include all the partonic channels at the Fixed-order (FO) up to NNLO with off-shell photon and $Z$ boson intermediate states. We restrict ourselves to center of mass energy of 13 TeV at the LHC, however our analysis can be extended to other energies as well as to other colliders.
%-------------------------------------------------------------------------
%\subsection{Fixed-order vs Resummed results: All-Channels}
%We begin our discussion by examining the NSV corrections at NNLO+NNLL$_{\rm SV+ NSV}$. 
We use the following electro-weak parameters for the vector boson masses and widths, Weinberg 
angle ($\theta_w$) and the fine structure constant ($\alpha$):}
\begin{align}\label{eq:parameters}
m_Z &= 91.1876 ~ \text{GeV},  ~~ \Gamma_Z=2.4952~ \text{GeV} \,,\nonumber\\&
 \text{sin}^2\theta_w=0.22343\,, ~~ \alpha=1/128 \,.
\end{align}
\textcolor{black}{The parton distribution functions are directly taken from the $\tt{lhapdf}$ \cite{Buckley:2014ana} routine. All results are obtained using the $\tt{MMHT2014}$ \cite{Harland-Lang:2014zoa} parton densities throughout. The strong coupling constant
is evolved to the renormalisation scale $\mu_R$ using the three-loop QCD beta function in the
$\overline{\text{MS}}$-scheme with $n_f=5$ active massless quark flavours. }

\textcolor{black}{
We begin with a discussion on the relative contributions of SV and NSV terms  in the fixed order results.  
The partonic coefficient function, $\Delta_q$ given in \eqref{eq:Deltaqfull}, comprises of functions namely, SV distributions, $\{\delta(1-z), \mathcal{D}_i(z)\}$ and NSV logarithms, $\ln ^k(1-z)$.   The cross section at the hadronic level is obtained by convoluting  CFs with the appropriate fluxes of the incoming partons.  Before we present the corresponding contributions of SV and NSV terms to hadronic
cross sections, we plot the convolutions of each distribution and the $\ln(1-z)$ term   with the flux of the incoming quark anti-quark pairs
%at $\mu_F=200$ GeV 
for various values of $\tau$.  In particular we plot the following integral as a function of $\tau$:}
\begin{eqnarray}
\mathcal{F}(\tau) = \int_{\tau}^1 {d z \over z } \tilde \Phi_{q\bar q}\left({\tau \over z}\right) \mathcal{G}(z),\quad \text{where} \quad \mathcal{G}(z) = \{\delta(1-z),\mathcal{D}_1(z),\ln(1-z)\}\,.
\end{eqnarray}
\textcolor{black}{
The plot given in Fig.\ref{fig:OnlyFunc} demonstrates the hierarchical behaviour of the terms in the 
threshold expansion at the numerical level, in particular, it reflects to the fact that at the threshold,   $\mathcal{F}$ gets larger contribution from the distributions compared to the sub-leading $\ln(1-z)$ terms. 
\begin{figure}[hbtp]
\centering
\includegraphics[scale=.5]{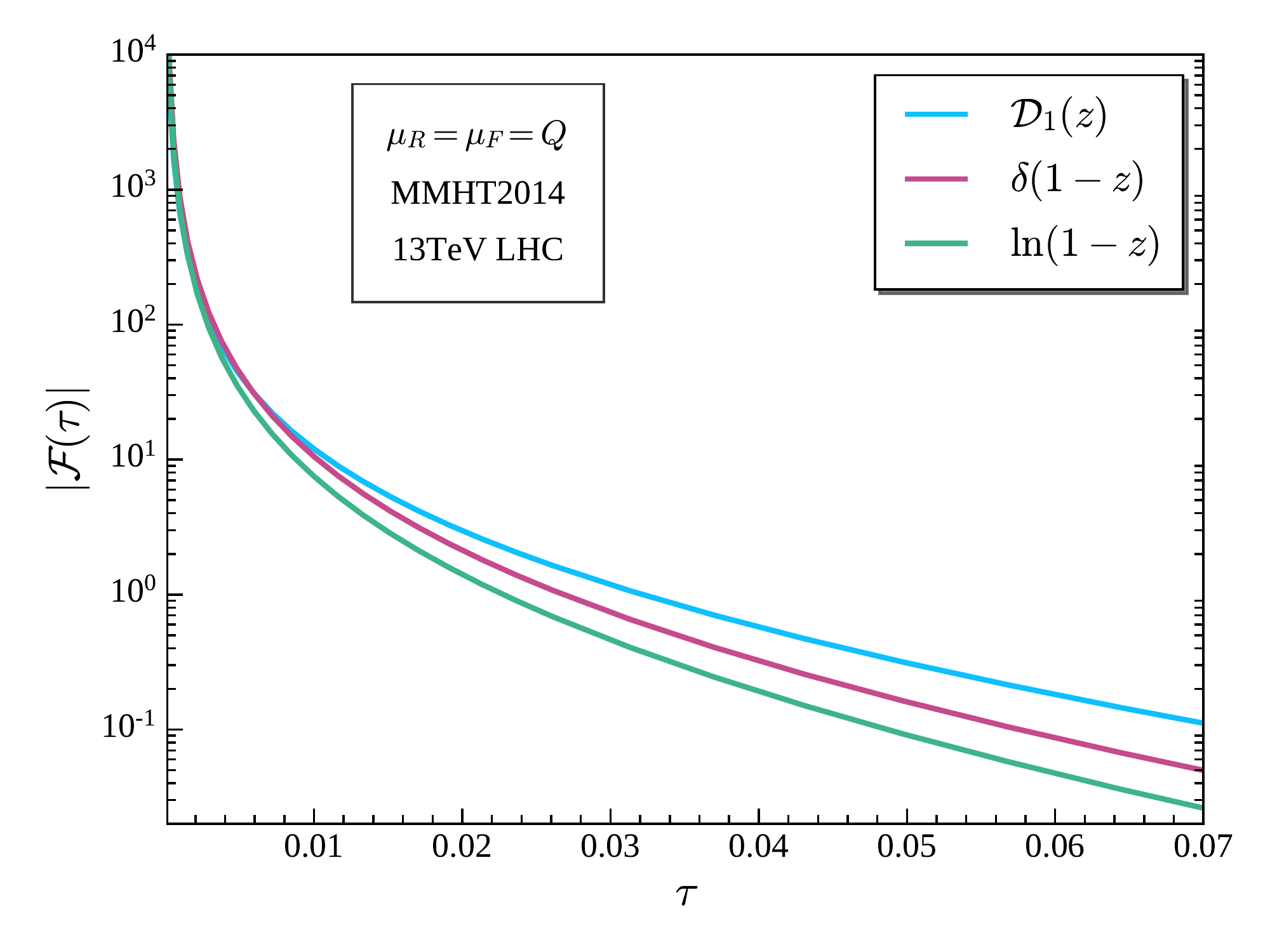}
\caption{Variation of SV and NSV functions with respect to $\tau = Q^2/S$.}
\label{fig:OnlyFunc}
\end{figure}
Interestingly, the hierarchy is reversed 
 when the corresponding coefficients from the perturbative results of CFs are taken into account along with these functions.   This is shown in Table \ref{tab:N2LO} for $\Delta_q^{(2)}$, i.e., at NNLO where, contrary to Fig.\ref{fig:OnlyFunc}, we find that the dominant contribution comes from the NSV logarithms which amount to 10.8\% of the Born contribution, whereas the SV distributions give -0.035\% . 
 We find that this trend is same at other known orders too.  For example at NLO, the total SV contributions amount to 9.01\% whereas NSV gives rise to a total of 20.5\% of the Born contribution. Similarly Table \ref{tab:N3LO} shows the contributions from SV and NSV logarithms at the N$^3$LO level.}
 
 \textcolor{black}{
From the above analysis, one observes that the NSV contributions become as important, if not more important than the SV contributions owing to their large coefficients. This was earlier pointed out in \cite{Anastasiou:2014lda} for the case of Higgs boson production through gluon fusion, that the common belief of threshold contribution as the dominant contribution is unreliable with each higher order corrections. Here, even in the case of Drell-Yan,  we find that NSV contributions become much more phenomenologically relevant with each powers of logarithmic corrections.}
%%%%%%%%%%%%%%%%% NNLO %%%%%%%%%%%%%%%%%%%%%%
%\begin{table}[h] 
%\begin{center}
%\begin{small}
%%\newcolumntype{P}[1]{>{\centering\arraybackslash}p{#1}}
%{\renewcommand{\arraystretch}{1.5}
%\begin{tabular}{|P{1.5cm} |P{1.8cm}||P{1.5cm}|P{1.5cm}|}
%\hline
%\rowcolor{lightgray}
%%\multicolumn{1}{|c||}{Order}
%     \multicolumn{1}{c|}{ }   
   % &\multicolumn{1}{c|}{ } 
  %  &\multicolumn{1}{c|}{Q = $\mu_R$ = $\mu_F=200$ (GeV)  }
 %   &\multicolumn{1}{c|}{Q = $\mu_R$ = $\mu_F=1000$ (GeV)  }
 %   \\ 
% \hline
% \hline
%%  \hline
%\multirow{4}{5cm}   {\hspace{0.5cm}SV} &  $\mathcal{D}_3$& \% & \%   \\
%& $\mathcal{D}_2$ &  \% &   \% \\
%& $\mathcal{D}_2$ &  \% &   \% \\
% & $\mathcal{D}_1$ & \% &   \% \\
% & $\mathcal{D}_0$ &   \% &   \% \\
% & $\delta(1-z)$ & \% & \% \\
%\hline
%\multicolumn{2}{|c||}{TOTAL}
%    & \multicolumn{1}{c|}{ \hspace{2cm}  \%}   
%    &\multicolumn{1}{c|}{\hspace{2cm}  \%}  
%    \\ 
%\hline
%\multirow{4}{5cm} {\hspace{0.5cm}NSV} &  $L_3$& \% & \%   \\
%& $L_2$ &  \% &   \% \\
%& $L_1$ &  \% &   \% \\
% & $L_0$ & \% &   \% \\

%\hline
%\multicolumn{2}{|c||}{TOTAL}
  %  & \multicolumn{1}{c|}{ \hspace{2cm}  \%}   
  %  &\multicolumn{1}{c|}{\hspace{2cm}  \%}  
 %   \\ 
%\hline
%\end{tabular}}
%\caption{\% contribution of SV distributions and NSV logarithms to the Born cross section at NNLO for Q= 200 GeV %and 1000 GeV. Here $L_i=\ln^i(1-z)$.}
%\label{Tab:Table1}
%\end{small}
%\end{center}
%\end{table}
%%%%%OLD TABLE%%%%%%%%
 \begin{table}[h] 
 \begin{center}
 \begin{small}
% %\newcolumntype{P}[1]{>{\centering\arraybackslash}p{#1}}
 {\renewcommand{\arraystretch}{1.5}
 \begin{tabular}{|P{1.3cm}||P{1.5cm} |P{1.8cm}||P{1.5cm}|P{1.5cm}|}
 \hline
 \rowcolor{lightgray}
 \multicolumn{1}{|c||}{$Q$ = $\mu_R$ = $\mu_F$ (GeV)}
     & \multicolumn{2}{c|}{ SV}   
     &\multicolumn{2}{c|}{ NSV}  
     \\ 
  \hline
  \hline
 %  \hline
 \multirow{5}{5cm} { \hspace{1cm} 200} &  $\mathcal{D}_3$ &  6.13\% & $\ln^3(1-z)$&  12.4\% \\
 & $\mathcal{D}_2$ &  1.49\% & $\ln^2(1-z)$&   7.83\% \\
  & $\mathcal{D}_1$ &  -3.24\% & $\ln^1(1-z)$&   -2.82\% \\
  & $\mathcal{D}_0$ & -4.74\% & $\ln^0(1-z)$&   -6.57\% \\
  & $\delta(1-z)$ & 0.003\% & & \\
  \hline
  %\hline
 %  \cline{2-5}
 %   \cline{2-5}
 %\rowcolor{lightgray} 
 \multicolumn{1}{|c||}{\hspace{0.2cm} TOTAL}
     & \multicolumn{2}{c|}{ \hspace{2cm} -0.035\%}   
     &\multicolumn{2}{c|}{\hspace{2cm}  10.8\%}  
     \\ 
 \hline
 %Drell-Yan &8.59\%&5.44\% & 9.82\% & 2.62\% & 1.49\%&-1.00\%\\
 %\hline
 \end{tabular}}
 \caption{\% contribution of SV distributions and NSV logarithms to the Born cross section at NNLO for $Q=200$ GeV.}
 \label{tab:N2LO}
 \end{small}
 \end{center}
 \end{table}
  %%%%%OLD TABLE%%%%%%%%
 \begin{table}[h] 
 \begin{center}
 \begin{small}
% %\newcolumntype{P}[1]{>{\centering\arraybackslash}p{#1}}
 {\renewcommand{\arraystretch}{1.5}
 \begin{tabular}{|P{1.3cm}||P{1.5cm} |P{1.8cm}||P{1.5cm}|P{1.5cm}|}
 \hline
 \rowcolor{lightgray}
 \multicolumn{1}{|c||}{$Q$ = $\mu_R$ = $\mu_F$ (GeV)}
     & \multicolumn{2}{c|}{ SV}   
     &\multicolumn{2}{c|}{ NSV}  
     \\ 
  \hline
  \hline
 %  \hline
 \multirow{7}{5cm} { \hspace{1cm} 200} &  $\mathcal{D}_5$ &  5.44\% & $\ln^5(1-z)$& 8.60\% \\
 &$\mathcal{D}_4$ & 2.62\% & $\ln^4(1-z)$& 9.82\% \\
 &$\mathcal{D}_3$ & -2.73\% & $\ln^3(1-z)$& -1.54\% \\
 & $\mathcal{D}_2$ & -4.25\% & $\ln^2(1-z)$&  -8.98\% \\
  & $\mathcal{D}_1$ & -1.94\% & $\ln^1(1-z)$& -6.14\% \\
  & $\mathcal{D}_0$ & -0.146\% & $\ln^0(1-z)$&  -1.28\% \\
  & $\delta(1-z)$ & 1.03\% & & \\
  \hline
  %\hline
 %  \cline{2-5}
 %   \cline{2-5}
 %\rowcolor{lightgray} 
 \multicolumn{1}{|c||}{\hspace{0.2cm} TOTAL}
     & \multicolumn{2}{c|}{ \hspace{2cm} 0.026\%}   
     &\multicolumn{2}{c|}{\hspace{2cm}  0.47\%}  
     \\ 
 \hline
 %Drell-Yan &8.59\%&5.44\% & 9.82\% & 2.62\% & 1.49\%&-1.00\%\\
 %\hline
 \end{tabular}}
 \caption{\% contribution of SV distributions and NSV logarithms to the Born cross section at N$^3$LO for $Q=200$ GeV.}
 \label{tab:N3LO}
 \end{small}
 \end{center}
 \end{table}

\textcolor{black}{Having established the relevance of formally subleading NSV logarithms in the fixed order results, we now turn to  assess the impact of their resummation on the  cross sections.
%and the corresponding perturbative uncertainities.
Besides the theoretical motivation of resumming the large enhancements arising from these logarithms, it would be interesting to see how phenomenologically important the resummed NSV logarithms are, in addition to the well established SV resummation for the Drell-Yan cross section.
We begin the analysis by addressing the following questions:}
% For the discussion of various contributions
% to the Drell-Yan cross section,
% we will show results for the $\rm {K}$-factor defined as,
% \begin{equation}
% %  K \left(\tau, \frac{Q^2}{\mu_R^2},\frac{\mu_F^2}{\mu_R^2}\right) = \frac{\frac{d\sigma}{dQ}\left(\tau,\frac{Q^2}{\mu_R^2},\frac{\mu_F^2}{\mu_R^2}\right)}{\frac{d\sigma^{\text{LO}}(\tau,1,1)}{dQ}} 
%   K \left(\tau, \frac{q^2}{\mu_R^2},\frac{\mu_F^2}{\mu_R^2}\right) = \frac{\frac{d\sigma}{dQ}\left(\tau,\frac{q^2}{\mu_R^2},\frac{\mu_F^2}{\mu_R^2}\right)}{\sigma^{(0)}_{DY}} 
% \end{equation}
% where th $\mu_R$ and $
% \mu_F$ are respectively renormalisation and factorisation scales.
%  }
% \textcolor{blue}{
% We observed that in fixed order predictions,  the contributions from NSV terms to hadronic cross section are larger than those from SV terms.  
% Having obtained the resummed results containing NSV terms, we would like to address the following questions:
% \begin{itemize}
% \item  It would be interesting to find out whether this continues to be the case for the resummed predictions.  
% \item In addition, will the inclusion of nsv terms in the resummed prediction improve the 
% stability of the predictions against the variations of  renormalisation and factorisation scales.
% \item analysis with a qqbar channels
% \item analysis with all channels
% \item after you convince everybody that sv+nsv is better, then
% we can give results for only sv+nsv along with other analysis like scale, schemes etc
% basically rearrangement!  please think about it
% %
% \item OR
% \end{itemize}
 \textcolor{black}{
\begin{itemize}
\item
%How large is the resummed corrections to the fixed order result? 
 In comparison to the fixed order corrections, how large are the SV+NSV resummed effects on the cross sections?
% \item 
%How sensitive is the SV+NSV resummed cross section to the %choices of factorisation ($\mu_F$) and  renormalisation  ($\mu_R$) scales?
 \item 
How do the resummed NSV terms change the predictions of SV resummed result?
\item What are the impacts of different schemes on the SV+NSV resummation?
% SV resummation depends on how $N$ independent part is treated and this can be exploited to improve the predictions.  We 
% will investigate this taking into account NSV resummation.
\end{itemize}
We will discuss each of these questions in great detail in subsequent sections. To begin with, let us look at the impact of SV+NSV resummed results in comparison to the fixed order results, which is the topic of the next section.
}
\subsection{Fixed-order vs Resummed results}\label{sec:Kfactor}
\begin{figure}[ht]
\begin{center}
\includegraphics[scale=0.5]{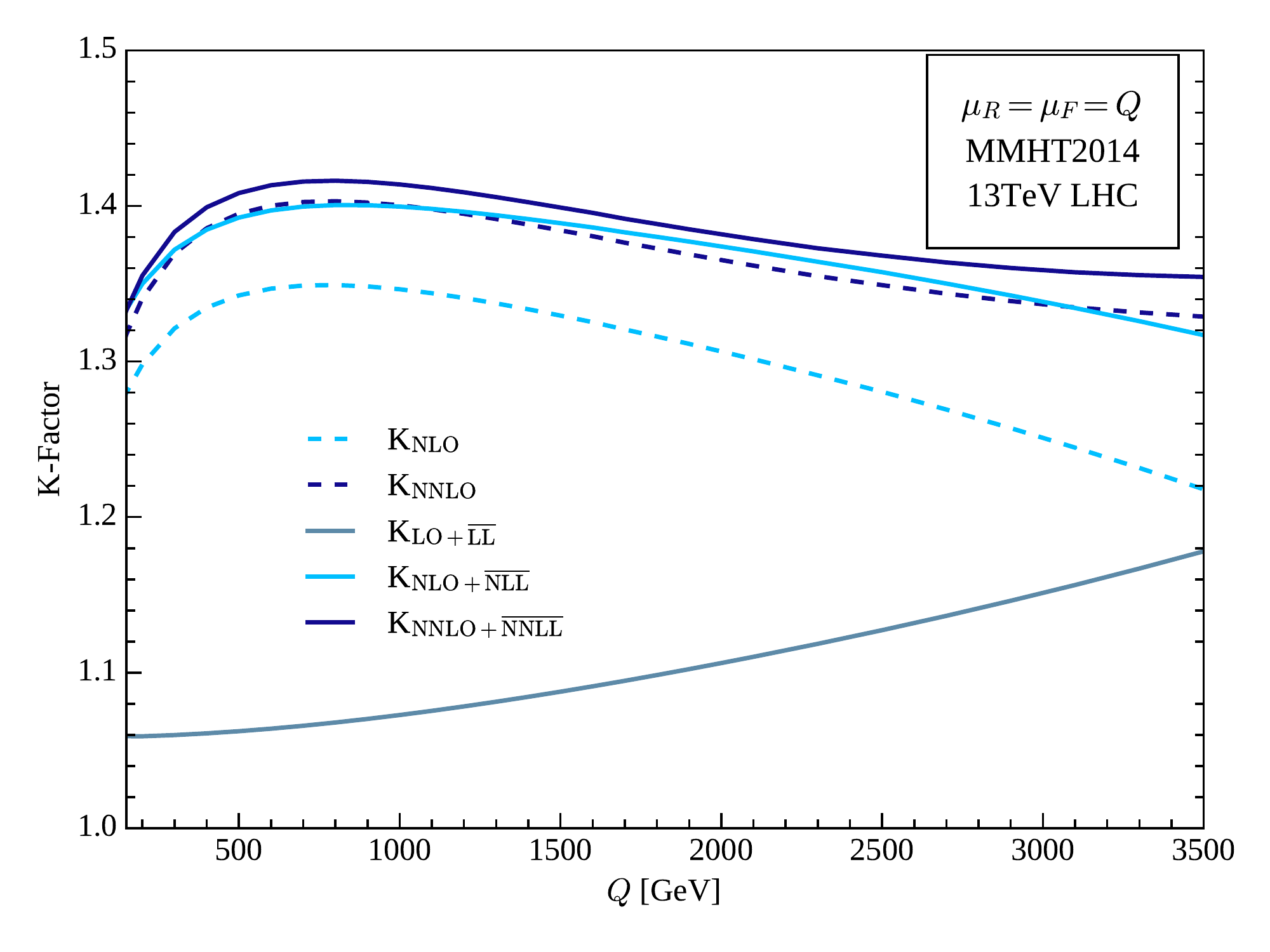}
\end{center}
\caption{\small{K-factors till NNLO+$\overline{\rm{NNLL}}$ level at the central scale $Q=\mu_R=\mu_F$}}
\label{KallN}
\end{figure}

{\color{black}
In the following we study how the inclusion of  SV+ NSV resummation modifies the predictions from fixed order results for inclusive DY di-lepton pair production. 
For this purpose, we get the matched predictions by appropriately including the leading, the next-to-leading and the next-to-next-to-leading resummed results with the corresponding fixed order results.  By investigating how sensitive is the SV+NSV resummed cross section to the choices of factorisation ($\mu_F$) and  renormalisation  ($\mu_R$) scales, we study their perturbative uncertainities.}{\color{black}
The quantitative impact of higher order effects can be obtained using $``$K-factor" defined by
\begin{equation}\label{eq:KDEf}
 \mathrm{K} \left(Q\right) = \dfrac{\dfrac{d\sigma}{dQ}\left(\mu_R=\mu_F=Q\right)}{\dfrac{d\sigma^{\text{LO}}}{dQ}(\mu_R=\mu_F=Q)} 
%   K \left(\tau, \frac{q^2}{\mu_R^2},\frac{\mu_F^2}{\mu_R^2}\right) = \frac{\frac{d\sigma}{dQ}\left(\tau,\frac{q^2}{\mu_R^2},\frac{\mu_F^2}{\mu_R^2}\right)}{\sigma^{(0)}_{DY}} 
\end{equation}
where we have set renormalisation ($\mu_R$) and 
factorisation ($\mu_F$) scales at $Q$. The K-factor for NLO+${ \overline {\rm NLL}}$ and NNLO+${ \overline {\rm NNLL}}$ are depicted in Fig.~\ref{KallN} along with the corresponding fixed order ones. 
%\textcolor{red}{Here we collectively denote the SV + NSV resummed logarithmic accuracy, i.e., leading logarithm (LL) by $\overline{\rm LL}$, next-to-leading logarithm (NLL) corrections  by $\overline{\rm NLL}$ and so on.}
}

\textcolor{black}{In Table \ref{Tab:Kfactor} we present the K factors resulting from both fixed as well as resummed contributions at three different values of $Q$, namely $Q=500,1000,2000$ GeV.  We find that there is an increment of $\{39.5\%,36.5\%\}$ in the cross section
when we go from LO to NNLO for the $Q$ values $\{500,2000\}$ GeV respectively. 
The inclusion of the resummed results, for the same values of $Q$, increases, the LO by $\{6.2\%,10.62\%\}$ when we include $\overline{\rm LL}$, the NLO by $\{3.7\%,5.2\%\}$ due to  $\overline{\rm NLL}$ and the NNLO by $\{0.94\%,1.2\%\}$ due to $\overline{\rm NNLL}$. This is reflected in  Fig. \ref{KallN}, where the resummed curves can be found to lie above their corresponding fixed order ones signifying the enhancement due to the resummed corrections.}

\textcolor{black}{Interestingly, we can also see from Table \ref{Tab:Kfactor}, that the K factors at $\rm NLO+\overline{\rm NLL}$  are closer to those of NNLO hinting that $\overline{\rm NLL}$ from quark part mimics the contributions from entire second order. The resummed curves at $\rm NLO+\overline{\rm NLL}$ and $\rm NNLO+\overline{\rm NNLL}$ are closer compared to the fixed order   ones, namely NLO and NNLO. This accounts for the fact that the addition of resummed effect improves the reliability of  perturbative predictions. The K factors at NNLO and $\rm NNLO+\overline{\rm NNLL}$ are more closer than those at NLO and $\rm NLO+\overline{\rm NLL}$. This is attributed to the fact that the resummed correction decreases as we go for higher order resummed contributions.}
\begin{table}[H] 
\begin{center}
\begin{small}
%\newcolumntype{P}[1]{>{\centering\arraybackslash}p{#1}}
{\renewcommand{\arraystretch}{1.7}
\begin{tabular}{|p{1.3cm}||P{1.5cm} ||P{1.5cm}|p{1.2cm}||P{1.5cm}|P{1.5cm}|}
\rowcolor{lightgray}
\multicolumn{1}{c||}{$\mu_R=\mu_F=Q$(GeV)}
    & \multicolumn{1}{c|}{ $\rm LO+\overline{\rm LL}$}   
    &\multicolumn{1}{c||}{ $\rm NLO$ }  
    &\multicolumn{1}{c|}{$\rm NLO+\overline{\rm NLL}$}  
    &\multicolumn{1}{c||}{$\rm NNLO$ } 
    & \multicolumn{1}{c|}{$\rm NNLO+\overline{\rm NNLL}$ }\\
%    & \multicolumn{1}{c|}{N$^2$LO$_{q\bar q}$+N$^2$LL(sv+nsv)}\\
 \hline
%  \hline
500 & 1.0624 & 1.3425&1.3925 & 1.3950   & 1.4082  \\
\hline
1000 &1.0728& 1.3464&1.3995  & 1.4004  & 1.4138  \\
\hline
2000 &1.1062  & 1.3064&1.3739  & 1.3652   & 1.3818  \\
\hline
%Drell-Yan &8.59\%&5.44\% & 9.82\% & 2.62\% & 1.49\%&-1.00\%\\
%\hline
\end{tabular}}
\caption{The K-factor values for resummed result in comparison to the fixed order ones. }
\label{Tab:Kfactor}
\end{small}
\end{center}
\end{table}

 \begin{figure}[ht]
\centering
\includegraphics[scale=0.6]{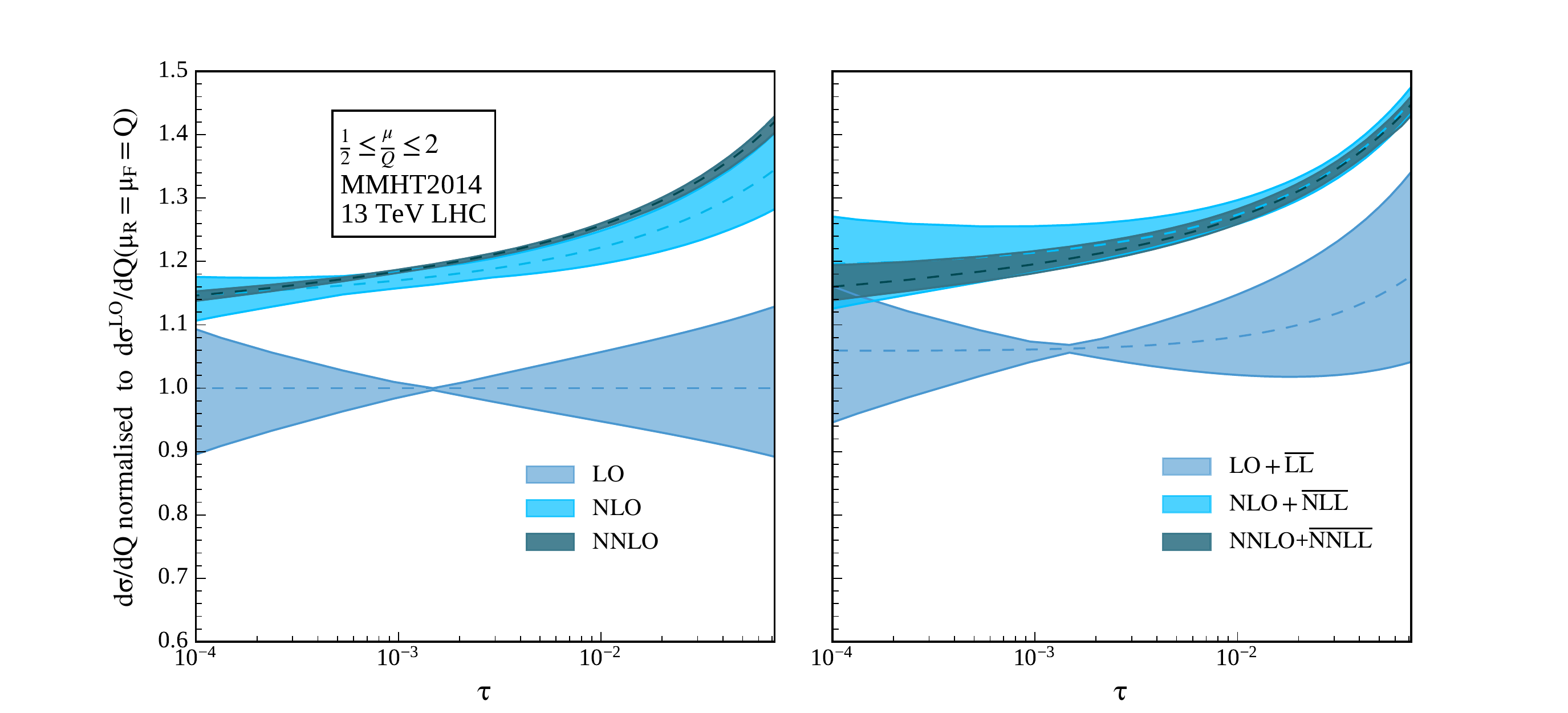}
\caption{7-point scale variation of the resummed result against fixed order around the central scale choice $(\mu_R,\mu_F) = (1,1)Q$ for $13$ Tev LHC. The dashed lines refer to the corresponding central scale $Q=\mu_R=\mu_F$ at each order. }
\label{7ptNall}
\end{figure} 
\subsubsection*{7-point scale uncertainities
 of the resummed results}\label{sec:7ptNexp}

\textcolor{black}{ Both fixed order as well as resummed results contain renormalisation and factorisation scales which are unphysical.  We now turn to assess the impact of these scales on our predictions. The dependence on these scales quantifies the corresponding errors due to their presence.  The standard approach to estimate this error is to use the canonical 7-point variation, where $\mu = \{ \mu_F, \mu_R \}$ is varied in the range $\frac{1}{2} \leq \frac{\mu}{Q} \leq 2$, keeping the ratio $\mu_R/\mu_F$ not larger than 2 and smaller than 1/2. }

The left panel of Fig. \ref{7ptNall} contains the invariant mass distributions obtained using fixed order
CFs as a function of $\tau$ and the bands are due to  7-point scale variation, while the right panel is obtained
using resummed results at various logarithmic accuracy.   
We find that the width of the resummed band at $\rm NLO +\overline{\rm NLL}$ is lesser than that of the corresponding fixed order ones from $Q = 1000$ GeV onwards but the width of the $\rm NNLO +\overline{\rm NNLL}$ doesn't show much improvement against the fixed order ones.
% the resummed predictions are more sensitive to the scales compared to fixed order results with an exception
% at larger value of $\tau$ for NLO+NLL.
The reason for this large scale uncertainity can be attributed to the fact that
the resummed predictions lack the off-diagonal counter part.   We will discuss more on this point in detail in subsequent analysis while considering the effect of both the scales independently.

%As seen earlier the resummation enhances the Drell-Yan corrections, however, they introduces larger scale uncertainities compared to the fixed order ones. From Fig \ref{7ptNall}, it can be seen that, the uncertainity band of $\rm NLO+\overline{\rm NLL}$ gets smaller to that of NLO at high energies, whereas the $\rm NNLO+\overline{\rm NNLL}$ band is wider than the corresponding NNLO ones. 
%In particular for $Q=2000$ GeV, the scale uncertainity for NNLO is found to be in between -.62\% to .37\%, whereas for $\rm NNLO+ \overline \rm NNLL$ it has increased to -0.78\% t0 .9\%).
%This behaviour is associated with the omission of off-diagonal resummation, whose contribution compensates the factorisation scale dependence. We will discuss more on these point in later sections while considering only the diagonal channel ($q\bar q$) contributions.}
\begin{table}[H] 
\begin{center}
\begin{small}
%\newcolumntype{P}[1]{>{\centering\arraybackslash}p{#1}}
{\renewcommand{\arraystretch}{1.9}
\begin{tabular}{|P{.8cm}|P{1.8cm}|P{1.8cm}|P{1.8cm}|P{1.8cm}|P{1.8cm}|P{1.5cm}|}
\rowcolor{lightgray}
\multicolumn{1}{c||}{$Q$}
    & \multicolumn{1}{c|}{ LO}   
    &\multicolumn{1}{c||}{LO+$\overline{\rm LL}$ }  
    &\multicolumn{1}{c|}{NLO}  
    &\multicolumn{1}{c||}{NLO+$\overline{\rm NLL}$} 
    
    & \multicolumn{1}{c|}{NNLO}
    & \multicolumn{1}{c|}{NNLO+$\overline{\rm {NNLL}}$}
    \\ 
 \hline
%  \hline
1000 & $2.3476^{+4.10\%}_{-3.92\%}$ & $2.5184^{+4.49\%}_{-4.25\%} $ & $3.1609^{+1.79\%}_{-1.69\%}$ & $3.2857^{+2.08\%}_{-1.18\%} $ & $3.2876^{+0.20\%}_{-0.31\%}$ &$ 3.3191^{+1.13\%}_{-0.86\%} $    \\
\hline
2000 &$  0.0501^{+8.50\%}_{-7.46\%}$ & $0.0554^{+9.10\%}_{-7.91\%} $& $0.0654^{+2.83\%}_{-2.98\%}$ & $0.0688^{+1.43\%}_{-1.23\%} $ &$0.0684^{+0.37\%}_{-0.62\%}$ & $ 0.0692^{+0.89\%}_{-0.78\%} $    \\
\hline
%Drell-Yan &8.59\%&5.44\% & 9.82\% & 2.62\% & 1.49\%&-1.00\%\\
%\hline
\end{tabular}}
\caption{Values of resummed cross section in $10^{-5}$ pb/GeV at various orders in comparison to the fixed order results at different central scales $Q=\mu_R = \mu_F = 1000 ~\text{and} ~ 2000$ GeV for 13 TeV LHC.}
\label{Tab:7pointAll}
\end{small}
\end{center}
\end{table}
{\color{black} 
In Table \ref{Tab:7pointAll} we quote both fixed order and resummed predictions at various logarithmic accuracies along with asymmetric errors resulting from 7-point scale variation for two values of $Q$, namely $Q=1000$ GeV and $Q=2000$ GeV.  
We find that there is a systematic enhancement of the cross sections as we increase the order of perturbation.  
For example, there is an increment of 24.2\% when going from $\rm LO+\rm \overline {LL}$ to $\rm NLO+\rm \overline {NLL}$ accuracy, which further improves by 0.58\% at $\rm NNLO+\rm \overline {NNLL}$ for $Q=2000$ GeV.  In addition,  the scale uncertainity gets reduced significantly while going from $\rm LO+\rm \overline {LL}$ to $\rm NNLO+\rm \overline {NNLL}$. This is also reflected in Fig. \ref{7ptNall} (right panel), where one finds the uncertainity band of $\rm NNLO+\rm \overline {NNLL}$  is contained within the $\rm NLO+\rm \overline {NLL}$ band throughout the considered $Q$-range. 
This was not the case for the fixed order predictions, where NNLO band was found to differ from the NLO one at high energies. This hints to the notable NSV contributions coming from the resummation effects in diagonal channels.   These conclusions  might change if we include resummed effects from off-diagonal channels which are currently not available.
}
\begin{figure}[ht]
\centering
\includegraphics[scale=0.6]{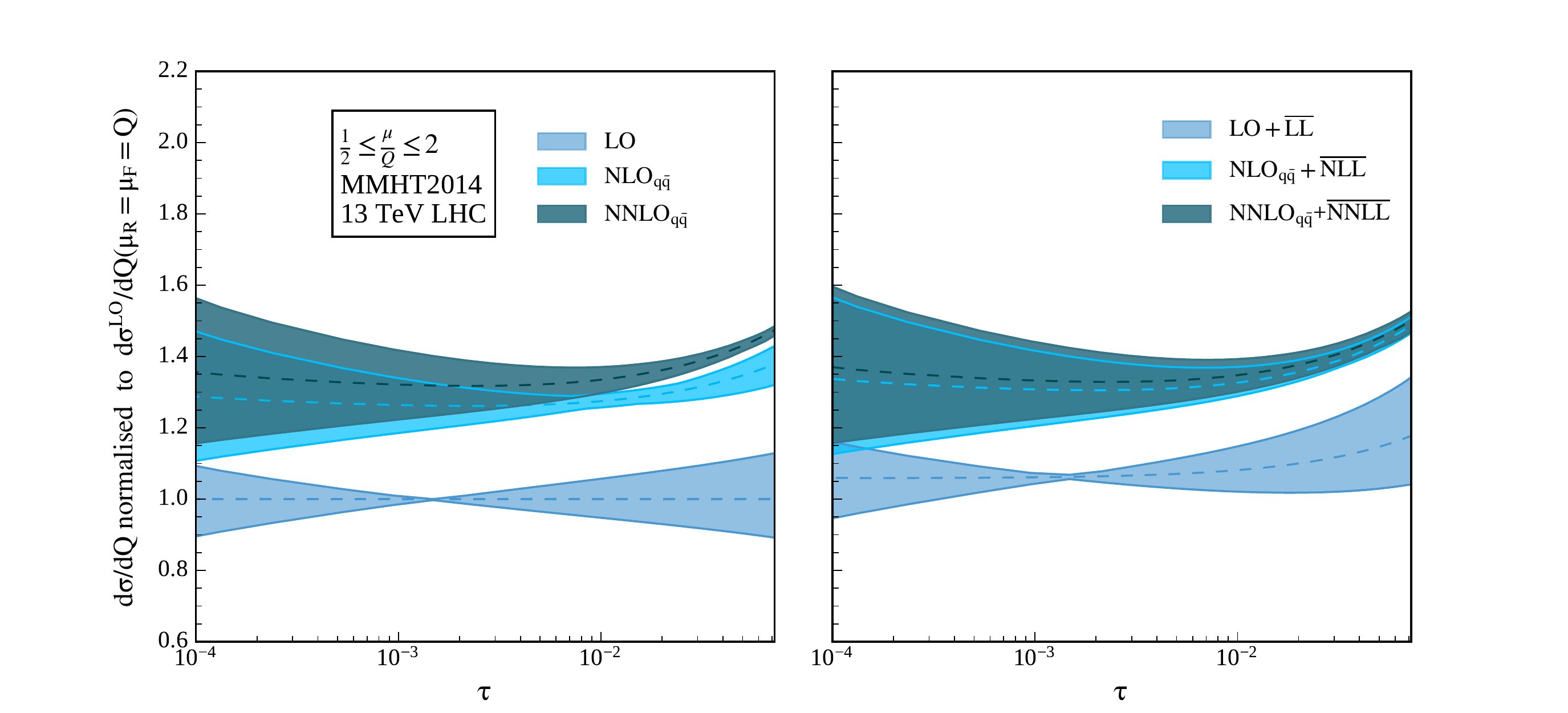}
\caption{7-point scale variation of the resummed result against fixed order around the central scale choice $(\mu_R,\mu_F) = (1,1)Q$ for $q \bar q$ channel.}
\label{fig6}
\end{figure}
\textcolor{black}{
In the above analysis, fixed order results used for the numerical predictions contained all the partonic channels while the resummed contributions are only from quark anti-quark initiated channels.  In the absence of resummed contributions,  under the 7-point scale variations, the scale dependence is expected to
go down as we increase the order of perturbation.  However, this may not be the case
if we include resummed effects only in quark anti-quark initiated channels. The quark gluon and gluon gluon initiated channels are also important as they contribute significantly to the cross section and more importantly they improve  the stability of perturbative predictions under the scale variations.   In order to understand the role of these 
partonic contributions, we drop them in the previous analysis  restricting to quark anti-quark initiated
contributions and then compare the outcomes given in Fig. \ref{fig6} and  in Table \ref{tab:ResumqQ}  against Fig. \ref{7ptNall} and the Table 
\ref{Tab:7pointAll}.
We find that there is a systematic enhancement of 28.19\% when going from $\rm LO+\overline{\rm LL}$ to $\rm NLO_{q \bar q}+\overline{\rm NLL}$ and 2.08\% from $\rm NLO_{q \bar q}+\overline{\rm NLL}$ to $\rm NNLO_{q \bar q}+\overline{\rm NNLL}$ for $Q=2000$ GeV. These increments are more compared to those in the case where all the channels are included in
the fixed order part. This stems from the fact that the cancellation which results due to the inclusion of other channels, in particular the $qg$-channel, is not considered here.
As a result of which 
the error bands in Fig.\ref{fig6} of 
$\rm NNLO_{q \bar q}+\overline{\rm NNLL}$ is wider than that of $\rm NLO_{q \bar q}+\overline{\rm NLL}$ in comparison to the resummed curves shown in Fig.\ref{7ptNall} (left panel).}
\begin{table}[!ht] 
\begin{center}
\begin{small}
%\newcolumntype{P}[1]{>{\centering\arraybackslash}p{#1}}
{\renewcommand{\arraystretch}{1.9}
\begin{tabular}{|p{1.2cm}||p{1.9cm}|P{1.5cm}||P{1.9cm}|P{1.5cm}|}
\rowcolor{lightgray}
\multicolumn{1}{c||}{$Q=\mu_R=\mu_F$(GeV)}
%    & \multicolumn{1}{c|}{ LO}   
%    &\multicolumn{1}{c||}{LO+$\overline{\rm LL}$ }  
    &\multicolumn{1}{c|}{NLO$_{q\bar q}$}  
    &\multicolumn{1}{c||}{NLO$_{q\bar q}$+$\overline{\rm NLL}$} 
    
    & \multicolumn{1}{c|}{NNLO$_{q\bar q}$}
    & \multicolumn{1}{c|}{NNLO$_{q\bar q}$+$\overline{\rm {NNLL}}$ }
\\
 \hline
%  \hline
1000 & $3.3204^{+1.91\%}_{-2.00\%}$  & $3.4452^{+4.18\%}_{-3.71\%}$   & $3.5260^{+3.46\%}_{-3.63\%}$  &$3.5576^{+4.16\%}_{-4.24\%}$ \\
\hline
2000 & $0.0676^{+1.65\%}_{-2.17\%}$  & $0.07102^{+2.33\%}_{-1.71\%}$   & $0.0717^{+1.63\%}_{-1.76\%}$  &$0.0725^{+2.37\%}_{-2.47\%}$ \\
\hline
%Drell-Yan &8.59\%&5.44\% & 9.82\% & 2.62\% & 1.49\%&-1.00\%\\
%\hline
\end{tabular}}
\caption{Values of resummed cross section at various orders in comparison to the fixed order results in $10^{-5}$ pb/GeV for $q\bar q$ channel at different central scales $Q=\mu_R = \mu_F = 1000 ~\text{and} ~ 2000$ GeV for 13 TeV LHC.}
\label{tab:ResumqQ}
\end{small}
\end{center}
\end{table}

\textcolor{black}{Hence in the 7-point variation, we find that the resummed result shows a systematic enhancement of the cross section as well as reduction of the uncertainties with the inclusion of each logarithmic corrections. But the scale uncertainties of the resummed result shows much improvement at the $\rm NLO+\rm \overline {NLL}$ than at the $\rm NNLO+\rm \overline {NNLL}$ level. To understand this and also the cause of the uncertainities better we now turn to analyze the effect of each scale individually on the resummed result. }
%%%%%%%%%%%%%%%%%%%%%%%%%%%%%%%%%%%%%%%%%%
%----------------------------------------------------
\subsubsection*{Uncertainities of the resummed results \textit{with respect to}  $\mu_R$ and $\mu_F$ }\label{sec:muRmuF}
\textcolor{black}{So far, we have studied the importance of both fixed order as well as resummed contributions using the K factor and the uncertainties arising from unphysical scales $\mu_R$ and $\mu_F$. The numbers in the Table \ref{Tab:Kfactor} demonstrate the importance of not only the fixed order contributions but also the effects coming from the resummed ones.  The resummed results comprises of both SV and NSV logarithms and the importance of large coefficients of the NSV terms was illustrated for the case of fixed order corrections in Table \ref{tab:N2LO}.  From the analysis of 7-point variation, we have shown that inclusion of leading, next-to-leading and next-to-next-to-leading collinear logarithms, in addition to the SV distributions, from the $q\bar q$ channel, to all orders in perturbation  theory through resummation significantly enhances the cross section.   But this enhancement comes with a price, namely the uncertainties resulting from  unphysical scales.  As we have seen in Fig. \ref{7ptNall} that while  the width of the resummed band at $\rm NLO +\overline{\rm NLL}$ is less than that of the corresponding fixed order ones from $Q = 1000$ GeV onwards, the width of the $\rm NNLO +\overline{\rm NNLL}$ doesn't show much improvement even at higher values of $Q$ as against the fixed order result.   In order to understand the reason behind this we aim to study the impact of the two scales separately. This is our next task.}  
%This will show us the behaviour of SV+NSV resummed result with respect to the individual scales and eventually reveal the role of two scales in the error bands of Fig \ref{7ptNall}.
\begin{figure}[!htp]
\centering
\includegraphics[scale=0.6]{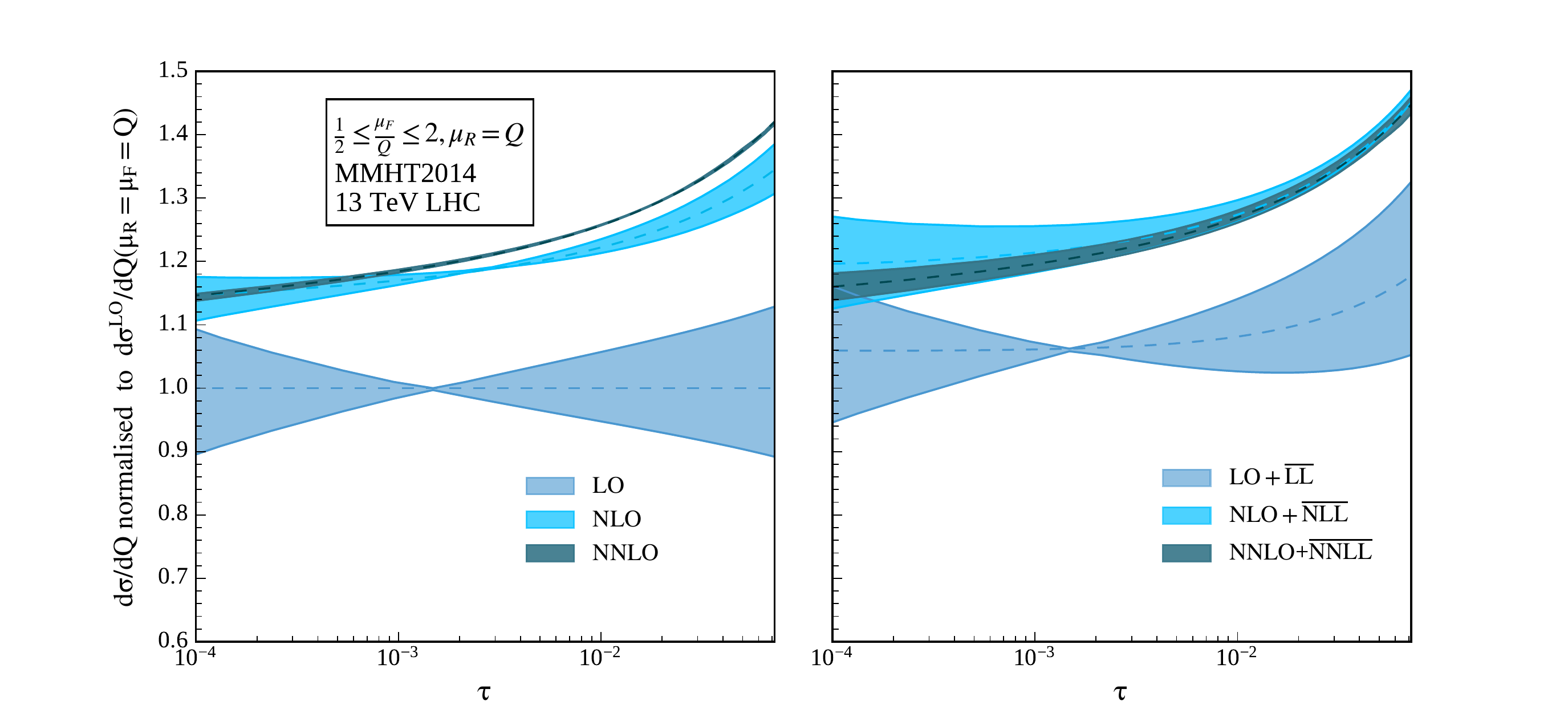}
\caption{$\mu_F$ scale variation of the resummed results against the fixed order with the scale $\mu_R$ held fixed at $Q$ for 13 TeV LHC.}
\label{muFall}
\end{figure} 

\textcolor{black}{The dependence of the cross section on $\mu_F$ is plotted in Fig. \ref{muFall}, as a function of $\tau$ with $\mu_R$ held fixed at $Q$. The bands are obtained by varying the scale $\mu_F$ by a factor of two up and down around the central scale $\mu_R=\mu_F=Q$. Here, the resummed bands look similar to that of Fig. \ref{7ptNall} (right panel), however the width of NLO+$\overline{\rm NLL}$ and NNLO+$\overline{\rm NNLL}$ bands become slightly thinner as compared to the 7-point scale variation. This suggest that the contribution to the width of the bands in Fig. \ref{7ptNall} mainly comes from the uncertainties arising from the $\mu_F$ variations. For instance, the uncertainity at NLO+$\overline{\rm NLL}$  with respect to $\mu_F$ variation   is 
$_{-0.52\%}^{+1.44\%}$ ~ whereas the uncertainties arising from the 7-point variation is $_{-1.23\%}^{+1.43\%}$ for $Q=2000$ GeV. Similarly at NNLO+$\overline{\rm NNLL}$, the uncertainity arising from $\mu_F$ variation is $_{-0.63\%}^{+0.79\%}$ and from 7-point variation is $_{-0.78\%}^{+0.89\%}$ for the same value of $Q$. 
Now let us compare the $\mu_F$ uncertainity of the resummed result with respect to their fixed order. We find that the width of the NLO band decreases with the inclusion of  $\overline{\rm NLL}$ from $Q=1600$ GeV on wards. But the $\mu_F$ uncertainity for NNLO increases at NNLO+$\overline{\rm NNLL}$. The reason behind this stems from the contribution coming from $qg$-channel. The one-loop correction from the $q\bar q$-channel is 22.09\% of the NLO cross section, whereas the correction at the same order from the $qg$-channel is about -5.04\% of the NLO cross section. Now at NLO+$\overline{\rm NLL}$,  we sum up the collinear logarithms from the diagonal channel which is also the dominating channel at NLO and hence the improvement  through resummation. But the scenario is different at NNLO level. The $a_s^2$ corrections from $q\bar q$ and $qg$-channel contribute to $4.86\%$  and $-2.47\%$ respectively to the NNLO cross section along with the other sub-dominating channels. However the sources of collinear logarithms at the threshold limit is only from $q\bar q$ and $qg$-channels. Hence at NNLO there is a bigger cancellation between $q\bar q$ and $qg$-channels than that at NLO. The cancellation at 
NNLO+$\overline{\rm NNLL}$ is not matched due to the unavailability of the $qg$ resummed collinear logarithms.  Thus the $\mu_F$ variation in Fig. \ref{muFall} reflects the role of the other channels and the need for $qg$ resummation for betterment of the result.} 
\begin{figure}[!ht]
\begin{center}
\includegraphics[scale=0.6]{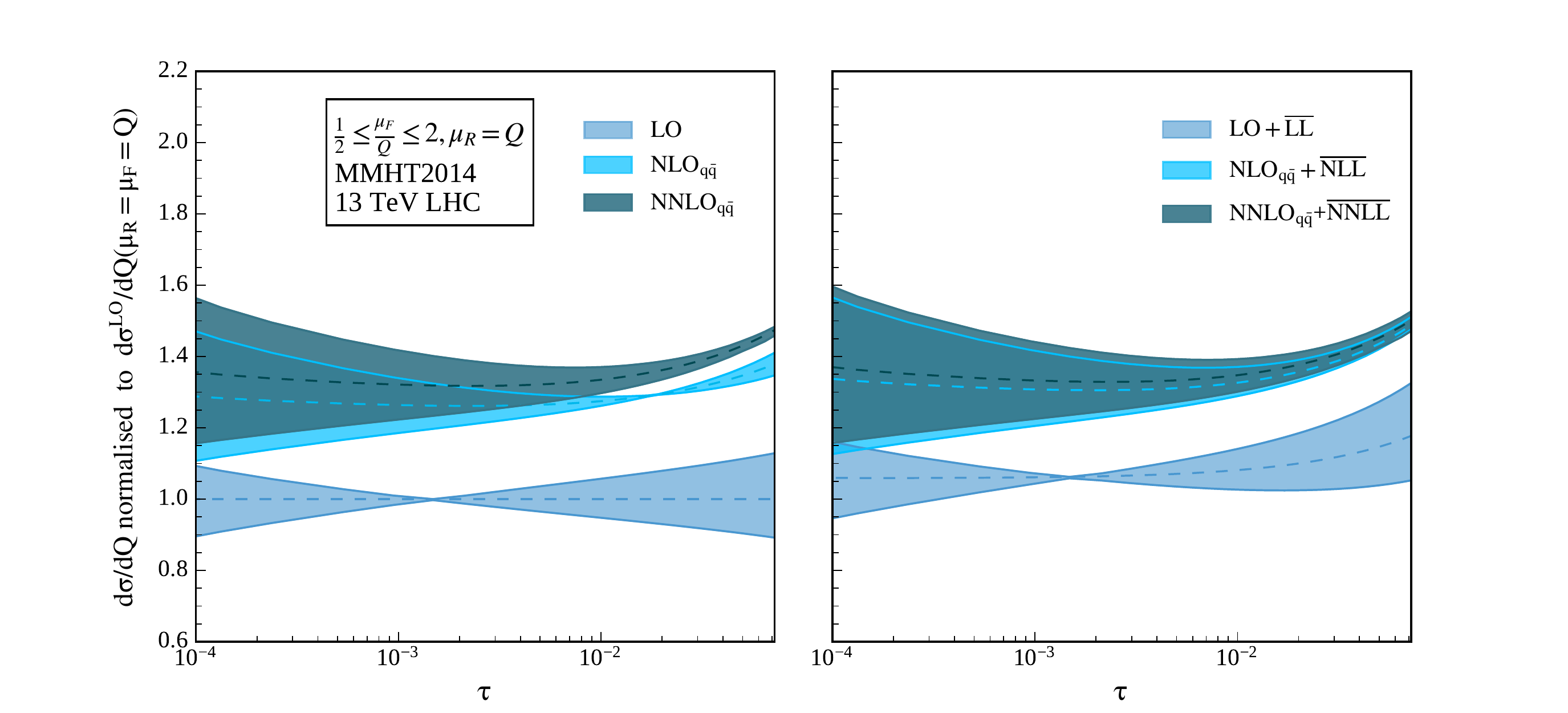}
\caption{$\mu_F$ scale variation of the resummed results against the fixed order with the scale $\mu_R$ held fixed at $Q$ for $q\bar q$-channel for 13 TeV LHC. }
\label{fig:muFq}
\end{center}
\end{figure} 

\textcolor{black}{Let us try to understand why the inclusion of resummed contributions in quark gluon initiated channels is important.  Note that in the
above analysis, we had taken all the partonic channels for the fixed order part and only quark anti-quark initiated channels for the resummed part.  We found that in the fixed order, the $\mu_F$ dependence from the PDFs and from quark anti-quark initiated as well
as from the quark-gluon initiated processes are expected to compensate each other according to renormalisation group equation with respect to factorisation scale.  However, in the resummed part, this will not happen due  the absence of resummed quark gluon counter part and this is the reason why one gets  larger  $\mu_F$ dependence in the predictions at NNLO+$\overline{\rm NNLL}$ level.  A symmetric analysis where we keep only quark anti-quark initiated channels
both in fixed and resummed contributions can demonstrate this better and hence the Fig. \ref{fig:muFq}.  As one can see easily,
the bands in both fixed order and resummed predictions are  wider compared to those in the Fig. \ref{muFall}} indicating
the importance of quark gluon initiated channel both in fixed as well as resummed parts.
%\\
\begin{figure}[!ht]
\centering
\includegraphics[scale=0.6]{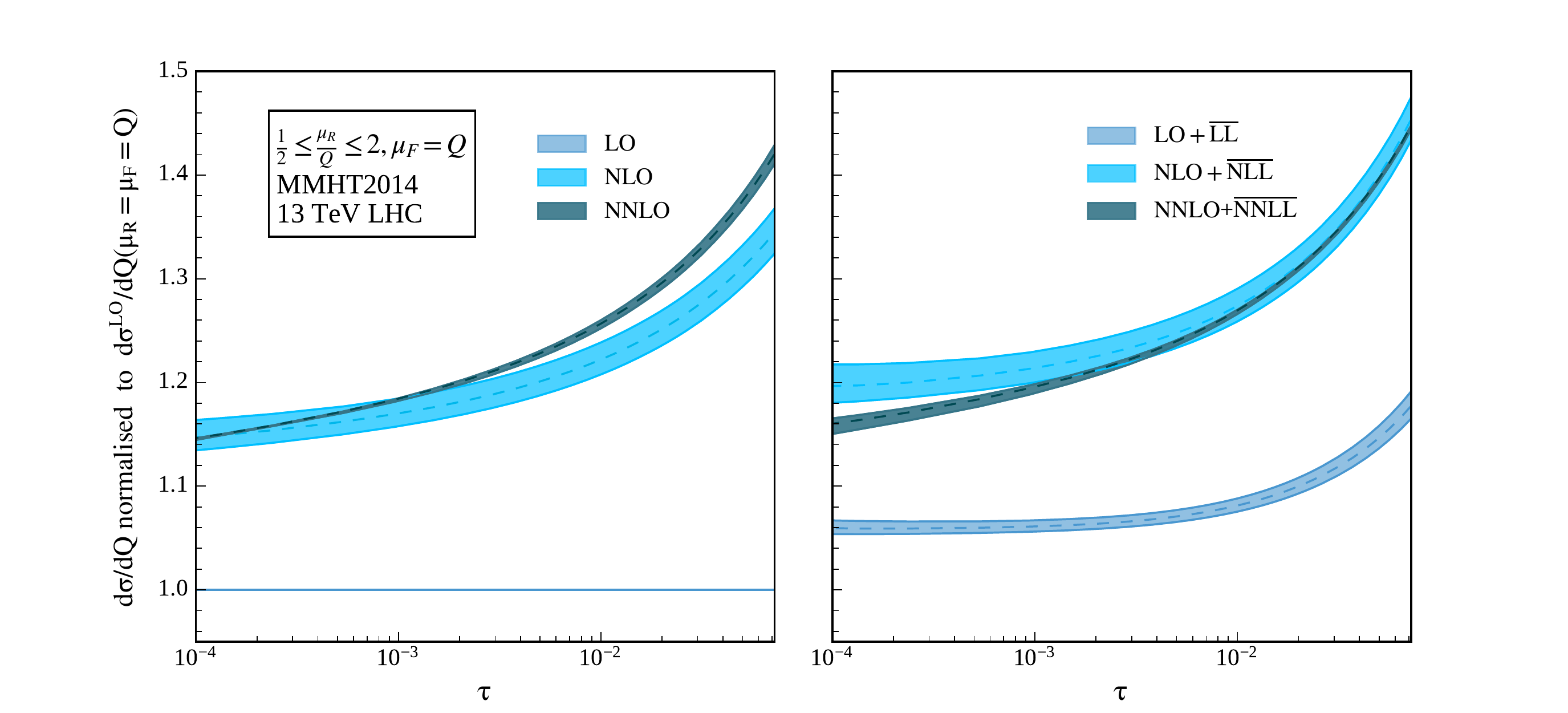}
\caption{$\mu_R$ scale variation of the resummed results against the fixed order with the scale $\mu_F$ held fixed at $Q$ for 13 TeV LHC.}
\label{fig3}
\end{figure} 
%It is well known that each partonic channel is invariant under $\mu_R$ variation and hence inclusion of more corrections within a channel is expected to reduce the scale uncertainity. On the other hand, different partonic channels mix under $\mu_F$ variation and hence inclusion of other channels is expected to reduce the scale uncertainity. Since we have achieved the resummation of only diagonal $q \bar q$ channel, it is more desirable to study the effects of NSV resummation by considering only $q \bar q$ channel throughout. This will be our focus in the next section. }\\
%-------------------------------------------------

\textcolor{black}{Fig. \ref{fig3} shows the dependence of the cross section on $\mu_R$ keeping $\mu_F$ fixed at $Q$. The bands are obtained by varying the scale $\mu_R$ by a factor of two up and down around the central scale $\mu_R=\mu_F=Q$. We observe that at NNLO+$\overline{\rm NNLL}$ the error band becomes substantially thinner as compared to Fig. \ref{muFall}. This is because each partonic
channel is invariant under $\mu_R$ variation when taken to all orders and hence inclusion of more corrections within a channel is expected to reduce the uncertainity.  
%However, different partonic channels mix under variation of factorisation scale $\mu_F$.
We find that the $\mu_R$ uncertainity at NLO+$\overline{\rm NLL}$ ranges between $_{-1.23\%}^{+ 1.35\%}$ whereas for NLO it is between $_{-1.28\%}^{+1.46\%}$ for $Q=2000$ GeV. And at NNLO+$\overline{\rm NNLL}$ the uncertainity is found to be  $_{-0.23\%}^{+0.02\%}$ and for NNLO it is $_{ -0.46\%}^{+  0.37\%}$ for the same value of $Q$. Hence from Fig. \ref{fig3} one can see that $\mu_R$ dependence goes down for $Q= 800$ GeV on wards.}

\begin{figure}[H]
\centering
\includegraphics[scale=0.6]{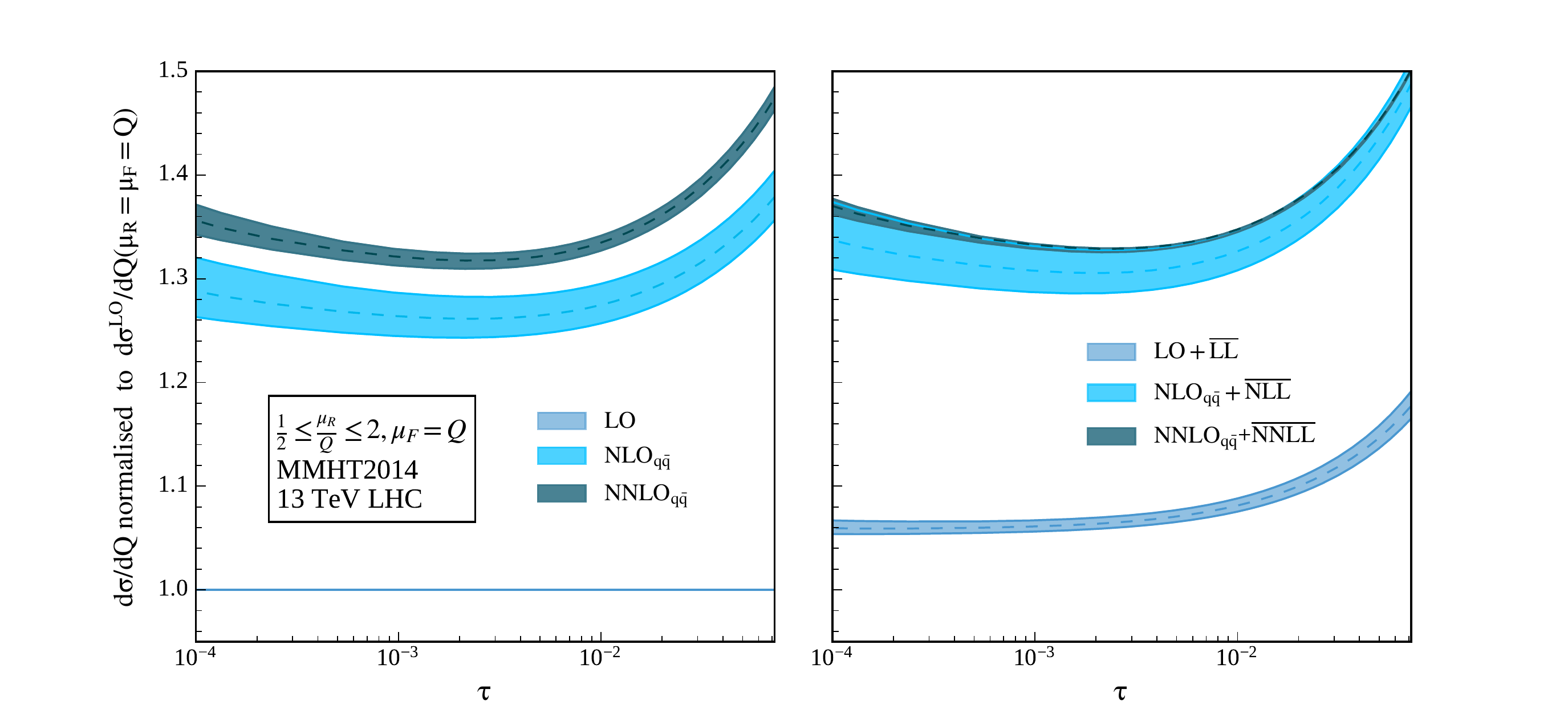}
\caption{$\mu_R$ scale variation of the resummed results against the fixed order with the scale $\mu_F$ held fixed at $Q$ for $q\bar q$-channel for 13 TeV LHC.}
\label{muRq}
\end{figure}

\textcolor{black}{Since different partonic channels do not mix under the variation of $\mu_R$, the symmetric analysis of keeping only quark anti-quark initiated channels will have similar behaviour as that of the case where other partonic channels are included in
the fixed order, which is given in Fig. \ref{muRq}. The $\mu_R$ uncertainity is significantly decreased as we go from NLO to NLO+$\overline{\rm NLL}$ to NNLO to NNLO+$\overline{\rm NNLL}$. We find the uncertainity at NNLO $_{ -0.57\%}^{+ 0.48\%}$ gets improved to $_{ -0.20\%}^{+0.006\%}$ for $Q = 1000$ GeV.  And this improvement continues to grow even for higher values of $Q$.
Therefore the inclusion of resummed result reduces the $\mu_R$ uncertainly remarkably as compared to the fixed order ones.}

Hence in conclusion to this section, we find that the uncertainity, which earlier manifested in the 7-point variation, shown in Fig. \ref{7ptNall}, was largely due to the $\mu_F$ uncertainity. Now as we know that different partonic channels mix under the variation of factorisation scale $\mu_F$ and so any uncertainity arising due to its variation only hints towards the ``uncompensated''  contributions from other channels. The fact that the resummed result at NLO+$\overline{\rm NLL}$ shows improvement as compared to the fixed order, emphasizes the importance of the resummed NSV logarithms. Similarly at the NNLO+$\overline{\rm NNLL}$ level the large cancellation between different channels, mainly $q\bar q$ and $qg$, equally hints towards the importance of the collinear logarithms of the $qg$ channel. Similarly for the $\mu_R$ variation, where the dependency is supposed to get better with the inclusion of more corrections within a partonic channel, we see a substantial improvement due to SV+NSV resummation in comparison to the fixed order result. But now that we have seen the effect of the combined resummed result on the fixed order  let us analyze  which part of the SV+NSV resummation, i.e., whether its the resummation of the distribution or of the NSV logarithms, plays the dominant role in any kind of improvement discussed so far.
% In Fig.\ref{muRq} (left panel), we study the distributions with the scale $\mu_R$ held fixed at $Q$ where the bands are obtained by varying the scale $\mu_F$ by a factor of 2 up and down for $13$ Tev LHC. We observe that the NLO$_{q \bar q}$ error band is wider as compared to the All-channel $\mu_F$ variation plot at small $Q$ values. But at large $Q$ values, the NLO$_{q \bar q}$ band becomes thinner as compared its All-channel counterpart. The NNLO band for the case of All-channel is found to be significantly thinner than the NNLO$_{q \bar q}$ band and this holds true even when we include NNLL. This hints to the existing fact that there is a large cancellation between $q \bar q$ and $qg$ channels. \\
%----------------------------------------------
%\begin{figure}[ht]
%\centering
%\includegraphics[scale=.6]{FigB/BandqMuFfix.pdf}\hfill
%\caption{Invariant mass distribution for a pair of leptons with the scale $\mu_F$ held fixed at $Q$. The bands are obtained by varying the scale $\mu_R$ by a factor of 2 up and down for $13$ Tev LHC.}
%\label{fig8}
%\end{figure}

% Figure \ref{muRq} (right panel) shows the invariant mass distribution with the scale $\mu_F$ held fixed at $Q$ where the bands are obtained by varying the scale $\mu_R$ by a factor of 2 up and down for $13$ Tev LHC. Here, the width of the bands decreases substantially as we go to NNLO$_{q \bar q}$+NNLL as it is observed in its All-channel counterpart. 

%-----------------------------------------------------
\subsection{SV resummation vs SV+NSV resummation}

{\color{black}
In the earlier section we have made a quantitative comparison of SV+NSV resummed results against the fixed order ones. We found that there is a significant enhancement of the cross section  and the $\mu_R$ scale uncertainity gets substantially improved with the inclusion of the resummed corrections. We also found that the uncertainties related to the  $\mu_F$ variation shows betterment at NLO+$\rm \overline{NLL}$ for higher $Q$ values but not at the NNLO+$\rm \overline{NNLL}$ level. Now, in this section we turn to a detailed analysis on the inclusion of NSV resummation over SV resummation so as to estimate the effect of resummed collinear logarithms from the $q\bar q$-channel.}

\begin{figure}[ht]
\centering
\includegraphics[scale=.6]{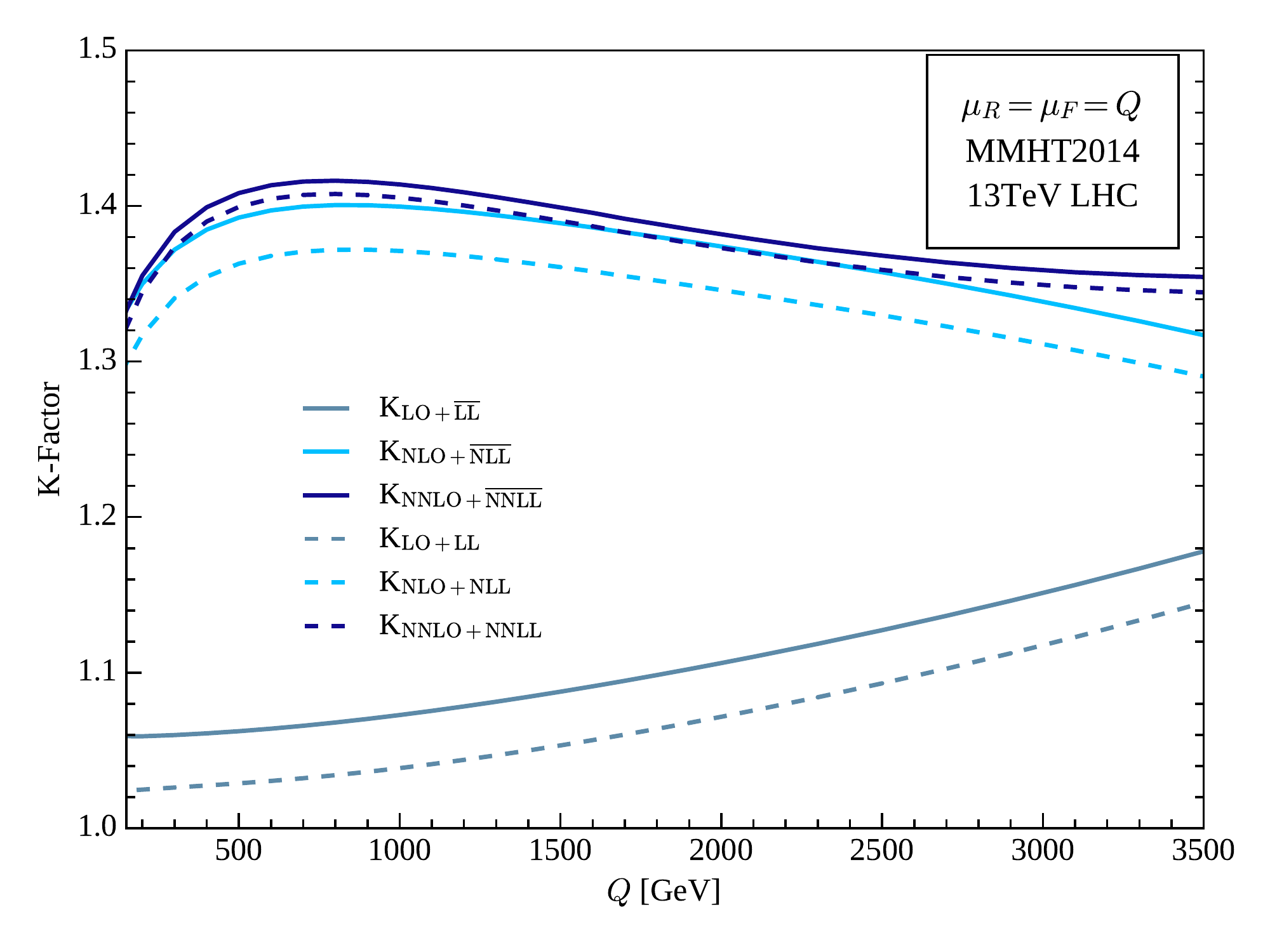}
\caption{A comparison of the K-factors of both SV and SV+NSV resummation for $13$ TeV LHC.}
\label{ksvnsv}
\end{figure}

{\color{black}
We begin with the K-factor, which is presented in Table \ref{tab:svnsvK}, to examine the impact of resummed NSV logarithms.  Keeping all the partonic channels in the fixed order, we find 
that the inclusion of the resummed NSV logarithms enhances the SV resummed corrections significantly throughout the considered $Q$ range. In particular, for $Q = 2000$ GeV, there is a considerable amount of increment of $2.08\%$ when we go from NLL to $\rm \overline{NLL}$ and 0.64\% from NNLL to $\rm \overline{NNLL}$.}\\
\begin{table}[H] 
\begin{center}
\begin{small}
%\newcolumntype{P}[1]{>{\centering\arraybackslash}p{#1}}
{\renewcommand{\arraystretch}{1.7}
\begin{tabular}{|P{1cm}||P{1.7cm} |P{1.7cm}||p{1.7cm}|P{1.7cm}|}
\rowcolor{lightgray}
\multicolumn{1}{c||}{$Q=\mu_R=\mu_F$}
      &\multicolumn{1}{c||}{NLO+NLL }  
    &\multicolumn{1}{c||}{NLO+$\overline{\rm NLL}$} 
    & \multicolumn{1}{c|}{NNLO+NNLL}
    & \multicolumn{1}{c|}{NNLO+$\overline{\rm {NNLL}}$ }
    \\ 
 \hline
%  \hline
1000 & 1.3711 & 1.3995 & 1.4053 & 1.4138    \\
\hline
2000 &1.3459 & 1.3739& 1.3729 & 1.3818  \\
\hline
%Drell-Yan &8.59\%&5.44\% & 9.82\% & 2.62\% & 1.49\%&-1.00\%\\
%\hline
\end{tabular}}
\caption{The K-factor values for NSV resummed result in comparison to the SV resummed predictions at various logarithmic accuracy.}
\label{tab:svnsvK}
\end{small}
\end{center}
\end{table}
\vspace{-0.5cm}
Fig. \ref{ksvnsv} demonstrates this for a wider range of $Q$ values.  We can also observe that the curves corresponding to SV+NSV resummed results at $\rm NLO+\overline{\rm NLL}$ and $\rm NNLO+\overline{\rm NNLL}$ are closer compared to the SV counter parts, accounting for the perturbative convergence when NSV effects are taken into consideration. It also shows that the resummed correction decreases when we go to higher logarithmic accuracy as  the resummed curves at the second order are closer than that of the first order ones.

\begin{figure}[!ht]
\centering
\includegraphics[scale=0.55]{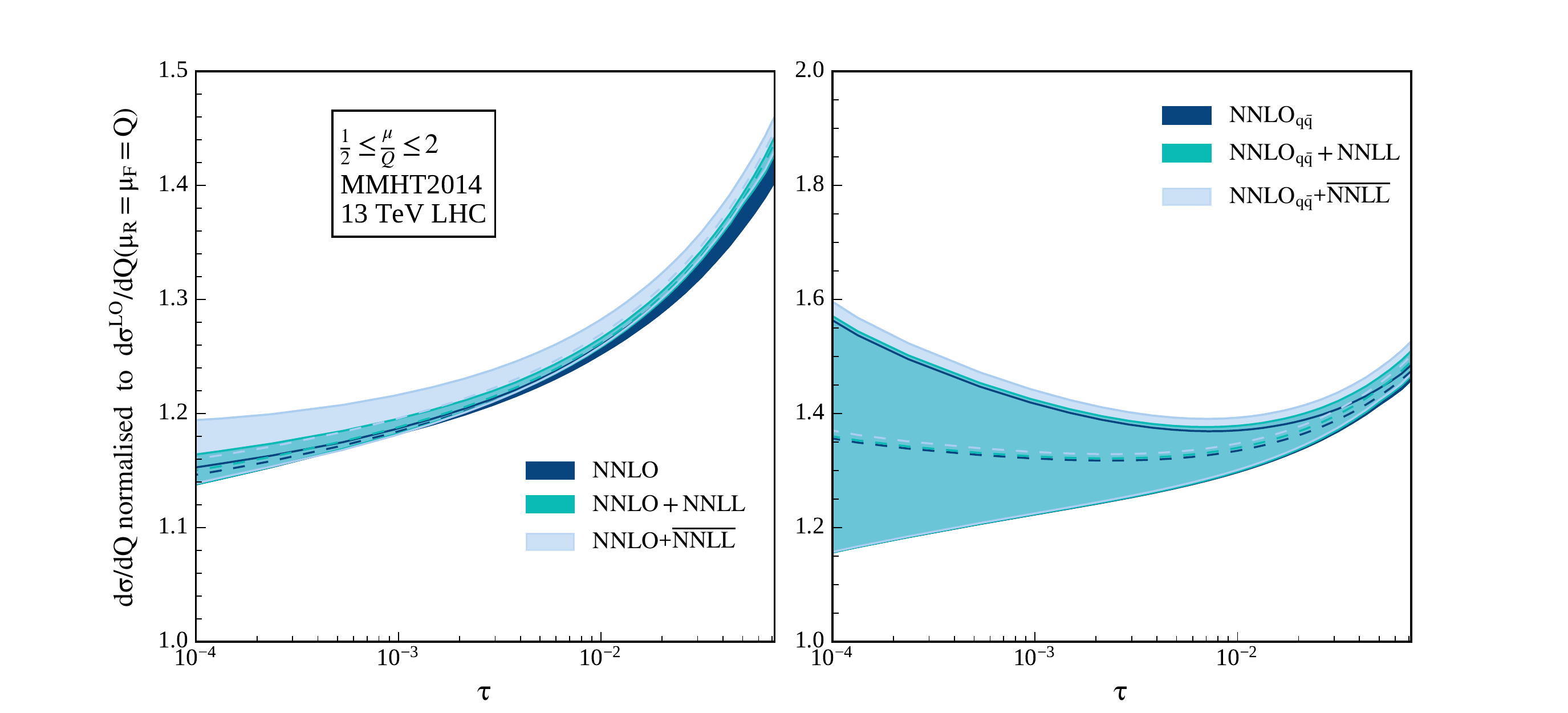}
\caption{7-point scale variations of NNLL and $\overline{\rm NNLL}$ matched to NNLO for all-channels (left panel) and $q\bar q$-channel (right panel).
}
\label{svnsv7}
\end{figure}

{\color{black}
We now analyze the scale uncertainties of the SV+NSV resummed results in comparison to the SV resummation. In Fig. \ref{svnsv7} we plot the SV and SV+NSV resummed results at various logarithmic accuracy as a function of $\tau$ taking into account the respective 7-point scale variations.
Note that the SV+NSV resummed predictions are more sensitive to the scales compared to the SV ones. For instance, for $Q=2000$ GeV, we find the scale uncertainity of SV resummed results, which was in between $_{-0.27\%}^{+0.32\%}$,  is enhanced to $_{-0.78\%}^{+0.89\%}$ when the resummed NSV corrections are added. 
This further hints towards our earlier findings in sec. \ref{sec:muRmuF}, that the absence of complete resummation of collinear logarithms, which includes both diagonal as well as off-diagonal contributions, causes the SV+NSV resummed predictions to be more sensitive to the unphysical scales. }
%\vspace{-0.3cm}
%\vspace{-0.1cm}

\textcolor{black}{However the SV resummation, which  gets contributions only from the diagonal channel, shows improvement in 7-point scale uncertainties for higher values of $Q$. 
Similar findings are observed even if we restrict ourselves to only diagonal ($q\bar q$) channel as shown in right panel of Fig. \ref{svnsv7}.} For comparison, we present the SV and SV+NSV resummed results along with the fixed order predictions for $Q=1000, 2000$ GeV with their respective percentage scale uncertainties in Table \ref{Tab:svnsv7point} at the second order.
\begin{table}[!htp] 
\begin{center}
\begin{small}
%\newcolumntype{P}[1]{>{\centering\arraybackslash}p{#1}}
{\renewcommand{\arraystretch}{1.9}
\begin{tabular}{|P{.8cm}||P{1.8cm} |P{1.8cm}||p{1.8cm}|}
\rowcolor{lightgray}
\multicolumn{1}{c||}{$Q= \mu_R=\mu_F$}
    &\multicolumn{1}{c|}{NNLO}  
    & \multicolumn{1}{c|}{NNLO +NNLL}
    & \multicolumn{1}{c|}{NNLO+$\overline{\rm {NNLL}}$}
    \\ 
 \hline
%  \hline
1000 & $3.2876^{+0.20\%}_{-0.31\%}$ & $3.2993^{+0.36\%}_{-0.29\%}$ & $3.3191^{+1,13\%}_{-0.86\%}$    \\
\hline
2000 &$  0.0684^{+0.37\%}_{-0.62\%}$ &$0.0687^{+0.32\%}_{-0.27\%}$ & $0.0692^{+0.89\%}_{-0.78\%}$     \\
\hline
%Drell-Yan &8.59\%&5.44\% & 9.82\% & 2.62\% & 1.49\%&-1.00\%\\
%\hline
\end{tabular}}
\caption{Values of SV and SV+NSV resummed cross section in $10^{-5}$ pb/GeV at second logarithmic accuracy in comparison to the fixed order results at different central scales .}
\label{Tab:svnsv7point}
\end{small}
\end{center}
\end{table}
%\vspace{-0.5cm}

{\color{black}
The width of the band is expected to reduce when we fix the factorisation scale and vary only the renormalisation scale, since the effect of latter gets cancelled within a given partonic channel. Hence we now turn to analyze the effect of each these scales separately on the resummed result.}

\begin{figure}[!ht]
\centering
\includegraphics[scale=0.55]{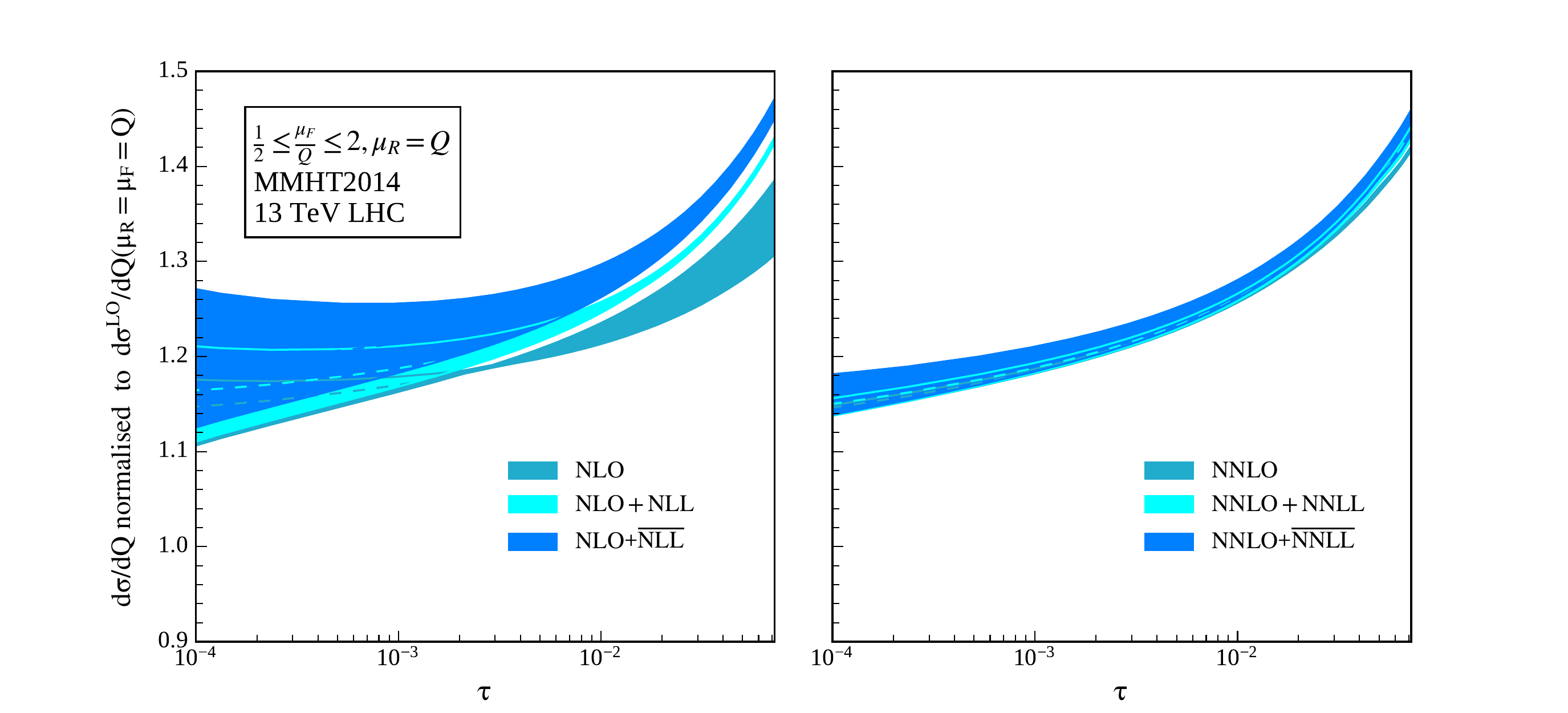}
\caption{$\mu_F$ variation between SV and SV+NSV resummed results matched to NLO (left panel) and NNLO (right panel) with the scale $\mu_R$ held fixed at $Q$.}
\label{svnsvmuF}
\end{figure}

{\color{black}
As seen in sec. \ref{sec:muRmuF},  the large uncertainties in the 7-point variations were mostly from the $\mu_F$ variation 
% in SV+NSV resummed results within a particular channel ($i.e,$ diagonal $q\bar q$ channel) as we do not have corresponding results for off-diagonal channels. Hence, there is no scope of cancellation of collinear logarithms among different channels. This is evident when
and hence we compare in Fig. \ref{svnsvmuF}  the $\mu_F$ sensitivity of SV and SV+NSV resummed cross sections as a function of $\tau$, keeping $\mu_R$ fixed at $Q$.
%against the Fig \ref{svnsv7} (left channel). 
 Note here that the bands are obtained by varying only $\mu_F$ by a factor of 2 up and down around the central scale $\mu_R = \mu_F =Q$. We find, 
% that the SV resummed result shows the least uncertainity in comparison to the fixed order as well as the SV+NSV resummed results at both NLO+NLL and NNLO+NNLL level. In particular,  we can see in Fig \ref{svnsvmuF} (left panels), 
that where the $\mu_F$ band of fixed order, i.e., at NLO, starts widening for higher values of $Q$, the resummed bands i.e., at NLO+NLL and NLO+$\overline{\rm NLL}$, shows a systematic reduction in the width for the same range of $Q$. But there is larger reduction in the SV resummed case in comparison to the SV+NSV. On the other hand the $\mu_F$ band of NNLO is better than both SV and SV+NSV for the considered $Q$ range. This clearly hints towards the level of cancellation between different partonic channels. At NNLO there is a significant cancellation between $q\bar q$ and $qg$-channel which leads to the reduction of the $\mu_F$ width from NLO to NNLO and the SV resummation, which resums the distributions present only in the diagonal channel, doesn't require any compensation from other channels. But when it comes to SV+NSV resummation there is a compensation required from the other partonic channels to enhance the $\mu_F$ stability.  Note that in all these analysis, we studied the impact of fixed order and  resummed CFs using
same PDF sets to desired logarithmic accuracy for both of them.  For studies related to $\mu_F$ variations, it is worthwhile
to consider resummed PDFs if they are available.}
%\vspace{-0.1cm}
% For a quantitative comparison, say at Q=2000 GeV, the uncertainity arising for NNLO+NNLL due to $\mu_F$ variation is $^{+0.32\%}_{-0.27\%}$, which is identical to the one  that comes from 7-point scale variation. However, for NNLO+$\rm \overline{NNLL}$, the uncertainity due to $\mu_F$ variation is slightly improved to $^{+0.80\%}_{-0.63\%}$ from those from 7-point variation, which is $^{+0.90\%}_{-0.79\%}$, for the same Q.
% Hence again, this indicates that we require resummed contributions from $qg$ channel for improving the uncertainity. }\\
% \textcolor{blue}{ The above needs to be corrected}\\

\begin{figure}[!ht]
\centering
\includegraphics[scale=0.55]{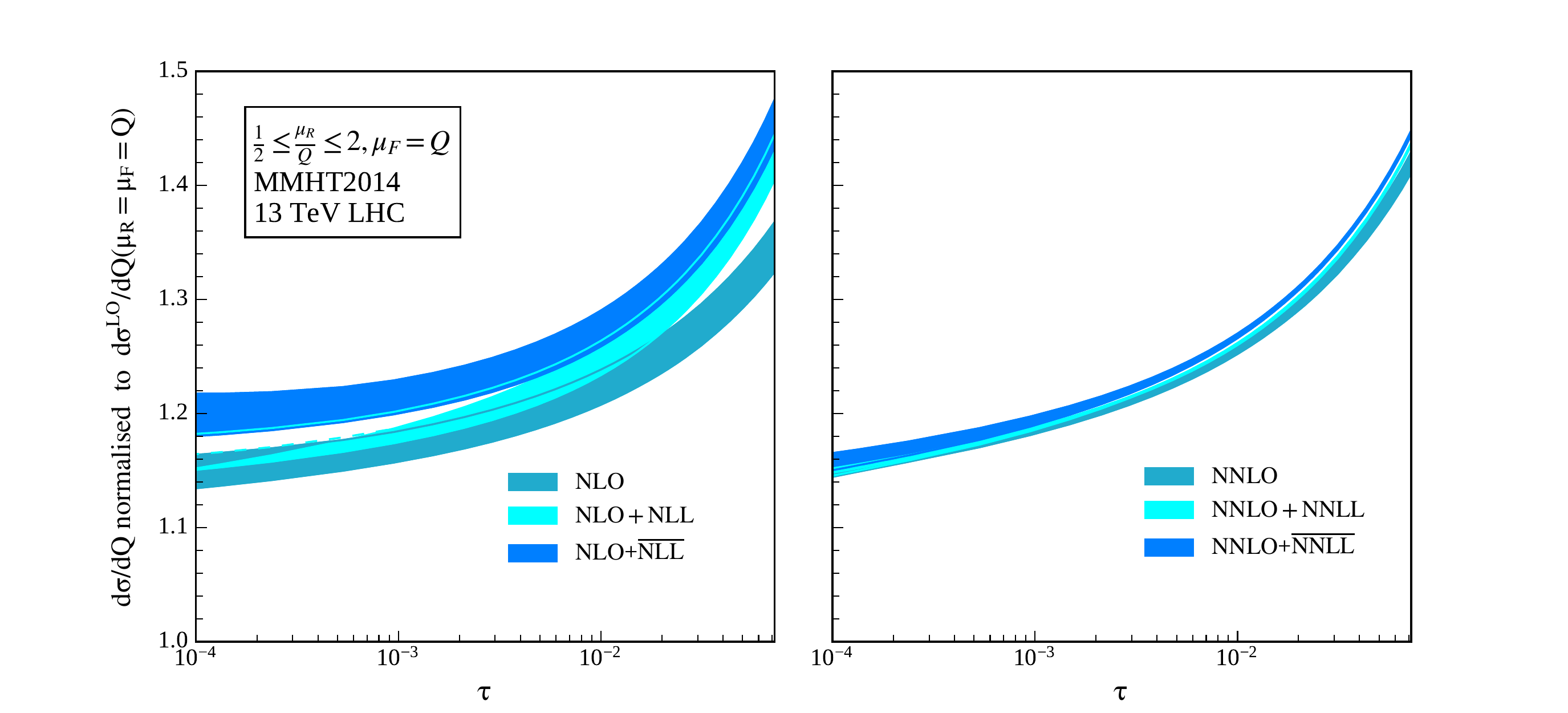}
\caption{$\mu_R$ variation between SV and SV+NSV resummed results matched to NLO(left panel) and NNLO(right panel) with the scale $\mu_F$ held fixed at $Q$.}
\label{svnsvmuR}
\end{figure}

\textcolor{black}{
 However, this is not the case if we keep the $\mu_F$ intact and study the sensitivity due to $\mu_R$ variation as the latter effects are supposed to cancel within a partonic channel. This is depicted in Fig. \ref{svnsvmuR}, where we vary the $\mu_R$ dependence of cross section as a function of $\tau$ with $\mu_F$ held fixed at $Q$. As can be seen from the plot, the width of the $\mu_R$ band is  significantly reduced in case of resummed bands in comparison to the fixed order band, leading to a reliable predictions from the resummation. And among the resummed bands, the uncertainity of SV+NSV resummed results are comparable to those of SV resummation.
% we see that at NLO level the SV band is comparable to the SV+NSV band but at NNLO level the improvement achieved by SV+NSV resummation is much more than the SV one. 
For instance, for $Q = 1000$ GeV, the uncertainity is $_{-1.11\%}^{+1.26\%}$ at NLO+NLL whereas it is $_{-1.15\%}^{+1.28\%}$ at NLO+$\rm \overline{NLL}$. Similarly, the uncertainity at NNLO+NNLL is $_{-0.12\%}^{+0.0\%}$ whereas at NNLO+$\rm \overline{NNLL}$ is it found to be $_{-0.30\%}^{+0.01\%}$ for the same value of $Q$.
% However, it can be noted that, the uncertainity of SV+NSV resummed results are comparable to those of SV resummation. In particular for Q=2500 GeV, the $\mu_R$ uncertainity for NNLO+NNLL is $^{+0.0\%}_{-0.15\%}$, while for NNLO+$\rm \overline{NNLL}$ it is $^{+0.017\%}_{-0.23\%}$. This is acceptable, since contribution of SV logarithms are only coming from diagonal channels, whereas for NSV there are off-diagonal contributions as well, which are not taken care in the present work. Note that, at the fixed order we take contributions from all partonic channels but the resummation effects are considered only for diagonal channels. 
}
%%%%%%%%%%%%%%%%%%%%%%%%%%%%%%%%%%%%%%%%%%%%%%%%%%%%%%%%%%%%%%%%%%%%%%%%%%%%%%%%%%%%%%%%%%%%%%%%%%%%%%%%%%%%%%%%%%%

\begin{figure}[!ht]
\centering
\includegraphics[scale=0.6]{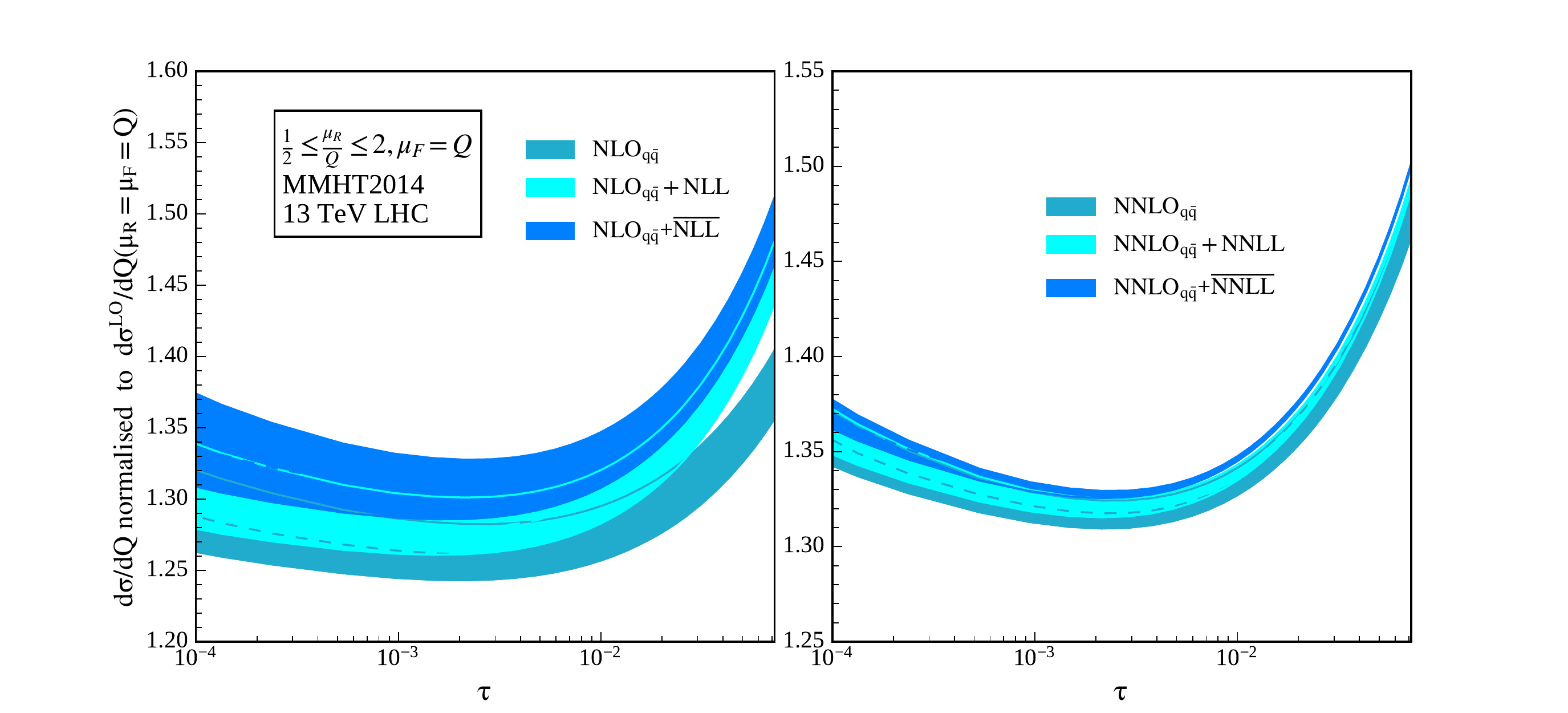}
\caption{$\mu_R$ variation between SV and SV+NSV matched to NNLO$_{q \bar q}$ with the scale $\mu_F$ held fixed at $Q$ in $q\bar q$-channel.}
\label{svnsvmuRq}
\end{figure}

In order to see the NSV effects more clearly, let us focus on the sensitivity of the predictions to the choice of the scale $\mu_R$ within   the $q\bar q$ channel as a function of $\tau$, which is depicted in Fig. \ref{svnsvmuRq}. Interestingly, the behaviour of $\rm {NNLO}_{q \overline q}+\rm \overline{NNLL}$ is significantly improved from the corresponding SV results, $\rm {NNLO}_{q \overline q}+\rm{NNLL}$, for a wide range of  $Q$.   There is a large increment in the central value, in addition to the significant reduction of the $\mu_R$ uncertainity. To see this quantitatively, we quote in Table \ref{Tab:svnsvqQ} both  the fixed order and SV and SV+NSV resummed predictions along with asymmetric errors resulting from $\mu_R$ variation with $\mu_F$ held fixed at $Q$, say $Q=1000$ and 2000 GeV. The uncertainties, in Table \ref{Tab:svnsvqQ},  are obtained with respect to the $\mu_R$ scale variation with $\mu_F$ held fixed at $Q$. This evidently shows that there is a considerable improvement when adding NSV resummation over the existing SV results and hence leading to more reliable predictions.

\textcolor{black}{To summarise, our analysis on the sensitivity of hadronic results to $\mu_F$ and $\mu_R$ scales where the resummed parts  from 
$q\bar q$ channel is taken into account, helps us to understand the role of various channels as well as on the PDFs that contribute. 
As far as the $\mu_R$ scale is concerned,  inclusion of SV as well as NSV resummed contributions in CFs alone helps to reduce the
sensitive to this scale.  However, this is not the case for $\mu_F$ as our numerical study shows the need for resummed
NSV contributions to CFs of $qg$-channel as well as the resummed PDFs.}
%
%Hence largely improves the reliability of Drell-Yan corrections. 
\begin{table}[!htp] 
\begin{center}
\begin{small}
%\newcolumntype{P}[1]{>{\centering\arraybackslash}p{#1}}
{\renewcommand{\arraystretch}{1.8}
\begin{tabular}{|P{.8cm}||P{1.8cm} |P{1.8cm}||p{1.8cm}|}
\rowcolor{lightgray}
\multicolumn{1}{c||}{$Q= \mu_R=\mu_F$}
    &\multicolumn{1}{c|}{${\rm NNLO}_{q\bar q}$}  
    & \multicolumn{1}{c|}{${\rm NNLO}_{q\bar q} +\rm NNLL$}
    & \multicolumn{1}{c|}{${\rm NNLO}_{q\bar q} + \overline{\rm {NNLL}}$}
    \\ 
 \hline
%  \hline
1000 & $3.5260^{+0.49\%}_{-0.58\%}$ & $3.5376^{+0.25\%}_{-0.39\%}$ & $3.5576^{+0.006\%}_{-0.20\%}$    \\
\hline
2000 &$  0.0717^{+0.54\%}_{-0.62\%}$ &$0.0721^{+0.19\%}_{-0.33\%}$ & $0.0725^{+0.0\%}_{-0.15\%}$    \\
\hline
%Drell-Yan &8.59\%&5.44\% & 9.82\% & 2.62\% & 1.49\%&-1.00\%\\
%\hline
\end{tabular}}
\caption{Comparison of SV and SV+NSV resummed cross section in $10^{-5}$ pb/GeV for $q\bar q$-channel at different central scales.}
\label{Tab:svnsvqQ}
\end{small}
\end{center}
\end{table}
\subsection{SV+NSV resummation in different schemes}
\begin{figure}[!ht]
\centering
\includegraphics[scale=0.55]{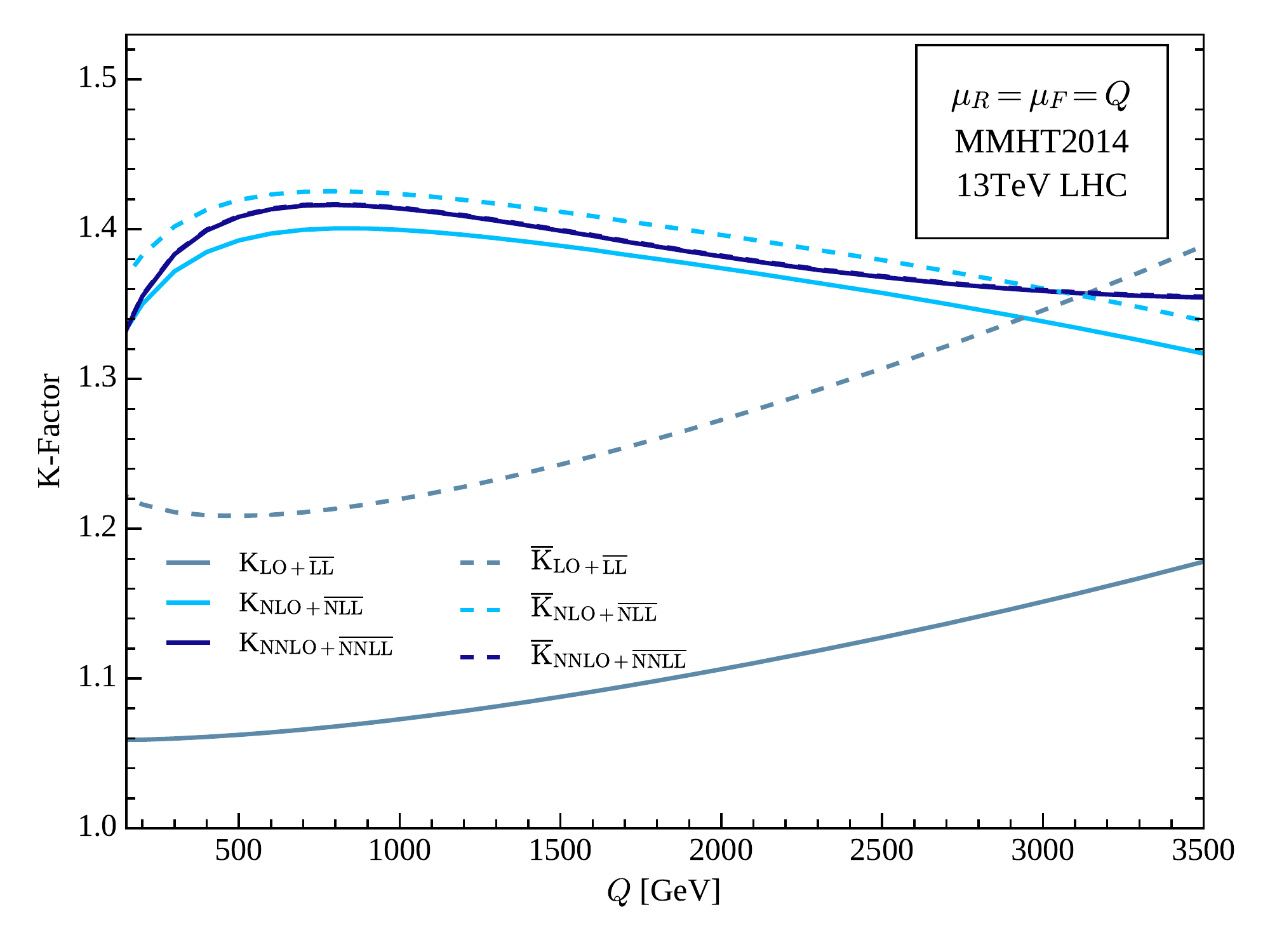}
\caption{Comparison between the $\rm {K}$-factors for the $N$  and $\overline N$ exponentiation at the central scale $Q=\mu_R=\mu_F$. }
\label{KNNb}
\end{figure}

\textcolor{black}{This section is devoted to the study of SV+NSV resummation effects in different schemes, namely $N$ and $\overline N$ exponentiations. The  details of both the schemes are discussed in sec. \ref{sec:resummation} and here we illustrate their numerical impact. The main point of difference between these two scheme arises from the ``additional"  resummation of $\gamma_E$, i.e., in $\overline N$.  We resum $\gamma_E$ terms along with $\ln N$ whereas for $N$ exponentiation we only resum the $\ln N$ terms. The effect of this ``additional" resummation of $\gamma_E$ has shown better convergence for SV resummation, now here it would be interesting to see what changes does SV+NSV resummation bring to the observations made earlier in \cite{Ajjath:2020rci}.\\
At first we begin with the analysis of $\rm {K}$ factor which is defined in \eqref{eq:KDEf}. To differentiate between these two scheme we denote $\rm {K}$ for $N$ and $\overline{\rm {K}}$\footnote{It is to be noted that both $\rm {K}$ and $\overline {\rm {K}}$ follows the same definition as given in \eqref{eq:KDEf}} for $\overline N$ exponentiation as given in Fig. \ref{KNNb}. Earlier in sec. \ref{sec:Kfactor}, we found that the $\rm {K}$ factor for $N$ exponentiation shows certain  hierarchy which grows from ${\rm LO+\overline{\rm LL}}$ to ${\rm NLO+\overline{\rm NLL}}$ to ${\rm NNLO+\overline{\rm NNLL}}$. Now for $\overline N$ we find, that the $\overline{\rm {K}}_{\rm NLO+\overline{\rm NLL}}$ is greater than $\overline{\rm {K}}_{\rm NNLO+\overline{\rm NNLL}}$. However at ${\rm NNLO+\overline{\rm NNLL}}$, we observe a striking feature, namely the $\rm {K}$ factors for both $N$ and $\overline N$ exactly overlap on each other for the considered $Q$ range. To quantify the overlap at ${\rm NNLO+\overline{\rm NNLL}}$, we quote that for $\overline{N}$ at $Q=2000$ GeV, the $\overline{\rm {K}}$ factor is 1.3823 while for $N$ it is 1.3818 (see Table \ref{Tab:Kfactor}). This implies that the correction at ${\rm NNLO+\overline{\rm NNLL}}$ is independent of any schemes. \\
Moreover, like $N$ exponentiation, we see an overall increment in the cross section for $\overline N$ with the inclusion of higher order logarithmic corrections. For instance at $Q = 2000$ GeV, the LO prediction is enhanced by $27.34\%$, the NLO by 6.88\%  and similarly the NNLO by 1.31\%. As a whole the resummed result at ${\rm NNLO+\overline{\rm NNLL}}$ increases the LO prediction by 38.32\% for the same value of $Q$. And the perturbative convergence between resummed curves in $\overline N$ is better  than $N$ exponentiation by a very small margin as can be seen from Fig. \ref{KNNb}.  }
\vspace{-0.5cm}
%\vspace{-0.5cm}

\begin{figure}[ht]
\centering
\includegraphics[scale=0.55]{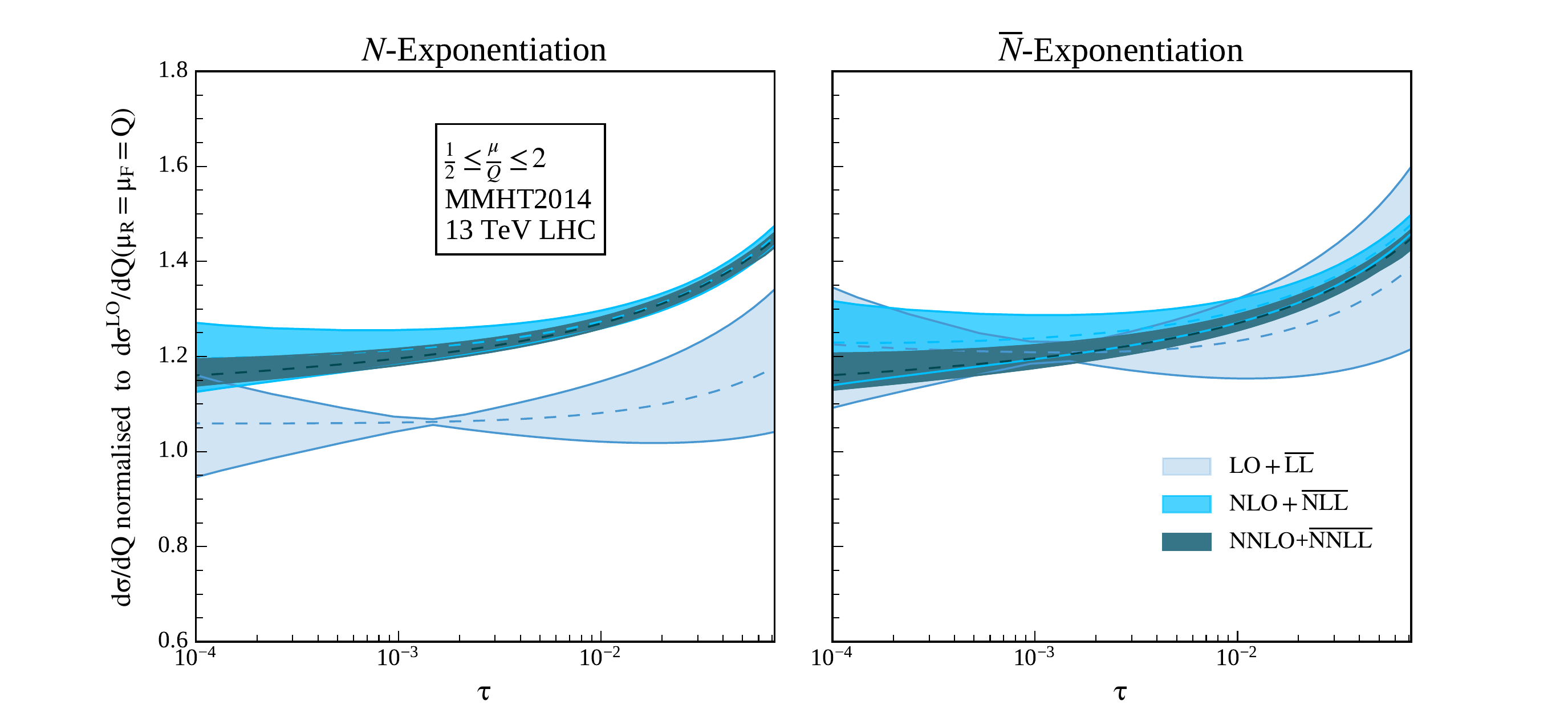}
\caption{A comparison between 7-point scale variation in both $N$ and $\overline{N}$ exponentiation around the central scale choice $(\mu_R,\mu_F) = (1,1)Q$ for $13$ Tev LHC.}
\label{fig:7ptNNB}
\end{figure}

\textcolor{black}{Now we proceed to study the uncertainties resulting from $\mu_R$ and $\mu_F$ for both $N$ and $\overline{N}$. This is presented in Fig. \ref{fig:7ptNNB}. We find that there is a systematic enhancement in the cross section and in scale reduction for both the schemes. However, in contrast to $N$ exponentiation, studied in sec. \ref{sec:7ptNexp}, we find that the bands of ${\rm NNLO+\overline{\rm NNLL}}$ is not contained within the band of ${\rm NLO+\overline{\rm NLL}}$ in $\overline{N}$-scheme, as the enhancement from ${\rm NLO+\overline{\rm NLL}}$ to ${\rm NNLO+\overline{\rm NNLL}}$ is negative. For instance, there is an increment of 9.6\% from LO +LL to ${\rm NLO+\overline{\rm NLL}}$ accuracy, which decreases by  $-0.85\%$  at  ${\rm NNLO+\overline{\rm NNLL}}$ for  $Q=2000$ GeV. This can also be seen from Fig. \ref{KNNb}. Although the width of ${\rm NNLO+\overline{\rm NNLL}}$ band is smaller than that of ${\rm NLO+\overline{\rm NLL}}$ yet the uncertainties associated with these two bands are more than $N$ exponentiation (see Table \ref{tab:shemes}). Hence in summary, we find that when we go from LO+LL to NNLO+${\overline {\mathrm NNLL}}$, the cross section increases more in $N$ scheme compared to $\overline N$ and the uncertainity in $N$ scheme is  smaller compared to $\overline{N}$.    Interestingly
at NNLO+${\overline {\rm NNLL}}$ level, their central values are very close to each other compared to previous order hinting better scheme independence as we increase the order of perturbation.  
%more while comparing the $N$ and $\overline{N}$ exponentiation schemes, the enhancement in cross section is more in former,
%than in $\overline{N}$ exponentiation scheme, 
%while the uncertainity is higher in latter
%$\overline{N}$ than in $N$ exponentiation scheme. 
%However the central value ($Q=\mu_R=\mu_F$) of the cross section  at ${\rm NNLO+\overline{\rm NNLL}}$ is very close in both the schemes.  
% \begin{itemize}
% \item The bands of NNLO+NNLL  is not contained within the band of NLO+NLL unlike NExp.
% \item The width of NNLO+NNLL band is comparable to NLO+NLL at higher Q values. For lower Q range NNLO+NNLL band is smaller than NLO+NLL. This feature is also contrary to Nexp. 
% \item The width of NLO+NLL is less than that of NLO at higher Q values, same as Nexp. However the width of  NNLO+NNLL is more to NNLO at higher Q values, unlike NExp. 
% \item In conclusion among the resummed results we find that NNLO+NNLL band is comparable to the NLO+NLL with respect to the scale variations.
% \end{itemize}
}

{\color{black}
Besides $N$ and $\overline N$ there are two other schemes as we discussed in sec. \ref{sec:resummation}, which are {\it Soft} and {\it All exponentiation}. While in {\it Soft} -scheme, we exponentiate only the Mellin moment of complete soft-collinear function, in {\it All} -scheme, the form factor contributions are also taken into exponentiation, additionally. We have explored the NSV resummation under these two schemes along with the $N$- and $\overline N$-schemes for two different $Q$-values, namely $Q=1000$ GeV and 2000 GeV and the results are enlisted in Table \ref{tab:shemes} for different logarithmic accuracies. Noting various values, we find that at LO+$\rm \overline{LL}$ and NLO+$\rm \overline{NLL}$, the $\overline{N}$ scheme gives the large corrections, say $0.0638\times 10^{-5}$ pb/GeV and $0.0699\times 10^{-5}$ pb/GeV respectively at $Q=2000$ GeV, whereas, at NNLO+$\rm \overline{NNLL}$ i.e., after adding more logarithmic corrections, the large contributions are arising from {\it All} -scheme.  However, the uncertainties coming from $\mu_R$ and $\mu_F$ variation are found to be least in $N$ and {\it All exponentiation} at any logarithmic accuracy for the considered $Q$-range.}
% We show fig.\ref{KallN6} where we took the ratio with respect to the Standard N exponentiation results at each order. It is to be noted that LO+LL results are same for all these three approaches by construction. At NNLL the Soft exponentiation gets additional $0.25\%$, $0.11\%$ and $0.10\%$ corrections compared to the Standard N approach at $Q = 200, Q=500$ and $Q=1500$ GeV, respectively. Similarly the All exponentiation  gets additional $0.47\%$, $0.22\%$ and $0.21\%$ corrections compared to the Standard N approach at $Q = 200, Q=500$ and $Q=1500$ GeV, respectively. 
% %-------------------------------------------
% We have quantified the impact of resummed results through $\rm {K}$-factor. In fig.\ref{KNNb}
% we present the resummed $\rm {K}$-factors up to NNLL. For All exponentiation case, we find that the $\rm {K}$-factors are $1.360$, $1.412$ and $1.402$ at NNLL at $Q=200$, $500$ and $1500$, respectively. The $\rm {K}$-factors for the Soft exponentiation case are found to be $1.357$, $1.409$ and $1.40$ at NNLL at $Q=200$, $500$ and $1500$, respectively.    
%---------------------------------
%-------------------------------------------
  \begin{table}[!htp] 
  \begin{center}
  \begin{small}
 % \newcolumntype{P}[1]{>{\centering\arraybackslash}p{#1}}
  {\renewcommand{\arraystretch}{2.1}
  \begin{tabular}{|P{2.3cm}||P{1cm}|P{2.0cm} |P{2.0cm}|p{2.0cm}|P{2.0cm}|}
  \rowcolor{lightgray}
   \multicolumn{1}{|c||}{Order}
  &\multicolumn{1}{c||}{$Q=\mu_R=\mu_F$}
       &\multicolumn{1}{c||}{\it N-Exp }  
      &\multicolumn{1}{c||}{$\overline{\it N}$-\it Exp} 
      & \multicolumn{1}{c|}{\it Soft-Exp}
      & \multicolumn{1}{c|}{\it All-Exp }
      \\ 
  % \hline
  %  \hline
 \multirow{2}{*} {$\rm LO+\overline{LL}$} &  1000~GeV& $2.5184^{+4.49\%}_{-4.25\%}$  & $2.8636^{+5.53\%}_{-5.07\%}$  & $2.5184^{+4.49\%}_{-4.25\%}$  & $2.5184^{+4.49\%}_{-4.25\%}$  \\
 \cline{2-6}
  &  2000~GeV & $0.0554^{+9.10\%}_{-7.92\%}$  & $0.0638^{+10.21\%}_{-8.73\%}$ &  $0.0554^{+9.10\%}_{-7.92\%}$ & $0.0554^{+9.10\%}_{-7.92\%}$ \\
  \hline
\multirow{2}{*} {$\rm NLO+\overline{NLL}$}  & 1000~GeV & $3.2857^{+2.08\%}_{-1.18\%}$   &  $3.3418^{+2.41\%}_{-2.13\%}$  &  $3.3027^{+2.12\%}_{-1.74\%}$ &  $3.3101^{+1.89\%}_{-1.24\%}$ \\
  \cline{2-6}
  &  2000~GeV & $0.0688^{+1.44\%}_{-1.24\%}$  & $0.0699^{+1.69\%}_{-1.51\%}$ & $0.0691^{+1.59\%}_{-1.16\%}$ & $0.0693^{+1.44\%}_{-1.31\%}$\\
  \hline
\multirow{2}{*} {$\rm NNLO+\overline{NNLL}$}  &  1000~GeV&  $3.3191^{+1.13\%}_{-0.87\%}$ &  $3.3204^{+1.58\%}_{-1.29\%}$  &  $3.3222^{+1.32\%}_{-0.91\%}$ &  $3.3264^{+1.13\%}_{-0.91\%}$ \\
 \cline{2-6}
  &  2000~GeV &  $0.0692^{+0.89\%}_{-0.79\%}$ & $0.0693^{+1.28\%}_{-1.15\%}$ & $0.0693^{+1.04\%}_{-0.82\%}$ & $0.0694^{+0.89\%}_{-0.82\%}$ \\
  \hline
\end{tabular}}
  \caption{A comparison of resummed cross-sections in $10^{-5}$ pb/GeV between different resummation schemes up to NNLO+$\overline{\rm NNLL}$.}
  \label{tab:shemes}
  \end{small}
  \end{center}
  \end{table}
% Next we study the uncertainties resulting from unphysical scales in these approaches.  We observe that different approaches for resummation provide a systematic scale reduction at lower invariant mass of the di-lepton pair. For example, in the Standard N exponentiation case, the scale uncertainity reduces from
% $-1.2\%-1.66\%$ at NLO+NLL to $-0.80\%-0.97\%$ at NNLO+NNLL for $Q=1500$ GeV. Similar pattern is seen for the Soft and All exponentiation as well as seen in fig.\ref{KallN8}. However, when
% we compare among themselves, we see that in the case of All exponentiation the scale
% uncertainity is reduced to $1.49\%$ at NLO+NLL compared to $1.66\%$ for N exponentiation
% and $1.69\%$ for Soft exponentiation at the same order at $Q=1500$ GeV. On the other hand, for the NNLO+NNLL case, the scale uncertainties are comparable in All and N exponentiation which are $0.971\%$ and $0.972\%$, respectively. However, for the NNLO+NNLL case also, the Soft exponentiation shows higher scale uncertainity among all three schemes which is $1.13\%$. 
%\newpage
\subsection{Numerical results for different collider energies}
In Tables \ref{table:200GeV}, \ref{table:500GeV} and \ref{table:1TeV}, we summarise the results for different
collider energies and different values of the invariant mass of di-leptons. We estimate the theoretical uncertainity by independently varying the scales $\mu = \{ \mu_F, \mu_R \}$ up and down, by a factor of two i.e. $\frac{1}{2} \leq \frac{\mu}{Q} \leq 2$. We find that the inclusion of SV NNLL increases the NNLO distribution by $0.24\%-0.55\%$ whereas the inclusion of NSV at $\overline {\rm {NNLL}}$ increases it by $0.72\%-1.65\%$, in the considered  collider energies and the $Q$ range. \\
\begin{center}
%The table \ref{table:1}
\begin{table}[h!]
\begin{small}
\centering
{\renewcommand{\arraystretch}{1.4}
 \begin{tabular}{||P{1.3cm}|| P{3.3cm}|P{3.3cm}|P{3.7cm} ||} 
 \hline
 $\sqrt{s}$ & NNLO  & NNLO+NNLL   & NNLO+$\overline{\rm {NNLL}}$   \\ [1.2ex] 
 \hline\hline 
 7 TeV   & ${2.56\times 10^{-2}}^{+0.29\%}_{-0.28\%}$   &  ${2.57\times 10^{-2}}^{+0.82\%}_{-0.62\%}$  &  ${2.59\times 10^{-2}}^{+2.19\%}_{-1.38\%}$ \\ [1.2ex] 
 8 TeV  &  ${3.06\times10^{-2}}^{+0.29\%}_{-0.32\%}$  &   ${3.07\times10^{-2}}^{+0.86\%}_{-0.65\%}$ &  ${3.09\times10^{-2}}^{+2.25\%}_{-1.41\%}$  \\   [1.2ex] 
 13 TeV &  ${5.63\times10^{-2}}^{+0.41\%}_{-0.45\%}$  &  ${5.65\times10^{-2}}^{+0.99\%}_{-0.76\%}$   & ${5.69\times10^{-2}}^{+2.43\%}_{-1.49\%}$\\   [1.2ex] 
 14 TeV &  ${6.16\times10^{-2}}^{+0.43\%}_{-0.47\%}$  &  ${6.18\times10^{-2}}^{+1.00\%}_{-0.78\%}$  & ${6.22\times10^{-2}}^{+2.45\%}_{-1.50\%}$\\   [1.2ex] 
 100TeV & ${5.13\times10^{-1}}^{+0.62\%}_{-1.18\%}$   &  ${5.15\times10^{-1}}^{+1.16\%}_{-1.44\%}$  & ${5.18\times10^{-1}}^{+2.79\%}_{-2.09\%}$\\  [1.2ex] 
 \hline
\end{tabular}}
\caption{Resummed results for invariant mass distribution of di-lepton pair against fixed order (in pb/GeV) at different centre of mass energies at the LHC for $Q= 200$ GeV. }
% The scale uncertainity has been estimated using 7-point scale variation around the central scale  $Q = \mu_R = \mu_F = 200 \text{GeV}$.}
\label{table:200GeV}
\end{small}
\end{table}
\end{center}
 \begin{center}

%The table \ref{table:}
\begin{table}[h!]
\begin{small}
\centering
{\renewcommand{\arraystretch}{1.4}
 \begin{tabular}{||P{1.3cm}|| P{3.3cm}|P{3.3cm}|P{3.7cm} ||} 
 \hline
 $\sqrt{s}$ & NNLO  & NNLO+NNLL   & NNLO+$\overline{\rm {NNLL}}$   \\ [1.2ex] 
% $\sqrt{s}$ & NNLO (pb/GeV) & NNLO+NNLL$_{SV}$ (pb/GeV)  & NNLO+NNLL$_{SV+NSV}$ (pb/GeV)  \\ [1.2ex] 
 \hline\hline 
 7 TeV  &  ${3.25\times 10^{-4}}^{+0.22\%}_{-0.35\%}$  &    ${3.26\times 10^{-4}}^{+0.404\%}_{-0.33\%}$  &  ${3.28\times 10^{-4}}^{+1.36\%}_{-1.02\%}$  \\ [1.2ex] 
 8 TeV  &  ${4.13\times 10^{-4}}^{+0.19\%}_{-0.32\%}$  &    ${4.14\times 10^{-4}}^{+0.412\%}_{-0.347\%}$  &  ${4.17\times 10^{-4}}^{+1.40\%}_{-1.03\%}$  \\   [1.2ex] 
 13 TeV &  ${9.04\times 10^{-4}}^{+0.21\%}_{-0.207\%}$  &  ${9.07\times 10^{-4}}^{+0.546\%}_{-0.40\%}$  & ${9.13\times 10^{-4}}^{+1.56\%}_{-1.08\%}$ \\   [1.2ex] 
 14 TeV &  ${10.09\times 10^{-4}}^{+0.213\%}_{-0.19\%}$  & ${10.13\times 10^{-4}}^{+0.565\%}_{-0.409\%}$   & ${10.19\times 10^{-4}}^{+1.59\%}_{-1.09\%}$ \\   [1.2ex] 
 100TeV &  ${1.214\times 10^{-2}}^{+0.35\%}_{-0.43\%}$  &  ${1.217\times 10^{-2}}^{+0.787\%}_{-0.650\%}$   & ${1.223\times 10^{-2}}^{+1.94\%}_{-1.23\%}$  \\  [1.2ex] 
 \hline
\end{tabular}}
\caption{Resummed results for invariant mass distribution of di-lepton pair against fixed order (in pb/GeV) at different centre of mass energies at the LHC for $Q= 500$ GeV.}
\label{table:500GeV}
\end{small}
\end{table}
\end{center}
\begin{center}
%The table \ref{table:3}

\begin{table}[h!]
\begin{small}
\centering
{\renewcommand{\arraystretch}{1.4}
\begin{tabular}{||P{1.3cm}|| P{3.3cm}|P{3.3cm}|P{3.7cm} ||} 
 \hline
 $\sqrt{s}$ & NNLO  & NNLO+NNLL   & NNLO+$\overline{\rm {NNLL}}$   \\ [1.2ex] 
% $\sqrt{s}$ & NNLO (pb/GeV) & NNLO+NNLL$_{SV}$ (pb/GeV)  & NNLO+NNLL$_{SV+NSV}$ (pb/GeV)  \\ [1.2ex] 
 \hline\hline 
 7 TeV  &  ${7.55\times 10^{-6}}^{+0.42\%}_{-0.704\%}$   &   ${7.59\times 10^{-6}}^{+0.32\%}_{-0.26\%}$   &  ${7.65\times 10^{-6}}^{+0.99\%}_{-0.87\%}$   \\ [1.2ex] 
 
 8 TeV  & ${1.092\times 10^{-5}}^{+0.36\%}_{-0.57\%}$    &  ${1.098\times 10^{-5}}^{+0.33\%}_{-0.27\%}$   &${1.11\times 10^{-5}}^{+1.02\%}_{-0.86\%}$    \\   [1.2ex] 
 
 13 TeV &  ${3.29\times 10^{-5}}^{+0.28\%}_{-0.32\%}$   &  ${3.30\times 10^{-5}}^{+0.36\%}_{-0.29\%}$    &  ${3.32\times 10^{-5}}^{+1.135\%}_{-0.87\%}$  \\   [1.2ex] 
 
 14 TeV &  ${3.79\times 10^{-5}}^{+0.19\%}_{-0.29\%}$   &  ${3.81\times 10^{-5}}^{+0.36\%}_{-0.30\%}$   &   ${3.83\times 10^{-5}}^{+1.15\%}_{-0.87\%}$\\   [1.2ex] 
 
 100TeV &  ${6.96\times 10^{-4}}^{+0.21\%}_{-0.19\%}$   &  ${6.98\times 10^{-4}}^{+0.59\%}_{-0.39\%}$   &  ${7.01\times 10^{-4}}^{+1.50\%}_{-0.94\%}$ \\  [1.2ex] 
 
 \hline
\end{tabular}}
\caption{Resummed results for invariant mass distribution of di-lepton pair against fixed order (in pb/GeV) at different centre of mass energies at the LHC for $Q= 1000$ GeV.}
\label{table:1TeV}
\end{small}
\end{table}
\end{center}

\section{Discussion and Conclusions} \label{sec:concl}
 Production of pair of leptons in Drell-Yan process is one of the cleanest processes at the LHC, also well studied both in SM and beyond SM.  The perturbative predictions taking into account radiative corrections
 from strong and electroweak interactions are known to unprecedented accuracy.  Third order perturbative results
 in QCD improved with the  SV enhanced resummed contributions and  with partial second
 order results in electroweak sector have already played important role in our understanding of the underlying dynamics
 and in addition, these results have set stringent constraints on the parameters of various
 beyond the SM scenarios.  In this article we have improved
 the predictions by including next-to-SV enhanced resummed contributions from strong interactions.  We have
 used the recent formalism that systematically resums such next-to-SV logarithms to all orders.
 We have restricted ourselves  to the production mechanism where only neutral gauge bosons such photon and $Z$ boson produce leptons.
 For our study, we have included fixed order results from QCD upto NNLO and resummed results upto $\overline {\rm NNLL}$ level.  The latter
 contains both resummed threshold SV contributions as well as next-to-SV resumed ones.  Using the so called matched results,
 we study the impact of resummed NSV contributions on the invariant mass of the pair of leptons.   As the fixed order predictions
 have already indicated that NSV logarithms dominate over SV distributions, the inclusion of resummed NSV terms
along with the SV resumed predictions are expected to show appreciable numerical effects that can not be ignored. We have studied their impact over a wide range of $Q$ values, i.e., 100 to 3500 GeV for 13 TeV LHC.
 We find that about 6\% to 11\% increase when we go from  LO to LO+$\overline {\rm LL}$ while at second order it stabilises to 1\% when we include $\overline {\rm NNLL}$ to NNLO. These predictions were obtained by setting both renormalisation $\mu_R$
 and factorisation $\mu_F$ scales at $Q$.  The sensitivity of our predictions with respect to these scales can be
studied using 7-point scale variation and single scale variation.
 In the  seven point scale variation, we find that the inclusion of
 resummed $\overline {\rm LL},\overline {\rm NLL}$ and $\overline{\rm NNLL}$ terms to fixed order LO, NLO and NNLO contributions
 increases the sensitivity of the predictions to these scales.
 A detailed investigation reveals that
 the resumed results are more sensitive to factorisation scale compared to renormalisation scale and in particular
 we find that NSV part of the resumed result is largely responsible for this.
 The reason for this is the absence of NSV resummed terms from quark gluon initiated processes in our analysis.
 These contributions would provide right logarithms of $\mu_F$ to compensate against those from the PDFs at the hadronic level.
 Note that different partonic channels mix under factorisation scale variations when they are convoluted with appropriate
 PDFs. Hence, absence of any partonic channel at any given order can increase the sensitivity to $\mu_F$ at the hadronic
 level.   We have also noticed that the $\mu_F$ sensitivity is large at second order compared to lower orders.
 This is easy to understand if we observe that the quark gluon initiated channels give larger negative
 contribution at NNLO level compared to NLO level.  In order to understand the role of quark-gluon
 initiated channel in the context of unphysical scales, we have computed  both SV and NSV resummed contributions with and without $qg$-channels and performed 7-point and single scale variations.  We find that at $a_s^2$, the
 corrections from $q \bar q$ and $qg$-channel are 4.86\% and $-2.47\%$ respectively
 to the NNLO cross section along with the other sub-dominating channels.  As these corrections are closer to each other at the fixed order level, we expect that resumed NSV from $qg$-channel is as important as the one from $q\overline q$ channel.
 
Unlike in the case of $\mu_F$, different channels that contribute do not mix under the variation of $\mu_R$ as each of them is renormalisation group invariant with respect to $\mu_R$.  Hence, we expect that the predictions based on single scale variation 
should be less sensitive to $\mu_R$ as we increase the order of perturbation.  Our numerical study confirms this both for fixed order as well as for resummed predictions.

We have also investigated the role of resummed NSV terms in the resummed predictions in comparison to the known SV resummation.  We find that there is about 2\% increment when we go from NLL to $\overline{\rm NLL}$ and about 0.64\% from NNLL to $\overline{\rm NNLL}$ demonstrating the importance of resummed NSV contributions from quark anti-quark initiated channels.   The 7-point scale variation gives larger scale uncertainity if we add resummed contributions from NSV terms.  This stems from the scale $\mu_F$ as resummed NSV part lack contributions from $qg$ channel.  The single scale variations where we keep one scale fixed and vary the other scale confirms our interpretation.

One finds that the SV part of the resummed predictions depends on how one treats the $N$-independent part of
the resummed results.  Such terms come from both form factors and SV part of the soft-collinear function.  
Different schemes are adopted to deal with such terms and they give different numerical predictions. We consider four schemes and study their numerical impact on the predictions taking into account NSV terms.   Interestingly we find that the difference in predictions from  $N$ and $\overline{N}$ schemes gets reduced as we increase the logarithmic accuracy signaling the scheme independence.  Finally,  we conclude our detailed numerical analysis by presenting predictions for invariant mass distributions at scales $Q=1000,2000$ GeV for different hadronic center of mass energies along with uncertainities resulting from unphysical scales.

In summary, our analysis provides the precise numerical predictions from resummed NSV terms upto $\rm NNLO + \overline {\rm{NNLL}}$ level for the first time to the invariant mass distribution of a pair of leptons produced at the LHC.

\section{Acknowledgements}
We thank Claude Duhr and Moch for useful discussion throughout this project. We also thank Claude Duhr and Bernhard Mistlberger for providing third order results for the inclusive reactions.  
We would like to thank L. Magnea and E. Laenen for their encouragement to work on this area. In addition we would also like to thank the computer administrative unit of IMSc  for their help and support.
%\newpage
%% Appendix
\appendix

\section{Anomalous dimensions} \label{app:anodim}

Here we present all the anomalous dimensions used in performing the resummation.
\subsection*{Cusp anomalous dimensions $A_i^q$} 
In the following we list the cusp anomalous dimensions $A_i^q$ till four-loop level:  
 \begin{align} 
\begin{autobreak} 
A^q_1 =
 4 ~\CF  ,   
\end{autobreak} 
\\ 
\begin{autobreak} 
A^q_2 =
\CF ~ \NF  ~  \bigg( 
- \frac{40}{9} \bigg)      
+ \CF ~ \CA  ~  \bigg( \frac{268}{9}
- 8 ~ \z2 \bigg)
,   
\end{autobreak} 
\\ 
\begin{autobreak} 
A_3^q =
 \CF ~ \NF^2  ~  \bigg( 
- \frac{16}{27} \bigg)      
+ \CF ~ \CA ~ \NF  ~  \bigg( 
- \frac{836}{27}
- \frac{112}{3} ~ \z3
+ \frac{160}{9} ~ \z2 \bigg)      
+ \CF ~ \CA^2  ~  \bigg( \frac{490}{3}
+ \frac{88}{3} ~ \z3
- \frac{1072}{9} ~ \z2
+ \frac{176}{5} ~ \z2^2 \bigg)      
+ \CF^2 ~ \NF  ~  \bigg( 
- \frac{110}{3}
+ 32 ~ \z3 \bigg)
, 
\end{autobreak} 
\\
\begin{autobreak} 
A^q_4 =
 \CF  \dFAoNA    \bigg( \frac{7040}{3}  \z5
+ \frac{256}{3}  \z3
- 768  \z3^2
- 256  \z2
- \frac{15872}{35}  \z2^3 \bigg)      
+ \CF  \NF  \dFFoNA    \bigg( - \frac{2560}{3}  \z5 - \frac{512}{3}  \z3
+ 512  \z2 \bigg)      
+ \CF  \NF^3    \bigg( 
- \frac{32}{81}
+ \frac{64}{27}  \z3 \bigg)      
+ \CF^2  \NF^2    \bigg( \frac{2392}{81}
- \frac{640}{9}  \z3
+ \frac{64}{5}  \z2^2 \bigg)      
+ \CF^3  \NF    \bigg( \frac{572}{9}
- 320  \z5
+ \frac{592}{3}  \z3 \bigg)      
+ \CA  \CF  \NF^2    \bigg( \frac{923}{81}
+ \frac{2240}{27}  \z3
- \frac{608}{81}  \z2
- \frac{224}{15}  \z2^2 \bigg)      
+ \CA  \CF^2  \NF    \bigg( 
- \frac{34066}{81}
+ 160  \z5
+ \frac{3712}{9}  \z3
+ \frac{440}{3}  \z2
- 128  \z2  \z3
- \frac{352}{5}  \z2^2 \bigg)      
+ \CA^2  \CF  \NF    \bigg( 
- \frac{24137}{81}
+ \frac{2096}{9}  \z5
- \frac{23104}{27}  \z3
+ \frac{20320}{81}  \z2
+ \frac{448}{3}  \z2  \z3
- \frac{352}{15}  \z2^2 \bigg)      
+ \CA^3  \CF    \bigg( \frac{84278}{81}
- \frac{3608}{9}  \z5
+ \frac{20944}{27}  \z3
- 16  \z3^2
- \frac{88400}{81}  \z2
- \frac{352}{3}  \z2  \z3
+ \frac{3608}{5}  \z2^2
- \frac{20032}{105}  \z2^3 \bigg)\,,
\end{autobreak} 
\end{align} 
where $n_f$ is the number of active quark flavours in the theory. The quadratic Casimirs $C_F$ and $C_A$ are given by  
$\frac{n_c^2-1}{2~ nc}$ and $n_c$ respectively.
The quartic Casimirs are given by
\begin{align}
% \frac{d_A^{abcd}d_A^{abcd}}{N_A} &= \frac{n_c^2 (n_c^2 + 36)}{24}, 
% \frac{d_A^{abcd}d_F^{abcd}}{N_A} = \frac{n_c (n_c^2 + 6)}{48}, \nn\\
% \frac{d_F^{abcd}d_A^{abcd}}{N_F} &=  \frac{(n_c^2-1)(n_c^2+6)}{48}, 
% \frac{d_F^{abcd}d_F^{abcd}}{N_F} = \frac{(n_c^2-1)(n_c^4 - 6 n_c^2 + 18)}{96n_c^3},
\frac{d_F^{abcd}d_A^{abcd}}{N_A} &=  \frac{n_c (n_c^2+6)}{48}, \qquad
\frac{d_F^{abcd}d_F^{abcd}}{N_A} = \frac{(n_c^4 - 6 n_c^2 + 18)}{96n_c^2},
\end{align} 
with $N_A = n_c^2 -1$ and $N_F = n_c$ where $n_c = 3$ for QCD.
\subsection*{Collinear anomalous dimensions $B_i^q$}

The collinear anomalous dimensions $B_i^q$ are given till three-loop as, 
\begin{align} 
\begin{autobreak} 
B_1^q =
 3~ \CF ,   
\end{autobreak} 
\\ 
\begin{autobreak} 
B_2^q =
 \CF ~ \NF  ~  \bigg( 
- \frac{1}{3}
- \frac{8}{3} ~ \z2 \bigg)      
+ \CF ~ \CA  ~  \bigg( \frac{17}{6}
- 12 ~ \z3
+ \frac{44}{3} ~ \z2 \bigg)      
+ \CF^2  ~  \bigg( \frac{3}{2}
+ 24 ~ \z3
- 12 ~ \z2 \bigg) ,   
\end{autobreak} 
\\ 
\begin{autobreak} 
B_3^q =
 \CF ~ \NF^2  ~  \bigg( 
- \frac{17}{9}
- \frac{16}{9} ~ \z3
+ \frac{80}{27} ~ \z2 \bigg)      
+ \CF ~ \CA ~ \NF  ~  \bigg( 20
+ \frac{200}{9} ~ \z3
- \frac{1336}{27} ~ \z2
+ \frac{4}{5} ~ \z2^2 \bigg)      
+ \CF ~ \CA^2  ~  \bigg( 
- \frac{1657}{36}
+ 40 ~ \z5
- \frac{1552}{9} ~ \z3
+ \frac{4496}{27} ~ \z2
- 2 ~ \z2^2 \bigg)      
+ \CF^2 ~ \NF  ~  \bigg( 
- 23
- \frac{136}{3} ~ \z3
+ \frac{20}{3} ~ \z2
+ \frac{232}{15} ~ \z2^2 \bigg)      
+ \CF^2 ~ \CA  ~  \bigg( \frac{151}{4}
+ 120 ~ \z5
+ \frac{844}{3} ~ \z3
- \frac{410}{3} ~ \z2
+ 16 ~ \z2 ~ \z3
- \frac{988}{15} ~  \z2^2 \bigg)      
+ \CF^3  ~  \bigg( \frac{29}{2}
- 240 ~ \z5
+ 68 ~ \z3
+ 18 ~ \z2
- 32 ~ \z2 ~ \z3
+ \frac{288}{5} ~ \z2^2 \bigg)\,.
\end{autobreak} 
\end{align}
\subsection*{Soft anomalous dimensions $f_i^q$}
The soft anomalous dimensions $f_i^q$ till three-loop are given as
\begin{align}
\begin{autobreak} 
f_1^q =  
0
,   
\end{autobreak} 
\\ 
\begin{autobreak} 
f_2^q =
 \CF ~ \NF  ~  \bigg( 
- \frac{112}{27}
+ \frac{4}{3} ~ \z2 \bigg)      
+ \CF ~ \CA  ~  \bigg( \frac{808}{27}
- 28 ~ \z3
- \frac{22}{3} ~ \z2 \bigg),   
\end{autobreak} 
\\ 
\begin{autobreak} 
f_3^q =
 \CF ~ \NF^2  ~  \bigg( 
- \frac{2080}{729}
+ \frac{112}{27} ~ \z3
- \frac{40}{27} ~ \z2 \bigg)      
+ \CF ~ \CA ~ \NF  ~  \bigg( 
- \frac{11842}{729}
+ \frac{728}{27} ~ \z3
+ \frac{2828}{81} ~ \z2
- \frac{96}{5} ~ \z2^2 \bigg)      
+ \CF ~ \CA^2  ~  \bigg( \frac{136781}{729}
+ 192 ~ \z5
- \frac{1316}{3} ~ \z3
- \frac{12650}{81} ~ \z2
+ \frac{176}{3} ~ \z2 ~  \z3
+ \frac{352}{5} ~ \z2^2 \bigg)      
+ \CF^2 ~ \NF  ~  \bigg( 
- \frac{1711}{27}
+ \frac{304}{9} ~ \z3
+ 4 ~ \z2
+ \frac{32}{5} ~ \z2^2 \bigg)\,.
\end{autobreak} 
\end{align} 
\subsection*{NSV anomalous dimensions $C_i^q$ \& $D_i^q$ }
The NSV anomalous dimensions $C_i^q$ and  $D_i^q$ till three-loop are given as

\begin{align}
\begin{autobreak}
C_1^q = 
0,
\end{autobreak}\\
\begin{autobreak}
C_2^q = 
16 C_F^2 ,
\end{autobreak}\\
\begin{autobreak}
C_3^q =  
C_F^2 C_A \bigg( \frac{2144}{9} - 64 \zeta_2 \bigg) + n_f C_F^2  \bigg( - \frac{320}{9} \bigg)\,.
\end{autobreak}\\
%\end{align}
%\subsection*{NSV anomalous dimensions $D_i^q$}
%
%The NSV anomalous dimensions $D_i^q$ till three-loop are given as
%\begin{align}
\begin{autobreak}
D_1^q = 
-4 C_F, 
\end{autobreak} \\
\begin{autobreak}
D_2^q =
12 C_F^2  + C_A C_F \bigg( \frac{-400}{9} 
                    
                    + 8 \zeta_2 \bigg) + C_F n_f \bigg( \frac{64}{9} \bigg), 
\end{autobreak} \\
\begin{autobreak}
D_3^q = 
C_F C_A^2   \bigg( - \frac{8582}{27} - \frac{88}{3} \zeta_3 + \frac{1336}{9} \zeta_2 -                              \frac{176}{5} \zeta_2^2 \bigg)

       + C_F^2 C_A   \bigg(  \frac{302}{3} - 48 \zeta_3 + \frac{104}{3} \zeta_2 \bigg)

       + C_F^3   \bigg( 6 + 96 \zeta_3 - 48 \zeta_2 \bigg)

       + n_f C_F C_A   \bigg( \frac{724}{9} + \frac{112}{3} \zeta_3 - \frac{208}{9} \zeta_2 \bigg)

          + n_f C_F^2   \bigg( 30 - 32 \zeta_3 - \frac{32}{3} \zeta_2 \bigg)
 
       + n_f^2 C_F   \bigg( - \frac{64}{27} \bigg)\,.
\end{autobreak}
\end{align}
%\\
%--------------------------- SV--------------------------------------%
\subsection*{SV Threshold Exponents }
The function $\overline{G}^q_{SV}\big(a_s(q^2(1-z)^2),\epsilon\big)$ given in \ref{calQc} is related to the threshold exponent $\textbf{D}^q\big(a_s(q^2(1-z)^2),\epsilon\big)$ via Eq.(46) of \cite{Ravindran:2006cg} where the universal $\mathbf{D}_i^q$ coefficients, till three-loop, are given as,

\begin{align} 
\begin{autobreak} 
\mathbf{D}^q_1 
= 0,   
\end{autobreak} 
\\ 
\begin{autobreak} 
\mathbf{D}^q_2 
= {C}_{F}
 \nf    \bigg( \frac{224}{27}
- \frac{32}{3}  \z2 \bigg)      
+ \Ca    \bigg( 
- \frac{1616}{27}
+ 56  \z3
+ \frac{176}{3}  \z2 \bigg),   
\end{autobreak} 
\\ 
\begin{autobreak} 
\mathbf{D}^q_3 
= { C}_{F}
 \nf^2    \bigg( 
- \frac{3712}{729}
+ \frac{320}{27}  \z3
+ \frac{640}{27}  \z2 \bigg)      
+ \Cf  \nf    \bigg( \frac{3422}{27}
- \frac{608}{9}  \z3
- 32  \z2
- \frac{64}{5}  \z2^2 \bigg)      
+ \Ca  \nf    \bigg( \frac{125252}{729}
- \frac{2480}{9}  \z3
- \frac{29392}{81}  \z2
+ \frac{736}{15}  \z2^2 \bigg)      
+ \Ca^2    \bigg( 
- \frac{594058}{729}
- 384  \z5
+ \frac{40144}{27}  \z3
+ \frac{98224}{81}  \z2
- \frac{352}{3}   \z2  \z3
- \frac{2992}{15}  \z2^2 \bigg) \,.
\end{autobreak} 
\end{align}

\section{NSV coefficients} \label{app:phic}

Here we present all the NSV coefficients $\varphi_{q,i}^{(k)}$ given in \ref{varphiexp}.
\begin{eqnarray}
   \varphi_{q,1}^{(0)} &=&
        4 C_F \,,
\nonumber\\
   \varphi_{q,1}^{(1)} &=& 0\,,
\nonumber\\
   \varphi_{q,2}^{(0)} &=&
        C_F  C_A    \bigg( {1402 \over 27} - 28  \zeta_3 - {112 \over 3}   \zeta_2 \bigg)
       + C_F^2    (  - 32  \zeta_2 )
       + n_f  C_F    \bigg(  - {328 \over 27} + {16 \over 3}  \zeta_2 \bigg)\,,
\nonumber\\
   \varphi_{q,2}^{(1)} &=&
        10 C_F  C_A  
       - 10 C_F^2   \,,
\nonumber\\
   \varphi_{q,2}^{(2)} &=&
       - 4 C_F^2  \,,
\nonumber\\
   \varphi_{q, 3}^{(0)} &=&  C_F C_A^2   \bigg( {727211 \over 729} + 192 \zeta_5 - {29876 \over 27} 
         \zeta_3 - {82868 \over 81} \zeta_2 + {176 \over 3} \zeta_2 \zeta_3 + 120 
         \zeta_2^2 \bigg)
\nonumber\\&&
       + C_F^2 C_A   \bigg(  - {5143 \over 27} - {2180 \over 9} \zeta_3 - {11584 \over 27} 
         \zeta_2 + {2272 \over 15} \zeta_2^2 \bigg)
       + C_F^3   \bigg( 23 + 48 \zeta_3 
\nonumber\\&&
       - {32 \over 3} \zeta_2 - {448 \over 15} 
         \zeta_2^2 \bigg)
       + n_f C_F C_A   \bigg(  - {155902 \over 729} + {1292 \over 9} \zeta_3 + {26312 \over 81} \zeta_2 - {368 \over 15} \zeta_2^2 \bigg)
\nonumber\\&&
       + n_f C_F^2   \bigg(  - {1309 \over 9} + {496 \over 3} \zeta_3 + {2536 \over 27} 
         \zeta_2 + {32 \over 5} \zeta_2^2 \bigg)
       + n_f^2 C_F   \bigg( {12656 \over 729} 
\nonumber\\&&
       - {160 \over 27} \zeta_3 - {704 \over 27} 
         \zeta_2 \bigg) \,,
\nonumber\\
   \varphi_{q, 3}^{(1)} &=&  C_F C_A^2   \bigg( {244 \over 9} + 24 \zeta_3 - {8 \over 9} \zeta_2 \bigg)
       + C_F^2 C_A  \bigg (  - {18436 \over 81} + {544 \over 3} \zeta_3 + {964 \over 9} 
         \zeta_2 \bigg)
\nonumber\\&&
       + C_F^3   \bigg(  - {64 \over 3} - 64 \zeta_3 + {80 \over 3} \zeta_2 \bigg)
       + n_f C_F C_A   \bigg(  - {256 \over 9} - {28 \over 9} \zeta_2 \bigg)
\nonumber\\&&
       + n_f C_F^2   \bigg( {3952 \over 81} - {160 \over 9} \zeta_2 \bigg) \,,
\nonumber\\
   \varphi_{q, 3}^{(2)} &=&  C_F C_A^2   \bigg( 34 - {10 \over 3} \zeta_2 \bigg)
       + C_F^2 C_A   \bigg(  - 96 + {52 \over 3} \zeta_2 \bigg)
       + C_F^3   \bigg( {16 \over 3} \bigg)
\nonumber\\&&
       + n_f C_F C_A  \bigg (  - {10 \over 3} \bigg)
       + n_f C_F^2   \bigg( {40 \over 3} \bigg) \,,
\nonumber\\
   \varphi_{q, 3}^{(3)} &=&  C_F^2 C_A   \bigg(  - {176 \over 27} \bigg) + n_f C_F^2   \bigg( {32 \over 27} \bigg) \,.
\end{eqnarray}

\section{Resummation Coefficients for the $N$ exponentiation} \label{app:Nexp}

\subsection{The $N$-independent coefficients $C_0^q$} \label{app:C0q}
The $N$-independent coefficients $C^q_{0i}$ in Eq. \ref{C0expand}, till three-loop, are given by
\begin{align}
\begin{autobreak}
    C^q_{00} =
         1
         \,,
\end{autobreak}\\
\begin{autobreak}
    
   C^q_{01} =

         C_F   \bigg\{
          - 16
          + 8 \zeta_2
          + 6 L_{qr}
          - 6 L_{fr}
          \bigg\}\,,
\end{autobreak}\\
\begin{autobreak}
   C^q_{02} =

         C_F n_f   \bigg\{
            \frac{127}{6}
          + 8 \zeta_3
          - \frac{112}{9} \zeta_2
          - \frac{34}{3} L_{qr}
          + 2 L_{qr}^2
          + \frac{2}{3} L_{fr}
          + \frac{16}{3} L_{fr} \zeta_2
          - 2 L_{fr}^2
          \bigg\}

       + C_F C_A   \bigg\{- \frac{1535}{12}
          + 28 \zeta_3
          + \frac{592}{9} \zeta_2
          - \frac{12}{5} \zeta_2^2
          + \frac{193}{3} L_{qr}
          - 24 L_{qr} \zeta_3
          - 11 L_{qr}^2
          - \frac{17}{3} L_{fr}
          + 24 L_{fr} \zeta_3
          - \frac{88}{3} L_{fr} \zeta_2 + 11 L_{fr}^2 \bigg\}

       + C_F^2   \bigg\{
            \frac{511}{4}
          - 60 \zeta_3
          - 70 \zeta_2
          + \frac{72}{5} \zeta_2^2
          - 93 L_{qr}
          + 48 L_{qr} \zeta_3
          + 24 L_{qr} \zeta_2
          + 18 L_{qr}^2
          + 93 L_{fr}
          - 48 L_{fr} \zeta_3
          - 24 L_{fr} \zeta_2
          - 36 L_{fr} L_{qr}
          + 18 L_{fr}^2
          \bigg\}\,,
\end{autobreak}\\
\begin{autobreak}
   C^q_{03} =

         C_F N_4 n_{fv}   \bigg\{
           8
          - \frac{160}{3} \zeta_5
          + \frac{28}{3} \zeta_3
          + 20 \zeta_2
          - \frac{4}{5} \zeta_2^2
          \bigg\}

       + C_F n_f^2   \bigg\{
          - \frac{7081}{243}
          - \frac{1264}{81} \zeta_3
          + \frac{2416}{81} \zeta_2
          + \frac{128}{27} \zeta_2^2
          + \frac{220}{9} L_{qr}
          + \frac{64}{9} L_{qr} \zeta_3
          - \frac{32}{3} L_{qr} \zeta_2
          - \frac{68}{9} L_{qr}^2
          + \frac{8}{9} L_{qr}^3
          + \frac{34}{9} L_{fr}
          + \frac{32}{9} L_{fr} \zeta_3
          - \frac{160}{27} L_{fr} \zeta_2
          + \frac{4}{9} L_{fr}^2
          + \frac{32}{9} L_{fr}^2 \zeta_2
          - \frac{8}{9} L_{fr}^3
          \bigg\}

       + C_F C_A n_f   \bigg\{
           \frac{110651}{243}
          - 8 \zeta_5
          - \frac{6016}{81} \zeta_3
          - \frac{28132}{81} \zeta_2
          + \frac{208}{3} \zeta_2 \zeta_3
          - \frac{5756}{135} \zeta_2^2
          - \frac{3052}{9} L_{qr}
          + \frac{208}{9} L_{qr} \zeta_3
          + \frac{320}{3} L_{qr} \zeta_2
          - \frac{8}{5} L_{qr} \zeta_2^2
          + \frac{850}{9} L_{qr}^2
          - 16 L_{qr}^2 \zeta_3
          - \frac{88}{9} L_{qr}^3
          - 40 L_{fr}
          - \frac{400}{9} L_{fr} \zeta_3
          + \frac{2672}{27} L_{fr} \zeta_2
          - \frac{8}{5} L_{fr} \zeta_2^2
          - \frac{146}{9} L_{fr}^2
          + 16 L_{fr}^2 \zeta_3
          - \frac{352}{9} L_{fr}^2 \zeta_2
          + \frac{88}{9} L_{fr}^3
          \bigg\}

       + C_F C_A^2   \bigg\{
          - \frac{1505881}{972}
          - 204 \zeta_5
          + \frac{82385}{81} \zeta_3
          - \frac{400}{3} \zeta_3^2
          + 843 \zeta_2
          - \frac{884}{3} \zeta_2 \zeta_3
          + \frac{14611}{135} \zeta_2^2
          + \frac{13264}{315} \zeta_2^3
          + \frac{3082}{3} L_{qr}
          + 80 L_{qr} \zeta_5
          - \frac{4952}{9} L_{qr} \zeta_3
          - 240 L_{qr} \zeta_2
          + \frac{68}{5} L_{qr} \zeta_2^2
          - \frac{2429}{9} L_{qr}^2
          + 88 L_{qr}^2 \zeta_3
          + \frac{242}{9} L_{qr}^3
          + \frac{1657}{18} L_{fr}
          - 80 L_{fr} \zeta_5
          + \frac{3104}{9} L_{fr} \zeta_3
          - \frac{8992}{27} L_{fr} \zeta_2
          + 4 L_{fr} \zeta_2^2
          + \frac{493}{9} L_{fr}^2
          - 88 L_{fr}^2 \zeta_3
          + \frac{968}{9} L_{fr}^2 \zeta_2
          - \frac{242}{9} L_{fr}^3
          \bigg\}

       + C_F^2 n_f   \bigg\{
          - \frac{421}{3}
          - \frac{608}{9} \zeta_5
          + \frac{6952}{27} \zeta_3
          + \frac{2632}{27} \zeta_2
          - \frac{896}{9} \zeta_2 \zeta_3
          - \frac{3568}{135} \zeta_2^2
          + 230 L_{qr}
          - \frac{368}{3} L_{qr} \zeta_3
          - \frac{176}{3} L_{qr} \zeta_2
          + \frac{112}{15} L_{qr} \zeta_2^2
          - 92 L_{qr}^2
          + 32 L_{qr}^2 \zeta_3
          + 12 L_{qr}^3
          - \frac{275}{3} L_{fr}
          + \frac{128}{3} L_{fr} \zeta_3
          - \frac{56}{3} L_{fr} \zeta_2
          + \frac{176}{15} L_{fr} \zeta_2^2
          + 72 L_{fr} L_{qr}
          + 32 L_{fr} L_{qr} \zeta_2
          - 12 L_{fr} L_{qr}^2
          + 20 L_{fr}^2
          - 32 L_{fr}^2 \zeta_3
          - 32 L_{fr}^2 \zeta_2
          - 12 L_{fr}^2 L_{qr}
          + 12 L_{fr}^3
          \bigg\}

       + C_F^2 C_A   \bigg\{
            \frac{74321}{36}
          - \frac{5512}{9} \zeta_5
          - \frac{34612}{27} \zeta_3
          + \frac{592}{3} \zeta_3^2
          - \frac{13186}{27} \zeta_2
          + \frac{3392}{9} \zeta_2 \zeta_3
          + \frac{24064}{135} \zeta_2^2
          - \frac{36944}{315} \zeta_2^3
          - \frac{3439}{2} L_{qr}
          + 240 L_{qr} \zeta_5
          + \frac{4664}{3} L_{qr} \zeta_3
          + \frac{632}{3} L_{qr} \zeta_2
          - 160 L_{qr} \zeta_2 \zeta_3
          - \frac{256}{15} L_{qr} \zeta_2^2
          + 551 L_{qr}^2
          - 320 L_{qr}^2 \zeta_3
          - 66 L_{qr}^3
          + \frac{2348}{3} L_{fr}
          - 240 L_{fr} \zeta_5
          - \frac{3344}{3} L_{fr} \zeta_3
          + \frac{908}{3} L_{fr} \zeta_2
          + 160 L_{fr} \zeta_2 \zeta_3
          - \frac{1328}{15} L_{fr} \zeta_2^2
          - 420 L_{fr} L_{qr}
          + 288 L_{fr} L_{qr} \zeta_3
          - 176 L_{fr} L_{qr} \zeta_2
          + 66 L_{fr} L_{qr}^2
          - 131 L_{fr}^2
          + 32 L_{fr}^2 \zeta_3
          + 176 L_{fr}^2 \zeta_2
          + 66 L_{fr}^2 L_{qr}
          - 66 L_{fr}^3
          \bigg\}

       + C_F^3   \bigg\{
          - \frac{5599}{6}
          + 1328 \zeta_5
          - 460 \zeta_3
          + 32 \zeta_3^2
          - \frac{130}{3} \zeta_2
          + 80 \zeta_2 \zeta_3
          - \frac{612}{5} \zeta_2^2
          + \frac{25696}{315} \zeta_2^3
          + \frac{1495}{2} L_{qr}
          - 480 L_{qr} \zeta_5
          - 992 L_{qr} \zeta_3
          + 24 L_{qr} \zeta_2
          + 320 L_{qr} \zeta_2 \zeta_3
          + \frac{48}{5} L_{qr} \zeta_2^2
          - 270 L_{qr}^2
          + 288 L_{qr}^2 \zeta_3
          + 36 L_{qr}^3
          - \frac{1495}{2} L_{fr}
          + 480 L_{fr} \zeta_5
          + 992 L_{fr} \zeta_3
          - 24 L_{fr} \zeta_2
          - 320 L_{fr} \zeta_2 \zeta_3
          - \frac{48}{5} L_{fr} \zeta_2^2
          + 540 L_{fr} L_{qr}
          - 576 L_{fr} L_{qr} \zeta_3
          - 108 L_{fr} L_{qr}^2
          - 270 L_{fr}^2
          + 288 L_{fr}^2 \zeta_3
          + 108 L_{fr}^2 L_{qr}
          - 36 L_{fr}^3
          \bigg\}\,.
\end{autobreak}
\end{align}
In the aforementioned equations, $n_{fv}$ is proportional to the charge weighted sum of the quark flavours and $N_4 = (n_c^2-4)/n_c$ \cite{Gehrmann:2010ue}. Here $L_{qr} = \ln \big(\frac{Q^2}{\mu_R^2}\big)$ and $L_{fr} = \ln \big(\frac{\mu_F^2}{\mu_R^2}\big)$.

\subsection{The $N$-independent coefficients $g_{0,i}^q $} \label{app:g0q}
 The $N$-independent coefficients $g_{0,i}^q $ in Eq. \ref{lng0}, till three-loop,
are given by
\begin{align}
\begin{autobreak}
g_{0,0}^q = 
     0\,,
\end{autobreak}\\
\begin{autobreak}
 g_{0,1}^q=

        C_F   \bigg\{
           8 \zeta_2
          - 8 \gamma_E L_{qr}
          + 8 \gamma_E L_{fr}
          + 8 \gamma_E^2
          \bigg\}\,,
\end{autobreak}\\
\begin{autobreak}
g_{0,2}^q=

         C_F n_f   \bigg\{
          - \frac{64}{9} \zeta_3
          - \frac{80}{9} \zeta_2
          + \frac{16}{3} L_{qr} \zeta_2
          - \frac{224}{27} \gamma_E
          + \frac{80}{9} \gamma_E L_{qr}
          - \frac{8}{3} \gamma_E L_{qr}^2
          - \frac{80}{9} \gamma_E L_{fr}
          + \frac{8}{3} \gamma_E L_{fr}^2
          - \frac{80}{9} \gamma_E^2
          + \frac{16}{3} \gamma_E^2 L_{qr}
          - \frac{32}{9} \gamma_E^3
          \bigg\}

       + C_F C_A   \bigg\{
           \frac{352}{9} \zeta_3
          + \frac{536}{9} \zeta_2
          - 16 \zeta_2^2
          - \frac{88}{3} L_{qr} \zeta_2
          + \frac{1616}{27} \gamma_E
          - 56 \gamma_E \zeta_3
          - \frac{536}{9} \gamma_E L_{qr}
          + 16 \gamma_E L_{qr} \zeta_2
          + \frac{44}{3} \gamma_E L_{qr}^2
          + \frac{536}{9} \gamma_E L_{fr}
          - 16 \gamma_E L_{fr} \zeta_2
          - \frac{44}{3} \gamma_E L_{fr}^2
          + \frac{536}{9} \gamma_E^2
          - 16 \gamma_E^2 \zeta_2
          - \frac{88}{3} \gamma_E^2 L_{qr}
          + \frac{176}{9} \gamma_E^3
          \bigg\}\,,
\end{autobreak}\\
\begin{autobreak}

   g_{0,3}^q =

        C_F n_f^2   \bigg\{
           \frac{1280}{81} \zeta_3
          + \frac{800}{81} \zeta_2
          - \frac{64}{45} \zeta_2^2
          - \frac{256}{27} L_{qr} \zeta_3
          - \frac{320}{27} L_{qr} \zeta_2
          + \frac{32}{9} L_{qr}^2 \zeta_2
          + \frac{3712}{729} \gamma_E
          + \frac{64}{9} \gamma_E \zeta_3
          - \frac{800}{81} \gamma_E L_{qr}
          + \frac{160}{27} \gamma_E L_{qr}^2
          - \frac{32}{27} \gamma_E L_{qr}^3
          - \frac{32}{27} \gamma_E L_{fr}
          - \frac{160}{27} \gamma_E L_{fr}^2
          + \frac{32}{27} \gamma_E L_{fr}^3
          + \frac{800}{81} \gamma_E^2
          - \frac{320}{27} \gamma_E^2 L_{qr}
          + \frac{32}{9} \gamma_E^2 L_{qr}^2
          + \frac{640}{81} \gamma_E^3
          - \frac{128}{27} \gamma_E^3 L_{qr}
          + \frac{64}{27} \gamma_E^4
          \bigg\}

       + C_F C_A n_f   \bigg\{
          - \frac{18496}{81} \zeta_3
          - \frac{16408}{81} \zeta_2
          + \frac{256}{9} \zeta_2 \zeta_3
          + \frac{256}{5} \zeta_2^2
          + \frac{2816}{27} L_{qr} \zeta_3
          + \frac{4624}{27} L_{qr} \zeta_2
          - \frac{64}{3} L_{qr} \zeta_2^2
          - \frac{352}{9} L_{qr}^2 \zeta_2
          - \frac{125252}{729} \gamma_E
          + \frac{1808}{27} \gamma_E \zeta_3
          + \frac{1648}{81} \gamma_E \zeta_2
          - \frac{32}{5} \gamma_E \zeta_2^2
          + \frac{16408}{81} \gamma_E L_{qr}
          - \frac{320}{9} \gamma_E L_{qr} \zeta_2
          - \frac{2312}{27} \gamma_E L_{qr}^2
          + \frac{32}{3} \gamma_E L_{qr}^2 \zeta_2
          + \frac{352}{27} \gamma_E L_{qr}^3
          - \frac{1672}{27} \gamma_E L_{fr}
          - \frac{224}{3} \gamma_E L_{fr} \zeta_3
          + \frac{320}{9} \gamma_E L_{fr} \zeta_2
          + \frac{2312}{27} \gamma_E L_{fr}^2
          - \frac{32}{3} \gamma_E L_{fr}^2 \zeta_2
          - \frac{352}{27} \gamma_E L_{fr}^3
          - \frac{16408}{81} \gamma_E^2
          + \frac{320}{9} \gamma_E^2 \zeta_2
          + \frac{4624}{27} \gamma_E^2 L_{qr}
          - \frac{64}{3} \gamma_E^2 L_{qr} \zeta_2
          - \frac{352}{9} \gamma_E^2 L_{qr}^2
          - \frac{9248}{81} \gamma_E^3
          + \frac{128}{9} \gamma_E^3 \zeta_2
          + \frac{1408}{27} \gamma_E^3 L_{qr}
          - \frac{704}{27} \gamma_E^4
          \bigg\}

       + C_F C_A^2   \bigg\{
           \frac{56960}{81} \zeta_3
          + \frac{62012}{81} \zeta_2
          - \frac{4576}{9} \zeta_2 \zeta_3
          - \frac{12656}{45} \zeta_2^2
          + \frac{352}{5} \zeta_2^3
          - \frac{7744}{27} L_{qr} \zeta_3
          - \frac{14240}{27} L_{qr} \zeta_2
          + \frac{352}{3} L_{qr} \zeta_2^2
          + \frac{968}{9} L_{qr}^2 \zeta_2
          + \frac{594058}{729} \gamma_E
          + 384 \gamma_E \zeta_5
          - \frac{24656}{27} \gamma_E \zeta_3
          - \frac{12784}{81} \gamma_E \zeta_2
          + \frac{352}{3} \gamma_E \zeta_2 \zeta_3
          - \frac{176}{5} \gamma_E \zeta_2^2
          - \frac{62012}{81} \gamma_E L_{qr}
          + 352 \gamma_E L_{qr} \zeta_3
          + \frac{2144}{9} \gamma_E L_{qr} \zeta_2
          - \frac{352}{5} \gamma_E L_{qr} \zeta_2^2
          + \frac{7120}{27} \gamma_E L_{qr}^2
          - \frac{176}{3} \gamma_E L_{qr}^2 \zeta_2
          - \frac{968}{27} \gamma_E L_{qr}^3
          + \frac{980}{3} \gamma_E L_{fr}
          + \frac{176}{3} \gamma_E L_{fr} \zeta_3
          - \frac{2144}{9} \gamma_E L_{fr} \zeta_2
          + \frac{352}{5} \gamma_E L_{fr} \zeta_2^2
          - \frac{7120}{27} \gamma_E L_{fr}^2
          + \frac{176}{3} \gamma_E L_{fr}^2 \zeta_2
          + \frac{968}{27} \gamma_E L_{fr}^3
          + \frac{62012}{81} \gamma_E^2
          - 352 \gamma_E^2 \zeta_3
          - \frac{2144}{9} \gamma_E^2 \zeta_2
          + \frac{352}{5} \gamma_E^2 \zeta_2^2
          - \frac{14240}{27} \gamma_E^2 L_{qr}
          + \frac{352}{3} \gamma_E^2 L_{qr} \zeta_2
          + \frac{968}{9} \gamma_E^2 L_{qr}^2
          + \frac{28480}{81} \gamma_E^3
          - \frac{704}{9} \gamma_E^3 \zeta_2
          - \frac{3872}{27} \gamma_E^3 L_{qr}
          + \frac{1936}{27} \gamma_E^4
          \bigg\}

       + C_F^2 n_f   \bigg\{
          - \frac{64}{3} \zeta_3
          - \frac{220}{3} \zeta_2
          + 64 \zeta_2 \zeta_3
          + 16 L_{qr} \zeta_2
          - \frac{3422}{27} \gamma_E
          + \frac{608}{9} \gamma_E \zeta_3
          + \frac{64}{5} \gamma_E \zeta_2^2
          + \frac{220}{3} \gamma_E L_{qr}
          - 64 \gamma_E L_{qr} \zeta_3
          - 8 \gamma_E L_{qr}^2
          - \frac{220}{3} \gamma_E L_{fr}
          + 64 \gamma_E L_{fr} \zeta_3
          + 8 \gamma_E L_{fr}^2
          - \frac{220}{3} \gamma_E^2
          + 64 \gamma_E^2 \zeta_3
          + 16 \gamma_E^2 L_{qr}
          - \frac{32}{3} \gamma_E^3
          \bigg\}\,.
\end{autobreak}
\end{align}
Here, $\gamma_E$ is the Euler-Mascheroni constant.

\subsection{The $N$-independent coefficients $\Tilde{g}^q_0$} \label{app:g0t}
The $N$-independent coefficients $\Tilde{g}^q_{0,i}$ in Eq. \ref{g0t}, till three-loop, are given by

\begin{align}
\begin{autobreak}

   \Tilde{g}^q_{0,0} =
         1
         \,,
\end{autobreak}\\
\begin{autobreak}
    
    \Tilde{g}^q_{0,1} =

           C_F   \bigg\{
          - 16
          + 16 \zeta_2
          + 6 L_{qr}
          - 6 L_{fr}
          - 8 \gamma_E L_{qr}
          + 8 \gamma_E L_{fr}
          + 8 \gamma_E^2
          \bigg\}\,,
\end{autobreak}\\
\begin{autobreak}
    \Tilde{g}^q_{0,2} =

          C_F n_f   \bigg\{
            \frac{127}{6}
          + \frac{8}{9} \zeta_3
          - \frac{64}{3} \zeta_2
          - \frac{34}{3} L_{qr}
          + \frac{16}{3} L_{qr} \zeta_2
          + 2 L_{qr}^2
          + \frac{2}{3} L_{fr}
          + \frac{16}{3} L_{fr} \zeta_2
          - 2 L_{fr}^2
          - \frac{224}{27} \gamma_E
          + \frac{80}{9} \gamma_E L_{qr}
          - \frac{8}{3} \gamma_E L_{qr}^2
          - \frac{80}{9} \gamma_E L_{fr}
          + \frac{8}{3} \gamma_E L_{fr}^2
          - \frac{80}{9} \gamma_E^2
          + \frac{16}{3} \gamma_E^2 L_{qr}
          - \frac{32}{9} \gamma_E^3 \bigg\}

       + C_F C_A   \bigg\{
          - \frac{1535}{12}
          + \frac{604}{9} \zeta_3
          + \frac{376}{3} \zeta_2
          - \frac{92}{5} \zeta_2^2
          + \frac{193}{3} L_{qr}
          - 24 L_{qr} \zeta_3
          - \frac{88}{3} L_{qr} \zeta_2
          - 11 L_{qr}^2
          - \frac{17}{3} L_{fr}
          + 24 L_{fr} \zeta_3
          - \frac{88}{3} L_{fr} \zeta_2
          + 11 L_{fr}^2
          + \frac{1616}{27} \gamma_E
          - 56 \gamma_E \zeta_3
          - \frac{536}{9} \gamma_E L_{qr}
          + 16 \gamma_E L_{qr} \zeta_2
          + \frac{44}{3} \gamma_E L_{qr}^2
          + \frac{536}{9} \gamma_E L_{fr}
          - 16 \gamma_E L_{fr} \zeta_2
          - \frac{44}{3} \gamma_E L_{fr}^2
          + \frac{536}{9} \gamma_E^2
          - 16 \gamma_E^2 \zeta_2
          - \frac{88}{3} \gamma_E^2 L_{qr}
          + \frac{176}{9} \gamma_E^3
          \bigg\}

       + C_F^2   \bigg\{
          + \frac{511}{4}
          - 60 \zeta_3
          - 198 \zeta_2
          + \frac{552}{5} \zeta_2^2
          - 93 L_{qr}
          + 48 L_{qr} \zeta_3
          + 72 L_{qr} \zeta_2
          + 18 L_{qr}^2
          + 93 L_{fr}
          - 48 L_{fr} \zeta_3
          - 72 L_{fr} \zeta_2
          - 36 L_{fr} L_{qr}
          + 18 L_{fr}^2
          + 128 \gamma_E L_{qr}
          - 128 \gamma_E L_{qr} \zeta_2
          - 48 \gamma_E L_{qr}^2
          - 128 \gamma_E L_{fr}
          + 128 \gamma_E L_{fr} \zeta_2
          + 96 \gamma_E L_{fr} L_{qr}
          - 48 \gamma_E L_{fr}^2
          - 128 \gamma_E^2
          + 128 \gamma_E^2 \zeta_2
          + 48 \gamma_E^2 L_{qr}
          + 32 \gamma_E^2 L_{qr}^2
          - 48 \gamma_E^2 L_{fr}
          - 64 \gamma_E^2 L_{fr} L_{qr}
          + 32 \gamma_E^2 L_{fr}^2
          - 64 \gamma_E^3 L_{qr}
          + 64 \gamma_E^3 L_{fr}
          + 32 \gamma_E^4
          \bigg\}\,,
\end{autobreak}\\
\begin{autobreak}
    \Tilde{g}^q_{0,3} =

           C_F N_4 n_{fv}   \bigg\{
            8
          - \frac{160}{3} \zeta_5
          + \frac{28}{3} \zeta_3
          + 20 \zeta_2
          - \frac{4}{5} \zeta_2^2
          \bigg\}

       + C_F n_f^2   \bigg\{
          - \frac{7081}{243}
          + \frac{16}{81} \zeta_3
          + \frac{1072}{27} \zeta_2
          + \frac{448}{135} \zeta_2^2
          + \frac{220}{9} L_{qr}
          - \frac{64}{27} L_{qr} \zeta_3
          - \frac{608}{27} L_{qr} \zeta_2
          - \frac{68}{9} L_{qr}^2
          + \frac{32}{9} L_{qr}^2 \zeta_2
          + \frac{8}{9} L_{qr}^3
          + \frac{34}{9} L_{fr}
          + \frac{32}{9} L_{fr} \zeta_3
          - \frac{160}{27} L_{fr} \zeta_2
          + \frac{4}{9} L_{fr}^2
          + \frac{32}{9} L_{fr}^2 \zeta_2
          - \frac{8}{9} L_{fr}^3
          + \frac{3712}{729} \gamma_E
          + \frac{64}{9} \gamma_E \zeta_3
          - \frac{800}{81} \gamma_E L_{qr}
          + \frac{160}{27} \gamma_E L_{qr}^2
          - \frac{32}{27} \gamma_E L_{qr}^3
          - \frac{160}{27} \gamma_E L_{fr}^2
          + \frac{32}{27} \gamma_E L_{fr}^3
          + \frac{800}{81} \gamma_E^2
          - \frac{320}{27} \gamma_E^2 L_{qr}
          + \frac{32}{9} \gamma_E^2 L_{qr}^2
          + \frac{640}{81} \gamma_E^3
          - \frac{128}{27} \gamma_E^3 L_{qr}
          + \frac{64}{27} \gamma_E^4
          \bigg\}

       + C_F C_A n_f   \bigg\{
            \frac{110651}{243}
          - 8 \zeta_5
          - \frac{24512}{81} \zeta_3
          - \frac{44540}{81} \zeta_2
          + \frac{880}{9} \zeta_2 \zeta_3
          + \frac{1156}{135} \zeta_2^2
          - \frac{3052}{9} L_{qr}
          + \frac{3440}{27} L_{qr} \zeta_3
          + \frac{7504}{27} L_{qr} \zeta_2
          - \frac{344}{15} L_{qr} \zeta_2^2
          + \frac{850}{9} L_{qr}^2
          - 16 L_{qr}^2 \zeta_3
          - \frac{352}{9} L_{qr}^2 \zeta_2
          - \frac{88}{9} L_{qr}^3
          - 40 L_{fr}
          - \frac{400}{9} L_{fr} \zeta_3
          + \frac{2672}{27} L_{fr} \zeta_2
          - \frac{8}{5} L_{fr} \zeta_2^2
          - \frac{146}{9} L_{fr}^2
          + 16 L_{fr}^2 \zeta_3
          - \frac{352}{9} L_{fr}^2 \zeta_2
          + \frac{88}{9} L_{fr}^3
          - \frac{125252}{729} \gamma_E
          + \frac{1808}{27} \gamma_E \zeta_3
          + \frac{1648}{81} \gamma_E \zeta_2
          - \frac{32}{5} \gamma_E \zeta_2^2
          + \frac{16408}{81} \gamma_E L_{qr}
          - \frac{320}{9} \gamma_E L_{qr} \zeta_2
          - \frac{2312}{27} \gamma_E L_{qr}^2
          + \frac{32}{3} \gamma_E L_{qr}^2 \zeta_2
          + \frac{352}{27} \gamma_E L_{qr}^3
          - \frac{1672}{27} \gamma_E L_{fr}
          - \frac{224}{3} \gamma_E L_{fr} \zeta_3
          + \frac{320}{9} \gamma_E L_{fr} \zeta_2
          + \frac{2312}{27} \gamma_E L_{fr}^2
          - \frac{32}{3} \gamma_E L_{fr}^2 \zeta_2
          - \frac{352}{27} \gamma_E L_{fr}^3
          - \frac{16408}{81} \gamma_E^2
          + \frac{320}{9} \gamma_E^2 \zeta_2
          + \frac{4624}{27} \gamma_E^2 L_{qr}
          - \frac{64}{3} \gamma_E^2 L_{qr} \zeta_2
          - \frac{352}{9} \gamma_E^2 L_{qr}^2
          - \frac{9248}{81} \gamma_E^3
          + \frac{128}{9} \gamma_E^3 \zeta_2
          + \frac{1408}{27} \gamma_E^3 L_{qr}
          - \frac{704}{27} \gamma_E^4
          \bigg\}

       + C_F C_A^2   \bigg\{
          - \frac{1505881}{972}
          - 204 \zeta_5
          + \frac{139345}{81} \zeta_3
          - \frac{400}{3} \zeta_3^2
          + \frac{130295}{81} \zeta_2
          - \frac{7228}{9} \zeta_2 \zeta_3
          - \frac{23357}{135} \zeta_2^2
          + \frac{7088}{63} \zeta_2^3
          + \frac{3082}{3} L_{qr}
          + 80 L_{qr} \zeta_5
          - \frac{22600}{27} L_{qr} \zeta_3
          - \frac{20720}{27} L_{qr} \zeta_2
          + \frac{1964}{15} L_{qr} \zeta_2^2
          - \frac{2429}{9} L_{qr}^2
          + 88 L_{qr}^2 \zeta_3
          + \frac{968}{9} L_{qr}^2 \zeta_2
          + \frac{242}{9} L_{qr}^3
          + \frac{1657}{18} L_{fr}
          - 80 L_{fr} \zeta_5
          + \frac{3104}{9} L_{fr} \zeta_3
          - \frac{8992}{27} L_{fr} \zeta_2
          + 4 L_{fr} \zeta_2^2
          + \frac{493}{9} L_{fr}^2
          - 88 L_{fr}^2 \zeta_3
          + \frac{968}{9} L_{fr}^2 \zeta_2
          - \frac{242}{9} L_{fr}^3
          + \frac{594058}{729} \gamma_E
          + 384 \gamma_E \zeta_5
          - \frac{24656}{27} \gamma_E \zeta_3
          - \frac{12784}{81} \gamma_E \zeta_2
          + \frac{352}{3} \gamma_E \zeta_2 \zeta_3
          - \frac{176}{5} \gamma_E \zeta_2^2
          - \frac{62012}{81} \gamma_E L_{qr}
          + 352 \gamma_E L_{qr} \zeta_3
          + \frac{2144}{9} \gamma_E L_{qr} \zeta_2
          - \frac{352}{5} \gamma_E L_{qr} \zeta_2^2
          + \frac{7120}{27} \gamma_E L_{qr}^2
          - \frac{176}{3} \gamma_E L_{qr}^2 \zeta_2
          - \frac{968}{27} \gamma_E L_{qr}^3
          + \frac{980}{3} \gamma_E L_{fr}
          + \frac{176}{3} \gamma_E L_{fr} \zeta_3
          - \frac{2144}{9} \gamma_E L_{fr} \zeta_2
          + \frac{352}{5} \gamma_E L_{fr} \zeta_2^2
          - \frac{7120}{27} \gamma_E L_{fr}^2
          + \frac{176}{3} \gamma_E L_{fr}^2 \zeta_2
          + \frac{968}{27} \gamma_E L_{fr}^3
          + \frac{62012}{81} \gamma_E^2
          - 352 \gamma_E^2 \zeta_3
          - \frac{2144}{9} \gamma_E^2 \zeta_2
          + \frac{352}{5} \gamma_E^2 \zeta_2^2
          - \frac{14240}{27} \gamma_E^2 L_{qr}
          + \frac{352}{3} \gamma_E^2 L_{qr} \zeta_2
          + \frac{968}{9} \gamma_E^2 L_{qr}^2
          + \frac{28480}{81} \gamma_E^3
          - \frac{704}{9} \gamma_E^3 \zeta_2
          - \frac{3872}{27} \gamma_E^3 L_{qr}
          + \frac{1936}{27} \gamma_E^4
          \bigg\}

       + C_F^2 n_f   \bigg\{
          - \frac{421}{3}
          - \frac{608}{9} \zeta_5
          + \frac{9448}{27} \zeta_3
          + \frac{9064}{27} \zeta_2
          - \frac{256}{3} \zeta_2 \zeta_3
          - \frac{36208}{135} \zeta_2^2
          + 230 L_{qr}
          - \frac{496}{3} L_{qr} \zeta_3
          - 272 L_{qr} \zeta_2
          + \frac{464}{5} L_{qr} \zeta_2^2
          - 92 L_{qr}^2
          + 32 L_{qr}^2 \zeta_3
          + 48 L_{qr}^2 \zeta_2
          + 12 L_{qr}^3
          - \frac{275}{3} L_{fr}
          + \frac{256}{3} L_{fr} \zeta_3
          + 40 L_{fr} \zeta_2
          + \frac{272}{5} L_{fr} \zeta_2^2
          + 72 L_{fr} L_{qr}
          - 12 L_{fr} L_{qr}^2
          + 20 L_{fr}^2
          - 32 L_{fr}^2 \zeta_3
          - 48 L_{fr}^2 \zeta_2
          - 12 L_{fr}^2 L_{qr}
          + 12 L_{fr}^3
          + 6 \gamma_E
          + \frac{608}{9} \gamma_E \zeta_3
          - \frac{3584}{27} \gamma_E \zeta_2
          + \frac{64}{5} \gamma_E \zeta_2^2
          - 288 \gamma_E L_{qr}
          - \frac{640}{9} \gamma_E L_{qr} \zeta_3
          + \frac{2816}{9} \gamma_E L_{qr} \zeta_2
          + \frac{536}{3} \gamma_E L_{qr}^2
          - \frac{256}{3} \gamma_E L_{qr}^2 \zeta_2
          - 32 \gamma_E L_{qr}^3
          + 288 \gamma_E L_{fr}
          + \frac{640}{9} \gamma_E L_{fr} \zeta_3
          - \frac{2816}{9} \gamma_E L_{fr} \zeta_2
          - \frac{608}{3} \gamma_E L_{fr} L_{qr}
          + 32 \gamma_E L_{fr} L_{qr}^2
          + 24 \gamma_E L_{fr}^2
          + \frac{256}{3} \gamma_E L_{fr}^2 \zeta_2
          + 32 \gamma_E L_{fr}^2 L_{qr}
          - 32 \gamma_E L_{fr}^3
          + \frac{2144}{9} \gamma_E^2
          + \frac{640}{9} \gamma_E^2 \zeta_3
          - \frac{2816}{9} \gamma_E^2 \zeta_2
          - \frac{3968}{27} \gamma_E^2 L_{qr}
          + 128 \gamma_E^2 L_{qr} \zeta_2
          - \frac{208}{9} \gamma_E^2 L_{qr}^2
          + \frac{64}{3} \gamma_E^2 L_{qr}^3
          - \frac{208}{27} \gamma_E^2 L_{fr}
          + \frac{128}{3} \gamma_E^2 L_{fr} \zeta_2
          + \frac{992}{9} \gamma_E^2 L_{fr} L_{qr}
          - \frac{64}{3} \gamma_E^2 L_{fr} L_{qr}^2
          - \frac{784}{9} \gamma_E^2 L_{fr}^2
          - \frac{64}{3} \gamma_E^2 L_{fr}^2 L_{qr}
          + \frac{64}{3} \gamma_E^2 L_{fr}^3
          - \frac{544}{27} \gamma_E^3
          - \frac{512}{9} \gamma_E^3 \zeta_2
          + \frac{1088}{9} \gamma_E^3 L_{qr}
          - 64 \gamma_E^3 L_{qr}^2
          - \frac{1088}{9} \gamma_E^3 L_{fr}
          + \frac{128}{3} \gamma_E^3 L_{fr} L_{qr}
          + \frac{64}{3} \gamma_E^3 L_{fr}^2
          - \frac{640}{9} \gamma_E^4
          + \frac{640}{9} \gamma_E^4 L_{qr}
          - \frac{256}{9} \gamma_E^4 L_{fr}
          - \frac{256}{9} \gamma_E^5
          \bigg\}

       + C_F^2 C_A   \bigg\{
          + \frac{74321}{36}
          - \frac{5512}{9} \zeta_5
          - \frac{51508}{27} \zeta_3
          + \frac{592}{3} \zeta_3^2
          - \frac{66544}{27} \zeta_2
          + \frac{3680}{3} \zeta_2 \zeta_3
          + \frac{258304}{135} \zeta_2^2
          - \frac{123632}{315} \zeta_2^3
          - \frac{3439}{2} L_{qr}
          + 240 L_{qr} \zeta_5
          + \frac{5368}{3} L_{qr} \zeta_3
          + 1552 L_{qr} \zeta_2
          - 352 L_{qr} \zeta_2 \zeta_3
          - \frac{2912}{5} L_{qr} \zeta_2^2
          + 551 L_{qr}^2
          - 320 L_{qr}^2 \zeta_3
          - 264 L_{qr}^2 \zeta_2
          - 66 L_{qr}^3
          + \frac{2348}{3} L_{fr}
          - 240 L_{fr} \zeta_5
          - \frac{4048}{3} L_{fr} \zeta_3
          - 100 L_{fr} \zeta_2
          + 352 L_{fr} \zeta_2 \zeta_3
          - \frac{1136}{5} L_{fr} \zeta_2^2
          - 420 L_{fr} L_{qr}
          + 288 L_{fr} L_{qr} \zeta_3
          + 66 L_{fr} L_{qr}^2
          - 131 L_{fr}^2
          + 32 L_{fr}^2 \zeta_3
          + 264 L_{fr}^2 \zeta_2
          + 66 L_{fr}^2 L_{qr}
          - 66 L_{fr}^3
          - \frac{25856}{27} \gamma_E
          + 896 \gamma_E \zeta_3
          + \frac{25856}{27} \gamma_E \zeta_2
          - 896 \gamma_E \zeta_2 \zeta_3
          + \frac{7006}{3} \gamma_E L_{qr}
          - \frac{7856}{9} \gamma_E L_{qr} \zeta_3
          - \frac{19904}{9} \gamma_E L_{qr} \zeta_2
          + \frac{2016}{5} \gamma_E L_{qr} \zeta_2^2
          - \frac{3320}{3} \gamma_E L_{qr}^2
          + 192 \gamma_E L_{qr}^2 \zeta_3
          + \frac{1696}{3} \gamma_E L_{qr}^2 \zeta_2
          + 176 \gamma_E L_{qr}^3
          - \frac{7006}{3} \gamma_E L_{fr}
          + \frac{7856}{9} \gamma_E L_{fr} \zeta_3
          + \frac{19904}{9} \gamma_E L_{fr} \zeta_2
          - \frac{2016}{5} \gamma_E L_{fr} \zeta_2^2
          + \frac{3824}{3} \gamma_E L_{fr} L_{qr}
          - 384 \gamma_E L_{fr} L_{qr} \zeta_3
          - 192 \gamma_E L_{fr} L_{qr} \zeta_2
          - 176 \gamma_E L_{fr} L_{qr}^2
          - 168 \gamma_E L_{fr}^2
          + 192 \gamma_E L_{fr}^2 \zeta_3
          - \frac{1120}{3} \gamma_E L_{fr}^2 \zeta_2
          - 176 \gamma_E L_{fr}^2 L_{qr}
          + 176 \gamma_E L_{fr}^3
          - \frac{17786}{9} \gamma_E^2
          + \frac{4832}{9} \gamma_E^2 \zeta_3
          + \frac{19904}{9} \gamma_E^2 \zeta_2
          - \frac{2016}{5} \gamma_E^2 \zeta_2^2
          + \frac{23288}{27} \gamma_E^2 L_{qr}
          + 256 \gamma_E^2 L_{qr} \zeta_3
          - 800 \gamma_E^2 L_{qr} \zeta_2
          + \frac{1912}{9} \gamma_E^2 L_{qr}^2
          - 128 \gamma_E^2 L_{qr}^2 \zeta_2
          - \frac{352}{3} \gamma_E^2 L_{qr}^3
          + \frac{2056}{27} \gamma_E^2 L_{fr}
          - 256 \gamma_E^2 L_{fr} \zeta_3
          - \frac{416}{3} \gamma_E^2 L_{fr} \zeta_2
          - \frac{6992}{9} \gamma_E^2 L_{fr} L_{qr}
          + 256 \gamma_E^2 L_{fr} L_{qr} \zeta_2
          + \frac{352}{3} \gamma_E^2 L_{fr} L_{qr}^2
          + \frac{5080}{9} \gamma_E^2 L_{fr}^2
          - 128 \gamma_E^2 L_{fr}^2 \zeta_2
          + \frac{352}{3} \gamma_E^2 L_{fr}^2 L_{qr}
          - \frac{352}{3} \gamma_E^2 L_{fr}^3
          + \frac{4480}{27} \gamma_E^3
          - 448 \gamma_E^3 \zeta_3
          + \frac{2816}{9} \gamma_E^3 \zeta_2
          - \frac{7520}{9} \gamma_E^3 L_{qr}
          + 256 \gamma_E^3 L_{qr} \zeta_2
          + 352 \gamma_E^3 L_{qr}^2
          + \frac{7520}{9} \gamma_E^3 L_{fr}
          - 256 \gamma_E^3 L_{fr} \zeta_2
          - \frac{704}{3} \gamma_E^3 L_{fr} L_{qr}
          - \frac{352}{3} \gamma_E^3 L_{fr}^2
          + \frac{4288}{9} \gamma_E^4
          - 128 \gamma_E^4 \zeta_2
          - \frac{3520}{9} \gamma_E^4 L_{qr}
          + \frac{1408}{9} \gamma_E^4 L_{fr}
          + \frac{1408}{9} \gamma_E^5
          \bigg\}

       + C_F^3   \bigg\{
          - \frac{5599}{6}
          + 1328 \zeta_5
          - 460 \zeta_3
          + 32 \zeta_3^2
          + \frac{2936}{3} \zeta_2
          - 400 \zeta_2 \zeta_3
          - \frac{5972}{5} \zeta_2^2
          + \frac{169504}{315} \zeta_2^3
          + \frac{1495}{2} L_{qr}
          - 480 L_{qr} \zeta_5
          - 992 L_{qr} \zeta_3
          - 720 L_{qr} \zeta_2
          + 704 L_{qr} \zeta_2 \zeta_3
          + \frac{1968}{5} L_{qr} \zeta_2^2
          - 270 L_{qr}^2
          + 288 L_{qr}^2 \zeta_3
          + 144 L_{qr}^2 \zeta_2
          + 36 L_{qr}^3
          - \frac{1495}{2} L_{fr}
          + 480 L_{fr} \zeta_5
          + 992 L_{fr} \zeta_3
          + 720 L_{fr} \zeta_2
          - 704 L_{fr} \zeta_2 \zeta_3
          - \frac{1968}{5} L_{fr} \zeta_2^2
          + 540 L_{fr} L_{qr}
          - 576 L_{fr} L_{qr} \zeta_3
          - 288 L_{fr} L_{qr} \zeta_2
          - 108 L_{fr} L_{qr}^2
          - 270 L_{fr}^2
          + 288 L_{fr}^2 \zeta_3
          + 144 L_{fr}^2 \zeta_2
          + 108 L_{fr}^2 L_{qr}
          - 36 L_{fr}^3
          - 1022 \gamma_E L_{qr}
          + 480 \gamma_E L_{qr} \zeta_3
          + 1584 \gamma_E L_{qr} \zeta_2
          - \frac{4416}{5} \gamma_E L_{qr} \zeta_2^2
          + 744 \gamma_E L_{qr}^2
          - 384 \gamma_E L_{qr}^2 \zeta_3
          - 576 \gamma_E L_{qr}^2 \zeta_2
          - 144 \gamma_E L_{qr}^3
          + 1022 \gamma_E L_{fr}
          - 480 \gamma_E L_{fr} \zeta_3
          - 1584 \gamma_E L_{fr} \zeta_2
          + \frac{4416}{5} \gamma_E L_{fr} \zeta_2^2
          - 1488 \gamma_E L_{fr} L_{qr}
          + 768 \gamma_E L_{fr} L_{qr} \zeta_3
          + 1152 \gamma_E L_{fr} L_{qr} \zeta_2
          + 432 \gamma_E L_{fr} L_{qr}^2
          + 744 \gamma_E L_{fr}^2
          - 384 \gamma_E L_{fr}^2 \zeta_3
          - 576 \gamma_E L_{fr}^2 \zeta_2
          - 432 \gamma_E L_{fr}^2 L_{qr}
          + 144 \gamma_E L_{fr}^3
          + 1022 \gamma_E^2
          - 480 \gamma_E^2 \zeta_3
          - 1584 \gamma_E^2 \zeta_2
          + \frac{4416}{5} \gamma_E^2 \zeta_2^2
          - 744 \gamma_E^2 L_{qr}
          + 384 \gamma_E^2 L_{qr} \zeta_3
          + 576 \gamma_E^2 L_{qr} \zeta_2
          - 368 \gamma_E^2 L_{qr}^2
          + 512 \gamma_E^2 L_{qr}^2 \zeta_2
          + 192 \gamma_E^2 L_{qr}^3
          + 744 \gamma_E^2 L_{fr}
          - 384 \gamma_E^2 L_{fr} \zeta_3
          - 576 \gamma_E^2 L_{fr} \zeta_2
          + 736 \gamma_E^2 L_{fr} L_{qr}
          - 1024 \gamma_E^2 L_{fr} L_{qr} \zeta_2
          - 576 \gamma_E^2 L_{fr} L_{qr}^2
          - 368 \gamma_E^2 L_{fr}^2
          + 512 \gamma_E^2 L_{fr}^2 \zeta_2
          + 576 \gamma_E^2 L_{fr}^2 L_{qr}
          - 192 \gamma_E^2 L_{fr}^3
          + 1024 \gamma_E^3 L_{qr}
          - 1024 \gamma_E^3 L_{qr} \zeta_2
          - 384 \gamma_E^3 L_{qr}^2
          - \frac{256}{3} \gamma_E^3 L_{qr}^3
          - 1024 \gamma_E^3 L_{fr}
          + 1024 \gamma_E^3 L_{fr} \zeta_2
          + 768 \gamma_E^3 L_{fr} L_{qr}
          + 256 \gamma_E^3 L_{fr} L_{qr}^2
          - 384 \gamma_E^3 L_{fr}^2
          - 256 \gamma_E^3 L_{fr}^2 L_{qr}
          + \frac{256}{3} \gamma_E^3 L_{fr}^3
          - 512 \gamma_E^4
          + 512 \gamma_E^4 \zeta_2
          + 192 \gamma_E^4 L_{qr}
          + 256 \gamma_E^4 L_{qr}^2
          - 192 \gamma_E^4 L_{fr}
          - 512 \gamma_E^4 L_{fr} L_{qr}
          + 256 \gamma_E^4 L_{fr}^2
          - 256 \gamma_E^5 L_{qr}
          + 256 \gamma_E^5 L_{fr}
          + \frac{256}{3} \gamma_E^6
          \bigg\}\,.
\end{autobreak}
\end{align}

\subsection{The SV resummed exponent $g^q_i$} \label{app:gN}
The resummation exponents $g^q_i$ in Eq. \ref{PsiSVN}, till three-loop, are given by   
\begin{align}
    
\begin{autobreak}

g^q_1 =

        \frac{1}{\beta_0} C_F   \bigg\{
           8
          + \frac{8}{\omega} L_\omega
          - 8 L_\omega
          \bigg\}\,,
\end{autobreak}\\
\begin{autobreak}
g^q_2 =

         \frac{\beta_1}{\beta_0^3}  C_F   \bigg\{
           4 \omega
          + 4 L_\omega
          + 2 L_\omega^2
          \bigg\}

       + \frac{1}{\beta_0^2} C_F n_f   \bigg\{
           \frac{40}{9} \omega
          + \frac{40}{9} L_\omega
          \bigg\}

       + \frac{1}{\beta_0^2} C_F C_A   \bigg\{
          - \frac{268}{9} \omega
          + 8 \omega \zeta_2
          - \frac{268}{9} L_\omega
          + 8 L_\omega \zeta_2
          \bigg\}

       + \frac{1}{\beta_0} C_F   \bigg\{
           4 \omega L_{fr}
          + 4 L_\omega L_{qr}
          - 8 L_\omega \gamma_E
          \bigg\}\,,
\end{autobreak}\\
\begin{autobreak}
   g^q_3 =

           \frac{1}{(1-\omega)} \bigg[   \frac{ \beta_1^2}{\beta_0^4} C_F   \bigg\{
           2 \omega^2
          + 4 L_\omega \omega
          + 2 L_\omega^2
          \bigg\}

       + \frac{\beta_2}{\beta_0^3}  C_F   \bigg\{
           4 \omega
          - 2 \omega^2
          \bigg\}

       + \frac{\beta_1}{\beta_0^3}  C_F n_f   \bigg\{
           \frac{40}{9} \omega
          + \frac{20}{9} \omega^2
          + \frac{40}{9} L_\omega
          \bigg\}

       + \frac{\beta_1}{\beta_0^3}  C_F C_A   \bigg\{
          - \frac{268}{9} \omega
          + 8 \omega \zeta_2
          - \frac{134}{9} \omega^2
          + 4 \omega^2 \zeta_2
          - \frac{268}{9} L_\omega
          + 8 L_\omega \zeta_2
          \bigg\}

       + \frac{1}{\beta_0^2} C_F n_f^2   \bigg\{
          - \frac{8}{27}
          \bigg\}

       + \frac{1}{\beta_0^2} C_F C_A n_f   \bigg\{
          - \frac{418}{27}
          - \frac{56}{3} \zeta_3
          + \frac{80}{9} \zeta_2
          \bigg\}

       + \frac{1}{\beta_0^2} C_F C_A^2   \bigg\{
           \frac{245}{3}
          + \frac{44}{3} \zeta_3
          - \frac{536}{9} \zeta_2
          + \frac{88}{5} \zeta_2^2
          \bigg\}

       + \frac{1}{\beta_0^2} C_F^2 n_f   \bigg\{
          - \frac{55}{3}
          + 16 \zeta_3
          \bigg\}

       + \frac{\beta_1}{\beta_0^2}  C_F   \bigg\{
           4 \omega L_{qr}
          - 8 \omega \gamma_E
          + 4 L_\omega L_{qr}
          - 8 L_\omega \gamma_E
          \bigg\}

       + \frac{1}{\beta_0} C_F n_f   \bigg\{
          - \frac{112}{27}
          + \frac{4}{3} \zeta_2
          + \frac{40}{9} L_{qr}
          - \frac{80}{9} \gamma_E
          \bigg\}

       + \frac{1}{\beta_0} C_F C_A   \bigg\{
           \frac{808}{27}
          - 28 \zeta_3
          - \frac{22}{3} \zeta_2
          - \frac{268}{9} L_{qr}
          + 8 L_{qr} \zeta_2
          + \frac{536}{9} \gamma_E
          - 16 \gamma_E \zeta_2
          \bigg\}

       + C_F   \bigg\{
           2 \zeta_2
          + 2 L_{qr}^2
          - 8 \gamma_E L_{qr}
          + 8 \gamma_E^2
          \bigg\} \bigg]

         + C_F   \bigg\{
          - 2 \zeta_2
          - 2 L_{qr}^2
          + 8 \gamma_E L_{qr}
          - 8 \gamma_E^2
          - 2 \omega L_{fr}^2
          \bigg\}
           
           + \frac{\beta_2}{\beta_0^3}  C_F   \bigg\{
           4 L_\omega
          \bigg\} 
          
           + \frac{1}{\beta_0} C_F n_f   \bigg\{
           \frac{112}{27}
          - \frac{4}{3} \zeta_2
          - \frac{40}{9} L_{qr}
          + \frac{80}{9} \gamma_E
          - \frac{40}{9} \omega L_{fr}
          \bigg\}

       + \frac{1}{\beta_0} C_F C_A   \bigg\{
          - \frac{808}{27}
          + 28 \zeta_3
          + \frac{22}{3} \zeta_2
          + \frac{268}{9} L_{qr}
          - 8 L_{qr} \zeta_2
          - \frac{536}{9} \gamma_E
          + 16 \gamma_E \zeta_2
          + \frac{268}{9} \omega L_{fr}
          - 8 \omega L_{fr} \zeta_2
          \bigg\}
          
          + \frac{1}{\beta_0^2} C_F^2 n_f   \bigg\{ \frac{55}{3}
          - 16 \zeta_3
          + \frac{55}{3} \omega
          - 16 \omega \zeta_3
          \bigg\}
           + \frac{1}{\beta_0^2} C_F n_f^2   \bigg\{
           \frac{8}{27}
          + \frac{8}{27} \omega \bigg\}

       + \frac{1}{\beta_0^2} C_F C_A n_f   \bigg\{\frac{418}{27}
          + \frac{56}{3} \zeta_3
          - \frac{80}{9} \zeta_2
          + \frac{418}{27} \omega
          + \frac{56}{3} \omega \zeta_3
          - \frac{80}{9} \omega \zeta_2
          \bigg\}

       + \frac{1}{\beta_0^2} C_F C_A^2   \bigg\{
          - \frac{245}{3}
          - \frac{44}{3} \zeta_3
          + \frac{536}{9} \zeta_2
          - \frac{88}{5} \zeta_2^2
          - \frac{245}{3} \omega
          - \frac{44}{3} \omega \zeta_3
          + \frac{536}{9} \omega \zeta_2
          - \frac{88}{5} \omega \zeta_2^2
          \bigg\}  \,,
\end{autobreak}
\end{align}
Here, ${L}_{\omega}=\ln(1-\omega)$ and $\omega = 2 \beta_0 a_s(\mu_R^2) \ln N$. 
And $\beta_i$'s are the QCD $\beta$ functions which are given by
\begin{align}
  \beta_0&={11 \over 3 } C_A - {2 \over 3 } n_f \, ,
           \nonumber \\[0.5ex]
  \beta_1&={34 \over 3 } C_A^2- 2 n_f C_F -{10 \over 3} n_f C_A \, ,
           \nonumber \\[0.5ex]
  \beta_2&={2857 \over 54} C_A^3 
           -{1415 \over 54} C_A^2 n_f
           +{79 \over 54} C_A n_f^2
           +{11 \over 9} C_F n_f^2
           -{205 \over 18} C_F C_A n_f
           + C_F^2 n_f \,.
\end{align}

\subsection{The NSV resummed exponents  $\overline{g}^q_i$} \label{app:gbN}
The resummation exponents  $\overline{g}^q_i$ in \ref{PsiNSVN}, till three-loop, are given by
\begin{align}
\begin{autobreak}

   \overline{g}^q_1 =

         \frac{1}{\beta_0} C_F   \bigg\{
           4 L_\omega
          \bigg\}\,,
\end{autobreak}    \\
\begin{autobreak}
   \overline{g}^q_2 =

         \frac{1}{(1-\omega)}\bigg[ \frac{1}{\beta_0^2} C_F C_A n_f   \bigg\{
          - \frac{40}{3} \omega
          - \frac{40}{3} L_\omega
          \bigg\}

       + \frac{1}{\beta_0^2} C_F C_A^2   \bigg\{
           \frac{136}{3} \omega
          + \frac{136}{3} L_\omega
          \bigg\}

       + \frac{1}{\beta_0^2} C_F^2 n_f   \bigg\{ - 8 \omega
          - 8 L_\omega \bigg\}

       + \frac{1}{\beta_0} C_F n_f   \bigg\{
           \frac{40}{9} \omega
          \bigg\}

       + \frac{1}{\beta_0} C_F C_A   \bigg\{
          - \frac{268}{9} \omega
          + 8 \omega \zeta_2
          \bigg\}

       + C_F   \bigg\{ - 8
          + 4 L_{qr}
          - 4 L_{fr}
          + 4 L_{fr} \omega
          - 8 \gamma_E
          \bigg\} \bigg]\,,
\end{autobreak}    \\
\begin{autobreak}
   \overline{g}^q_3 =

        \frac{1}{(1-\omega)^2} \bigg[ \frac{1}{\beta_0^3} C_F C_A^2 n_f^2   \bigg\{
           \frac{200}{9} \omega^2
          - \frac{200}{9} L_\omega^2
          \bigg\}

       + \frac{1}{\beta_0^3} C_F C_A^3 n_f   \bigg\{
          - \frac{1360}{9} \omega^2
          + \frac{1360}{9} L_\omega^2
          \bigg\}

       + \frac{1}{\beta_0^3} C_F C_A^4   \bigg\{
           \frac{2312}{9} \omega^2
          - \frac{2312}{9} L_\omega^2
          \bigg\}

       + \frac{1}{\beta_0^3} C_F^2 C_A n_f^2   \bigg\{
            \frac{80}{3} \omega^2
          - \frac{80}{3} L_\omega^2
          \bigg\}

       + \frac{1}{\beta_0^3} C_F^2 C_A^2 n_f   \bigg\{ - \frac{272}{3} \omega^2
          + \frac{272}{3} L_\omega^2 \bigg\}

       + \frac{1}{\beta_0^3} C_F^3 n_f^2   \bigg\{
            8 \omega^2
          - 8 L_\omega^2
          \bigg\}

       + \frac{1}{\beta_0^2} C_F C_A n_f^2   \bigg\{
            \frac{400}{27} \omega
          - \frac{31}{3} \omega^2
          + \frac{400}{27} L_\omega
          \bigg\}

       + \frac{1}{\beta_0^2} C_F C_A^2 n_f   \bigg\{
          - \frac{4040}{27} \omega
          + \frac{80}{3} \omega \zeta_2
          + \frac{1145}{9} \omega^2
          - \frac{40}{3} \omega^2 \zeta_2
          - \frac{4040}{27} L_\omega
          + \frac{80}{3} L_\omega \zeta_2
          \bigg\}

       + \frac{1}{\beta_0^2} C_F C_A^3   \bigg\{
            \frac{9112}{27} \omega
          - \frac{272}{3} \omega \zeta_2
          - \frac{2471}{9} \omega^2
          + \frac{136}{3} \omega^2 \zeta_2
          + \frac{9112}{27} L_\omega
          - \frac{272}{3} L_\omega \zeta_2
          \bigg\}

       + \frac{1}{\beta_0^2} C_F^2 n_f^2   \bigg\{
           \frac{80}{9} \omega
          - \frac{62}{9} \omega^2
          + \frac{80}{9} L_\omega
          \bigg\}

       + \frac{1}{\beta_0^2} C_F^2 C_A n_f   \bigg\{
          - \frac{536}{9} \omega
          + 16 \omega \zeta_2
          + \frac{473}{9} \omega^2
          - 8 \omega^2 \zeta_2
          - \frac{536}{9} L_\omega
          + 16 L_\omega \zeta_2
          \bigg\}

       + \frac{1}{\beta_0^2} C_F^3 n_f   \bigg\{
          - 2 \omega^2
          \bigg\}

       + \frac{1}{\beta_0} C_F n_f^2   \bigg\{ \frac{16}{27} \omega
          - \frac{8}{27} \omega^2  \bigg\}

       + \frac{1}{\beta_0} C_F C_A n_f   \bigg\{
           \frac{836}{27} \omega
          + \frac{112}{3} \omega \zeta_3
          - \frac{160}{9} \omega \zeta_2
          - \frac{418}{27} \omega^2
          - \frac{56}{3} \omega^2 \zeta_3
          + \frac{80}{9} \omega^2 \zeta_2
          - \frac{80}{3} L_\omega
          + \frac{40}{3} L_\omega L_{qr}
          - \frac{80}{3} L_\omega \gamma_E
          \bigg\}

       + \frac{1}{\beta_0} C_F C_A^2   \bigg\{
          - \frac{490}{3} \omega
          - \frac{88}{3} \omega \zeta_3
          + \frac{1072}{9} \omega \zeta_2
          - \frac{176}{5} \omega \zeta_2^2
          + \frac{245}{3} \omega^2
          + \frac{44}{3} \omega^2 \zeta_3
          - \frac{536}{9} \omega^2 \zeta_2
          + \frac{88}{5} \omega^2 \zeta_2^2
          + \frac{272}{3} L_\omega
          - \frac{136}{3} L_\omega L_{qr}
          + \frac{272}{3} L_\omega \gamma_E
          \bigg\}

       + \frac{1}{\beta_0} C_F^2 n_f   \bigg\{
           \frac{110}{3} \omega
          - 32 \omega \zeta_3
          - \frac{55}{3} \omega^2
          + 16 \omega^2 \zeta_3
          - 16 L_\omega
          + 8 L_\omega L_{qr}
          - 16 L_\omega \gamma_E
          \bigg\}

       + C_F n_f   \bigg\{
           \frac{352}{27}
          - \frac{88}{9} L_{qr}
          + \frac{4}{3} L_{qr}^2
          + \frac{40}{9} L_{fr}
          - \frac{80}{9} L_{fr} \omega
          + \frac{40}{9} L_{fr} \omega^2
          - \frac{4}{3} L_{fr}^2
          + \frac{8}{3} L_{fr}^2 \omega
          - \frac{4}{3} L_{fr}^2 \omega^2
          + \frac{176}{9} \gamma_E
          - \frac{16}{3} \gamma_E L_{qr}
          + \frac{16}{3} \gamma_E^2
          \bigg\}

       + C_F C_A   \bigg\{- \frac{2416}{27}
          + 28 \zeta_3
          + 16 \zeta_2
          + \frac{532}{9} L_{qr}
          - 8 L_{qr} \zeta_2
          - \frac{22}{3} L_{qr}^2
          - \frac{268}{9} L_{fr}
          + 8 L_{fr} \zeta_2
          + \frac{536}{9} L_{fr} \omega
          - 16 L_{fr} \omega \zeta_2
          - \frac{268}{9} L_{fr} \omega^2
          + 8 L_{fr} \omega^2 \zeta_2
          + \frac{22}{3} L_{fr}^2
          - \frac{44}{3} L_{fr}^2 \omega
          + \frac{22}{3} L_{fr}^2 \omega^2
          - \frac{1064}{9} \gamma_E
          + 16 \gamma_E \zeta_2
          + \frac{88}{3} \gamma_E L_{qr}
          - \frac{88}{3} \gamma_E^2  \bigg\} \bigg] \,,
\end{autobreak}    
\end{align}

\subsection{The NSV resummed exponents $h^q_{ij}$} \label{app:hN}
The resummation constants $h^q_{ij}$ in Eq. \ref{PsiNSVN}, till three-loop, are given by
\begin{align}
\begin{autobreak}

   h^q_{00}  =

        \frac{1}{ \beta_0} C_F   \bigg\{
          - 8 L_\omega
          \bigg\}\,,
\end{autobreak}  \\
\begin{autobreak}

   h^q_{01}  = 
         0\,,
\end{autobreak}  \\
\begin{autobreak}

   h^q_{10}  =

       \frac{1}{(1-\omega)} \bigg[ \frac{1}{ \beta_0^2} \beta_1 C_F   \bigg\{
          - 8 \omega
          - 8 L_\omega
          \bigg\}

       +\frac{1}{ \beta_0} C_F n_f   \bigg\{
          - \frac{80}{9} \omega
          \bigg\}

       +\frac{1}{ \beta_0} C_F C_A   \bigg\{\frac{536}{9} \omega- 16 \omega \zeta_2 \bigg\}

       +\frac{1}{ \beta_0} C_F^2   \bigg\{
          - 24 \omega
          + 32 \gamma_E \omega
          \bigg\}

       + C_F   \bigg\{
           8
          + 8 \omega
          - 8 L_{qr}
          + 8 L_{fr}
          - 8 L_{fr} \omega
          + 16 \gamma_E
          \bigg\} \bigg] \,,
\end{autobreak}  \\
\begin{autobreak}

   h^q_{11} =

       +\frac{1}{ \beta_0} C_F^2   \bigg\{
          32 \omega
          \bigg\}

       + \frac{1}{(1-\omega)^2} \bigg[ \frac{1}{\beta_0} C_F^2   \bigg\{
          -  4 \omega
          \bigg\} \bigg] \,,
\end{autobreak}  \\
\begin{autobreak}

   h^q_{20}  =

          \frac{1}{(1-\omega)^2} \bigg[  \frac{1}{ \beta_0^3} \beta_1^2 C_F   \bigg\{
          - 4 \omega^2
          + 4 L_\omega^2
          \bigg\}

       +\frac{1}{ \beta_0^2} \beta_2 C_F   \bigg\{
           4 \omega^2
          \bigg\}

       +\frac{1}{ \beta_0^2} \beta_1 C_F n_f   \bigg\{
           \frac{80}{9} \omega
          - \frac{40}{9} \omega^2
          + \frac{80}{9} L_\omega
          \bigg\}

       +\frac{1}{ \beta_0^2} \beta_1 C_F C_A   \bigg\{
          - \frac{536}{9} \omega
          + 16 \omega \zeta_2
          + \frac{268}{9} \omega^2
          - 8 \omega^2 \zeta_2
          - \frac{536}{9} L_\omega
          + 16 L_\omega \zeta_2
          \bigg\}

       +\frac{1}{ \beta_0^2} \beta_1 C_F^2   \bigg\{
           24 \omega
          - 12 \omega^2
          - 32 \gamma_E \omega
          + 16 \gamma_E \omega^2
          + 24 L_\omega
          - 32 L_\omega \gamma_E
          \bigg\}

       +\frac{1}{ \beta_0} C_F n_f^2   \bigg\{- \frac{32}{27} \omega
          + \frac{16}{27} \omega^2 \bigg\}

       +\frac{1}{ \beta_0} C_F C_A n_f   \bigg\{
          - \frac{1672}{27} \omega
          - \frac{224}{3} \omega \zeta_3
          + \frac{320}{9} \omega \zeta_2
          + \frac{836}{27} \omega^2
          + \frac{112}{3} \omega^2 \zeta_3
          - \frac{160}{9} \omega^2 \zeta_2
          \bigg\}

       +\frac{1}{ \beta_0} C_F C_A^2   \bigg\{
           \frac{980}{3} \omega
          + \frac{176}{3} \omega \zeta_3
          - \frac{2144}{9} \omega \zeta_2
          + \frac{352}{5} \omega \zeta_2^2
          - \frac{490}{3} \omega^2
          - \frac{88}{3} \omega^2 \zeta_3
          + \frac{1072}{9} \omega^2 \zeta_2
          - \frac{176}{5} \omega^2 \zeta_2^2
          \bigg\}

       +\frac{1}{ \beta_0} C_F^2 n_f   \bigg\{
          - 44 \omega
          + 64 \omega \zeta_3
          + \frac{64}{3} \omega \zeta_2
          + 22 \omega^2
          - 32 \omega^2 \zeta_3
          - \frac{32}{3} \omega^2 \zeta_2
          - \frac{640}{9} \gamma_E \omega
          + \frac{320}{9} \gamma_E \omega^2
          \bigg\}

       +\frac{1}{ \beta_0} C_F^2 C_A   \bigg\{
          - \frac{604}{3} \omega
          + 96 \omega \zeta_3
          - \frac{208}{3} \omega \zeta_2
          + \frac{302}{3} \omega^2
          - 48 \omega^2 \zeta_3
          + \frac{104}{3} \omega^2 \zeta_2
          + \frac{4288}{9} \gamma_E \omega
          - 128 \gamma_E \omega \zeta_2
          - \frac{2144}{9} \gamma_E \omega^2
          + 64 \gamma_E \omega^2 \zeta_2
          \bigg\}

       +\frac{1}{ \beta_0} C_F^3   \bigg\{
          - 12 \omega
          - 192 \omega \zeta_3
          + 96 \omega \zeta_2
          + 6 \omega^2
          + 96 \omega^2 \zeta_3
          - 48 \omega^2 \zeta_2
          \bigg\}

       +\frac{1}{ \beta_0} \beta_1 C_F   \bigg\{
          - 16 L_\omega
          + 8 L_\omega L_{qr}
          - 16 L_\omega \gamma_E
          \bigg\}

       + C_F n_f   \bigg\{
          - \frac{656}{27}
          + \frac{32}{3} \zeta_2
          - \frac{80}{9} \omega
          + \frac{40}{9} \omega^2
          + \frac{80}{9} L_{qr}
          - \frac{80}{9} L_{fr}
          + \frac{160}{9} L_{fr} \omega
          - \frac{80}{9} L_{fr} \omega^2
          - \frac{160}{9} \gamma_E
          \bigg\}

       + C_F C_A   \bigg\{
           \frac{2804}{27}
          - 56 \zeta_3
          - \frac{224}{3} \zeta_2
          + \frac{536}{9} \omega
          - 16 \omega \zeta_2
          - \frac{268}{9} \omega^2
          + 8 \omega^2 \zeta_2
          - \frac{536}{9} L_{qr}
          + 16 L_{qr} \zeta_2
          + \frac{536}{9} L_{fr}
          - 16 L_{fr} \zeta_2
          - \frac{1072}{9} L_{fr} \omega
          + 32 L_{fr} \omega \zeta_2
          + \frac{536}{9} L_{fr} \omega^2
          - 16 L_{fr} \omega^2 \zeta_2
          + \frac{892}{9} \gamma_E
          - 32 \gamma_E \zeta_2
          \bigg\}

       + C_F^2   \bigg\{
          - 8 \zeta_2
          + 24 L_{qr}
          - 24 L_{fr}
          + 48 L_{fr} \omega
          - 24 L_{fr} \omega^2
          - 28 \gamma_E
          - 32 \gamma_E L_{qr}
          + 32 \gamma_E L_{fr}
          - 64 \gamma_E L_{fr} \omega
          + 32 \gamma_E L_{fr} \omega^2
          + 56 \gamma_E^2
          \bigg\}

       + \beta_0 C_F   \bigg\{
           16 \zeta_2
          - 16 L_{qr}
          + 4 L_{qr}^2
          + 8 L_{fr}
          - 16 L_{fr} \omega
          + 8 L_{fr} \omega^2
          - 4 L_{fr}^2
          + 8 L_{fr}^2 \omega
          - 4 L_{fr}^2 \omega^2
          + 32 \gamma_E
          - 16 \gamma_E L_{qr}
          + 16 \gamma_E^2
          \bigg\} \bigg] \,,
\end{autobreak}  \\
\begin{autobreak}

   h^q_{21}  =

           \frac{1}{(1-\omega)^2} \bigg[  \frac{1}{ \beta_0^2} \beta_1 C_F^2   \bigg\{
          - 32 \omega
          + 16 \omega^2
          - 32 L_\omega
          \bigg\}

       +\frac{1}{ \beta_0} C_F^2 n_f   \bigg\{
          - \frac{640}{9} \omega
          + \frac{320}{9} \omega^2
          \bigg\}

       +\frac{1}{ \beta_0} C_F^2 C_A   \bigg\{
           \frac{4288}{9} \omega
          - 128 \omega \zeta_2
          - \frac{2144}{9} \omega^2
          + 64 \omega^2 \zeta_2
          \bigg\}

       + C_F C_A   \bigg\{
          - 20
          \bigg\}

       + C_F^2   \bigg\{
           20
          - 32 L_{qr}
          + 32 L_{fr}
          - 64 L_{fr} \omega
          + 32 L_{fr} \omega^2
          + 48 \gamma_E
          \bigg\} \bigg] \,,
\end{autobreak}  \\
\begin{autobreak}

   h^q_{22}  =

          \frac{1}{(1-\omega)^3} \bigg[  \frac{1}{ \beta_0} C_F^2 n_f   \bigg\{
          - \frac{32}{27} \omega
          \bigg\}

       +\frac{1}{ \beta_0} C_F^2 C_A   \bigg\{
           \frac{176}{27} \omega
          \bigg\} \bigg] \,.
\end{autobreak}  
\begin{autobreak}

\end{autobreak} 
\end{align}

%-----------------------------------
\section{Resummation Coefficients for the $\overline{N}$ exponentiation} \label{app:NBexp}
%%%%%%%%%%%%%%%%%%%%%%%%%%%%%%%%%%%%%%%%%%%%%%%%%%%%%%%%%%%%%%%%%%%%%%%%%%%%%
For the case of  $\overline{N}$ exponentiation, all the ${N}$-dependent resummed exponents namely $g^q_i(\overline{\omega})$, $\overline{g}^q_i(\overline{\omega})$, and $h^q_{i}(\overline{\omega})$ given in 
Eqs \ref{eq:gnb} and \ref{hnb}, respectively can be obtained from the corresponding exponents in standard $N$-approach through setting all the $\gamma_E$ terms to zero and replacing all the $\ln N$ terms by $\ln \overline{N}$ as well as all the $\mathcal{O}(1)$ $\omega$ terms by $\overline{\omega}$ as mentioned in 
Sec. \ref{sec:resummation}. The $N$-independent constants $\bar{\tilde {g}}^q_0$ given in Eq. (\ref{g0nb}) can be obtained from their counterparts in standard $N$-approach by simply putting the $\gamma_E$ terms equal to zero.

%-----------------------------------
\section{Resummation Coefficients for the {\textbf {\textit {Soft exponentiation}}}} \label{app:softexp}
%%%%%%%%%%%%%%%%%%%%%%%%%%%%%%%%%%%%%%%%%%%%%%%%%%%%%%%%%%%%%%%%%%%%%%%%%%%%%
For the case of {\it soft  exponentiation}, all the terms coming from the soft-collinear function $\rm \Phi_q$ are exponentiated and hence this means all the contribution to the finite ($N$-independent) piece from the soft-collinear function is also being exponentiated. The resummation coefficients for the {\it soft  exponentiation} denoted by $\Tilde{g}_{0i}^{q, \rm Soft}$ and $g_i^{q,\rm Soft}$ in Eqs (\ref{eq:g0bsoft}) and (\ref{eq:gnbsoft}), respectively can be obtained from their counterparts in standard $N$ scheme as described below.

The $N$-independent constants $\Tilde{g}_{0i}^{q, \rm Soft}$ in Eq. (\ref{eq:g0bsoft}) can be put in the following form:
\begin{align}
\Tilde{g}_{01}^{q,\rm Soft} &= \Tilde{g}_{01}^q +   \Delta^{q,\rm Soft}_{\Tilde{g}_{01}} \,,  \nn\\
\Tilde{g}_{02}^{q,\rm Soft} &= \Tilde{g}_{02}^q +   \Delta^{q,\rm Soft}_{\Tilde{g}_{02}} \,,  \nn\\
\Tilde{g}_{03}^{q,\rm Soft} &= \Tilde{g}_{03}^q +  \Delta^{q,\rm Soft}_{\Tilde{g}_{03}} \,,  
\end{align}
where the coefficients $ \Delta^{q,\rm Soft}_{\Tilde{g}_{0i}}$ are given by,

\begin{align}
    
\begin{autobreak}
 \Delta^{q,\rm Soft}_{\Tilde{g}_{01}} =

        C_F   \bigg\{
          - 16
          + 14 \zeta_2
          + 6 L_{qr}
          - 2 L_{qr}^2
          - 6 L_{fr}
          \bigg\}\,,
\end{autobreak}\\
\begin{autobreak}
   \Delta^{q,\rm Soft}_{\Tilde{g}_{02}} =

        C_F n_f   \bigg\{
           \frac{4085}{162}
          + \frac{4}{9} \zeta_3
          - \frac{182}{9} \zeta_2
          - \frac{418}{27} L_{qr}
          + \frac{16}{3} L_{qr} \zeta_2
          + \frac{38}{9} L_{qr}^2
          - \frac{4}{9} L_{qr}^3
          + \frac{2}{3} L_{fr}
          + \frac{16}{3} L_{fr} \zeta_2
          - 2 L_{fr}^2
          \bigg\}

       + C_F C_A   \bigg\{
          - \frac{51157}{324}
          + \frac{626}{9} \zeta_3
          + \frac{1061}{9} \zeta_2
          - \frac{32}{5} \zeta_2^2
          + \frac{2545}{27} L_{qr}
          - 52 L_{qr} \zeta_3
          - \frac{88}{3} L_{qr} \zeta_2
          - \frac{233}{9} L_{qr}^2
          + 4 L_{qr}^2 \zeta_2
          + \frac{22}{9} L_{qr}^3
          - \frac{17}{3} L_{fr}
          + 24 L_{fr} \zeta_3
          - \frac{88}{3} L_{fr} \zeta_2
          + 11 L_{fr}^2
          \bigg\}

       + C_F^2   \bigg\{
          \frac{511}{4}
          - 60 \zeta_3
          - 166 \zeta_2
          + \frac{402}{5} \zeta_2^2
          - 93 L_{qr}
          + 48 L_{qr} \zeta_3
          + 60 L_{qr} \zeta_2
          + 50 L_{qr}^2
          - 28 L_{qr}^2 \zeta_2
          - 12 L_{qr}^3
          + 2 L_{qr}^4
          + 93 L_{fr}
          - 48 L_{fr} \zeta_3
          - 60 L_{fr} \zeta_2
          - 36 L_{fr} L_{qr}
          + 12 L_{fr} L_{qr}^2
          + 18 L_{fr}^2
          \bigg\}\,,
\end{autobreak}\\
\begin{autobreak}
    \Delta^{q,\rm Soft}_{\Tilde{g}_{03}}=

        C_F N_4 n_{fv}   \bigg\{
           8
          - \frac{160}{3} \zeta_5
          + \frac{28}{3} \zeta_3
          + 20 \zeta_2
          - \frac{4}{5} \zeta_2^2
          \bigg\}

       + C_F n_f^2   \bigg\{  - \frac{190931}{6561}
          - \frac{832}{243} \zeta_3
          + \frac{3224}{81} \zeta_2
          + \frac{344}{135} \zeta_2^2
          + \frac{19676}{729} L_{qr}
          + \frac{32}{27} L_{qr} \zeta_3
          - \frac{608}{27} L_{qr} \zeta_2
          - \frac{812}{81} L_{qr}^2
          + \frac{32}{9} L_{qr}^2 \zeta_2
          + \frac{152}{81} L_{qr}^3
          - \frac{4}{27} L_{qr}^4
          + \frac{34}{9} L_{fr}
          + \frac{32}{9} L_{fr} \zeta_3
          - \frac{160}{27} L_{fr} \zeta_2
          + \frac{4}{9} L_{fr}^2
          + \frac{32}{9} L_{fr}^2 \zeta_2
          - \frac{8}{9} L_{fr}^3   \bigg\}

       + C_F C_A n_f   \bigg\{
           \frac{3400342}{6561}
          - \frac{8}{3} \zeta_5
          - \frac{8576}{27} \zeta_3
          - \frac{403498}{729} \zeta_2
          + \frac{296}{3} \zeta_2 \zeta_3
          - \frac{140}{27} \zeta_2^2
          - \frac{309838}{729} L_{qr}
          + \frac{1448}{9} L_{qr} \zeta_3
          + \frac{23336}{81} L_{qr} \zeta_2
          - \frac{392}{15} L_{qr} \zeta_2^2
          + \frac{11752}{81} L_{qr}^2
          - 16 L_{qr}^2 \zeta_3
          - 48 L_{qr}^2 \zeta_2
          - \frac{1948}{81} L_{qr}^3
          + \frac{16}{9} L_{qr}^3 \zeta_2
          + \frac{44}{27} L_{qr}^4
          - 40 L_{fr}
          - \frac{400}{9} L_{fr} \zeta_3
          + \frac{2672}{27} L_{fr} \zeta_2
          - \frac{8}{5} L_{fr} \zeta_2^2
          - \frac{146}{9} L_{fr}^2
          + 16 L_{fr}^2 \zeta_3
          - \frac{352}{9} L_{fr}^2 \zeta_2
          + \frac{88}{9} L_{fr}^3
          \bigg\}

       + C_F C_A^2   \bigg\{
          - \frac{51082685}{26244}
          - \frac{868}{9} \zeta_5
          + \frac{505087}{243} \zeta_3
          - \frac{2272}{9} \zeta_3^2
          + \frac{1193026}{729} \zeta_2
          - \frac{2336}{3} \zeta_2 \zeta_3
          - \frac{4303}{135} \zeta_2^2
          + \frac{38272}{945} \zeta_2^3
          + \frac{1045955}{729} L_{qr}
          + 272 L_{qr} \zeta_5
          - \frac{34928}{27} L_{qr} \zeta_3
          - \frac{68552}{81} L_{qr} \zeta_2
          + \frac{176}{3} L_{qr} \zeta_2 \zeta_3
          + \frac{340}{3} L_{qr} \zeta_2^2
          - \frac{37364}{81} L_{qr}^2
          + 176 L_{qr}^2 \zeta_3
          + \frac{1504}{9} L_{qr}^2 \zeta_2
          - \frac{88}{5} L_{qr}^2 \zeta_2^2
          + \frac{5738}{81} L_{qr}^3
          - \frac{88}{9} L_{qr}^3 \zeta_2
          - \frac{121}{27} L_{qr}^4
          + \frac{1657}{18} L_{fr}
          - 80 L_{fr} \zeta_5
          + \frac{3104}{9} L_{fr} \zeta_3
          - \frac{8992}{27} L_{fr} \zeta_2
          + 4 L_{fr} \zeta_2^2
          + \frac{493}{9} L_{fr}^2
          - 88 L_{fr}^2 \zeta_3
          + \frac{968}{9} L_{fr}^2 \zeta_2
          - \frac{242}{9} L_{fr}^3
          \bigg\}

       + C_F^2 n_f   \bigg\{
          - \frac{56963}{486}
          - \frac{832}{9} \zeta_5
          + \frac{26080}{81} \zeta_3
          + \frac{27410}{81} \zeta_2
          - \frac{296}{3} \zeta_2 \zeta_3
          - \frac{5852}{27} \zeta_2^2
          + \frac{6947}{27} L_{qr}
          - \frac{1208}{9} L_{qr} \zeta_3
          - \frac{8120}{27} L_{qr} \zeta_2
          + \frac{1328}{15} L_{qr} \zeta_2^2
          - \frac{14948}{81} L_{qr}^2
          + \frac{136}{9} L_{qr}^2 \zeta_3
          + \frac{1040}{9} L_{qr}^2 \zeta_2
          + \frac{1676}{27} L_{qr}^3
          - \frac{152}{9} L_{qr}^3 \zeta_2
          - \frac{100}{9} L_{qr}^4
          + \frac{8}{9} L_{qr}^5
          - \frac{3131}{27} L_{fr}
          + 88 L_{fr} \zeta_3
          + 32 L_{fr} \zeta_2
          + \frac{656}{15} L_{fr} \zeta_2^2
          + \frac{872}{9} L_{fr} L_{qr}
          - \frac{80}{3} L_{fr} L_{qr}^2
          - \frac{32}{3} L_{fr} L_{qr}^2 \zeta_2
          + \frac{8}{3} L_{fr} L_{qr}^3
          + 20 L_{fr}^2
          - 32 L_{fr}^2 \zeta_3
          - 44 L_{fr}^2 \zeta_2
          - 12 L_{fr}^2 L_{qr}
          + 4 L_{fr}^2 L_{qr}^2
          + 12 L_{fr}^3
          \bigg\}

       + C_F^2 C_A   \bigg\{
           \frac{824281}{324}
          - \frac{5512}{9} \zeta_5
          - \frac{52564}{27} \zeta_3
          + \frac{592}{3} \zeta_3^2
          - \frac{406507}{162} \zeta_2
          + \frac{3380}{3} \zeta_2 \zeta_3
          + \frac{184474}{135} \zeta_2^2
          - \frac{11824}{63} \zeta_2^3
          - \frac{14269}{6} L_{qr}
          + 240 L_{qr} \zeta_5
          + 2252 L_{qr} \zeta_3
          + \frac{48536}{27} L_{qr} \zeta_2
          - 696 L_{qr} \zeta_2 \zeta_3
          - \frac{6776}{15} L_{qr} \zeta_2^2
          + \frac{208099}{162} L_{qr}^2
          - \frac{5644}{9} L_{qr}^2 \zeta_3
          - \frac{6752}{9} L_{qr}^2 \zeta_2
          + \frac{344}{5} L_{qr}^2 \zeta_2^2
          - \frac{10340}{27} L_{qr}^3
          + 104 L_{qr}^3 \zeta_3
          + \frac{1052}{9} L_{qr}^3 \zeta_2
          + \frac{598}{9} L_{qr}^4
          - 8 L_{qr}^4 \zeta_2
          - \frac{44}{9} L_{qr}^5
          + \frac{25988}{27} L_{fr}
          - 240 L_{fr} \zeta_5
          - 1364 L_{fr} \zeta_3
          - 44 L_{fr} \zeta_2
          + 304 L_{fr} \zeta_2 \zeta_3
          - \frac{3608}{15} L_{fr} \zeta_2^2
          - \frac{5396}{9} L_{fr} L_{qr}
          + 456 L_{fr} L_{qr} \zeta_3
          + \frac{500}{3} L_{fr} L_{qr}^2
          - 48 L_{fr} L_{qr}^2 \zeta_3
          + \frac{104}{3} L_{fr} L_{qr}^2 \zeta_2
          - \frac{44}{3} L_{fr} L_{qr}^3
          - 131 L_{fr}^2
          + 32 L_{fr}^2 \zeta_3
          + 242 L_{fr}^2 \zeta_2
          + 66 L_{fr}^2 L_{qr}
          - 22 L_{fr}^2 L_{qr}^2
          - 66 L_{fr}^3
          \bigg\}

       + C_F^3   \bigg\{
          - \frac{5599}{6}
          + 1328 \zeta_5
          - 460 \zeta_3
          + 32 \zeta_3^2
          + \frac{4339}{6} \zeta_2
          - 280 \zeta_2 \zeta_3
          - \frac{4152}{5} \zeta_2^2
          + \frac{109612}{315} \zeta_2^3
          + \frac{1495}{2} L_{qr}
          - 480 L_{qr} \zeta_5
          - 992 L_{qr} \zeta_3
          - 534 L_{qr} \zeta_2
          + 608 L_{qr} \zeta_2 \zeta_3
          + \frac{1308}{5} L_{qr} \zeta_2^2
          - \frac{1051}{2} L_{qr}^2
          + 408 L_{qr}^2 \zeta_3
          + 440 L_{qr}^2 \zeta_2
          - \frac{804}{5} L_{qr}^2 \zeta_2^2
          + 222 L_{qr}^3
          - 96 L_{qr}^3 \zeta_3
          - 120 L_{qr}^3 \zeta_2
          - 68 L_{qr}^4
          + 28 L_{qr}^4 \zeta_2
          + 12 L_{qr}^5
          - \frac{4}{3} L_{qr}^6
          - \frac{1495}{2} L_{fr}
          + 480 L_{fr} \zeta_5
          + 992 L_{fr} \zeta_3
          + 534 L_{fr} \zeta_2
          - 608 L_{fr} \zeta_2 \zeta_3
          - \frac{1308}{5} L_{fr} \zeta_2^2
          + 540 L_{fr} L_{qr}
          - 576 L_{fr} L_{qr} \zeta_3
          - 216 L_{fr} L_{qr} \zeta_2
          - 294 L_{fr} L_{qr}^2
          + 96 L_{fr} L_{qr}^2 \zeta_3
          + 120 L_{fr} L_{qr}^2 \zeta_2
          + 72 L_{fr} L_{qr}^3
          - 12 L_{fr} L_{qr}^4
          - 270 L_{fr}^2
          + 288 L_{fr}^2 \zeta_3
          + 108 L_{fr}^2 \zeta_2
          + 108 L_{fr}^2 L_{qr}
          - 36 L_{fr}^2 L_{qr}^2
          - 36 L_{fr}^3
          \bigg\}\,.
          \end{autobreak}
          \end{align}

%As a result, the $\Tilde{g}_0^q$ constants and the resummed exponents $g_i^q$ in the Standard $N$-approach change. We write the changes in $g_i^{q, \rm Soft}$ below in terms of the  exponents in Standard $N$, 
%
The $N$-dependent coefficients $g_i^{q,\rm Soft}$ in Eq. (\ref{eq:gnbsoft}) can be obtained as follows:

\begin{align}
g_1^{q,\rm Soft} &= g_1^q \,,  \nn\\
g_2^{q,\rm Soft} &= g_2^q + \as ~\Delta^{q,\rm Soft}_{g_2} \,,  \nn\\
g_3^{q,\rm Soft} &= g_3^q + \as~ \Delta^{q,\rm Soft}_{g_3} \,,  \nn\\
%g_4^{q,\rm Soft} &= g_4^q + \as ~\Delta^{q,\rm Soft}_{g_4} \,,  \nn\\
\end{align}

where the coefficients $\Delta^{q,\rm Soft}_{g_i}$ are given as, 
\begin{align}
    
\begin{autobreak}
   \Delta^{\rm Soft}_{g_2} =

        C_F   \bigg\{
            2 \zeta_2
          + 2 L_{qr}^2
          - 8 \gamma_E L_{qr}
          + 8 \gamma_E L_{fr}
          + 8 \gamma_E^2
          \bigg\}\,,
\end{autobreak}\\
\begin{autobreak}
   \Delta^{\rm Soft}_{g_3} =

        C_F n_f   \bigg\{
          - \frac{328}{81}
          + \frac{4}{9} \zeta_3
          - \frac{10}{9} \zeta_2
          + \frac{112}{27} L_{qr}
          - \frac{20}{9} L_{qr}^2
          + \frac{4}{9} L_{qr}^3
          - \frac{224}{27} \gamma_E
          + \frac{80}{9} \gamma_E L_{qr}
          - \frac{8}{3} \gamma_E L_{qr}^2
          - \frac{80}{9} \gamma_E L_{fr}
          + \frac{8}{3} \gamma_E L_{fr}^2
          - \frac{80}{9} \gamma_E^2
          + \frac{16}{3} \gamma_E^2 L_{qr}
          - \frac{32}{9} \gamma_E^3
          \bigg\}

       + C_F C_A   \bigg\{
           \frac{2428}{81}
          - \frac{22}{9} \zeta_3
          + \frac{67}{9} \zeta_2
          - 12 \zeta_2^2
          - \frac{808}{27} L_{qr}
          + 28 L_{qr} \zeta_3
          + \frac{134}{9} L_{qr}^2
          - 4 L_{qr}^2 \zeta_2
          - \frac{22}{9} L_{qr}^3
          + \frac{1616}{27} \gamma_E
          - 56 \gamma_E \zeta_3
          - \frac{536}{9} \gamma_E L_{qr}
          + 16 \gamma_E L_{qr} \zeta_2
          + \frac{44}{3} \gamma_E L_{qr}^2
          + \frac{536}{9} \gamma_E L_{fr}
          - 16 \gamma_E L_{fr} \zeta_2
          - \frac{44}{3} \gamma_E L_{fr}^2
          + \frac{536}{9} \gamma_E^2
          - 16 \gamma_E^2 \zeta_2
          - \frac{88}{3} \gamma_E^2 L_{qr}
          + \frac{176}{9} \gamma_E^3
          \bigg\}\,.
\end{autobreak}
\begin{autobreak}
\end{autobreak}
\end{align}

%--------------------------------------------------------------------

%%%%%%%%%%%%%%%%%%%%%%%%%%%%%%%%%%%%%%%%%%%%%%%%%%%%%%%%%%%%%%%%%%%%%%%%%%%%%
\section{Resummation Coefficients for the {\textbf {\textit {All exponentiation}}}} \label{app:Allexp}
%%%%%%%%%%%%%%%%%%%%%%%%%%%%%%%%%%%%%%%%%%%%%%%%%%%%%%%%%%%%%%%%%%%%%%%%%%%%%
For the case of {\it All exponentiation}, the complete $\Tilde{g}_0^q$ is being exponentiated along with the large-$N$ pieces. This brings into modification only for the resummed exponent compared to the `Standard ${N}$ exponentiation'. We write the modified resummed exponents denoted by $g_i^{q, \rm All}$ in Eq. (\ref{gnall}) in terms of exponents in standard ${N}$  as,
\begin{align}
g_1^{q,\rm All} &= g_1^q \,,  \nn\\
g_2^{q,\rm All} &= g_2^q +  \as~  \Delta^{q,\rm All}_{g_2} \,,  \nn\\
g_3^{q,\rm All} &= g_3^q +  \as~  \Delta^{q,\rm All}_{g_3} \,,  \nn\\
%g_4^{q,\rm All} &= g_4^q +  \as ~ \Delta^{q,\rm All}_{g_4} \,,
\end{align}
where $ \Delta^{q,\rm All}_{g_i}$ terms are found from exponentiating the complete $\Tilde{g}_0^q$ prefactor and they are given as,
\begin{align}
  \Delta^{q,\rm All}_{g_2} &= \Tilde{g}_{01}^{q} \,, \nn \\
   \Delta^{q,\rm All}_{g_3} &=  \bigg(-\frac{(\Tilde{g}_{01}^{q})^{2}}{2} +  \Tilde{g}_{02}^{q} \bigg) \,. %\nn\\
%   \Delta^{q,\rm All}_{g_4} &= \bigg(\frac{(\Tilde{g}_{01}^{q})^{3}}{3} - \Tilde{g}_{01}^{q} \Tilde{g}_{02}^{q} +  \Tilde{g
%   }_{03}^{q} \bigg) \,,
\end{align}
where the coefficients $\Tilde{g}_{0i}^q$ are given in \ref{app:g0t}.

%%%References
\bibliographystyle{JHEP}
\bibliography{nsvdy_resum}
\end{document}